\documentclass{aa} 
\usepackage{graphicx}
\usepackage{graphics}
\usepackage{amssymb}
\usepackage{txfonts}
\usepackage{array}
\usepackage{ulem}
\usepackage{longtable,lscape}
\usepackage{array} 
\usepackage{tabularx}
\usepackage{natbib}
\bibpunct{(}{)}{;}{a}{}{,}   
\begin{document}
\title{A  near-infrared  catalogue of  the Galactic  novae in  the VVV
  survey area\thanks{Based on observations  taken within the ESO VISTA
    Public Survey VVV, Programme ID 179.B-2002}}

\author{
     R.~K.~Saito\inst{1,2,3}
\and D.~Minniti\inst{1,3,4,5}
\and R.~Angeloni\inst{1,3,6}
\and M.~Catelan\inst{1,3}
\and J.~C.~Beamin\inst{1,3}
\and J.~Borissova\inst{2,3}
\and I.~D\'ek\'any\inst{1,3}
\and E.~Kerins\inst{7}
\and R.~Kurtev\inst{2}
\and R. E. Mennickent\inst{8} 
}

\offprints{R. K. Saito: rsaito@astro.puc.cl} 

\institute{Departamento  de Astronom\'ia  y  Astrof\'isica, Pontificia
  Universidad  Cat\'olica  de  Chile,  Av.   Vicu\~na  Mackenna  4860,
  782-0436 Macul, Santiago, Chile
\and Departamento  de   F\'{i}sica  y  Astronom\'{i}a,   Facultad  de
  Ciencias, Universidad  de Valpara\'{i}so, Ave.  Gran Breta\~na 1111,
  Playa Ancha, Casilla 5030, Valpara\'{i}so, Chile
\and The Milky Way Millennium Nucleus, Av. Vicu\~{n}a Mackenna 4860,
  782-0436 Macul, Santiago, Chile
\and Vatican Observatory, Vatican City State V-00120, Italy
\and  Departamento  de  Ciencias  Fisicas, Universidad  Andres  Bello,
Av. Republica 252, Santiago, Chile
\and    Department    of    Electrical   Engineering,    Center    for
Astro-Engineering,  Pontificia Universidad  Cat\'olica  de Chile,  Av.
Vicu\~na Mackenna 4860, 782-0436 Macul, Santiago, Chile
\and Jodrell  Bank Centre for Astrophysics,  University of Manchester,
Oxford Road, Manchester M13 9PL, UK
\and  Departamento  de  Astronom\'{i}a, Universidad  de  Concepci\'on,
Casilla 160-C, Concepci\'on, Chile
}
  
   \date{Received ; Accepted }


\abstract
{Near-infrared  data  of Classical  Novae  contain useful  information
  about the ejected  gas mass and the thermal  emission by dust formed
  during  eruption, and  provide independent  methods to  classify the
  objects according to the colour of their progenitors, and the fading
  rate and features  seen after eruption.  The VISTA  Variables in the
  V\'{i}a L\'actea survey (VVV) is a near-IR ESO Public Survey mapping
  the Milky Way  bulge and southern plane.  Data  taken during $2010 -
  2011$ covered the  entire area in the $JHK_{\rm  s}$ bands plus some
  epochs in $K_{\rm s}$-band of the ongoing VVV variability campaign.}
{We  used the  VVV data  to create  a near-IR  catalogue of  the known
  Galactic novae  in the 562  sq. deg. area  covered by VVV.   We also
  compiled the information about  novae from the variability tables of
  the VVV variability campaign.}
{We used the novae list  provided by VSX/AAVSO catalogue to search for
  all objects within  the VVV area.  From the 140  novae, we were able
  to  retrieve the  $JHK_{\rm  s}$  colours of  93  objects.  We  also
  checked in the ongoing VVV variability campaign for the light-curves
  of novae that erupted in the last years.}
{ The VVV near-IR catalogue of novae contains $JHK_{\rm s}$ photometry
  of 93 objects  completed as of December 2012.  VVV allows to monitor
  objects within up to $\Delta K_{\rm s}\sim10$~mag range.  VVV images
  can also  be used to discover  and study novae by  searching for the
  expanding shell.  Since objects  are seen at different distances and
  reddening  levels, the  colour-magnitude and  colour-colour diagrams
  show the novae spread in magnitude as well as in colour.  Dereddened
  colours and reddening-free indices were used with caution and cannot
  be  a good approach  in all  cases since  the distance  and spectral
  features prevent  more conclusive results for  some extreme objects.
  Light-curves for some recent novae are presented.}
{Thanks  to its  high spatial  resolution  in the  near-IR, and  large
  $K_{\rm s}$-range, the VVV survey can be a major contributor for the
  search and  study of novae  in the most crowded  and high-extinction
  regions of the Milky Way. The VVV survey area contains $\sim35\%$ of
  all known novae in the Galaxy.}

\authorrunning{Saito et al.}  
\titlerunning{A  near-IR catalogue of  the Galactic  novae in  the VVV
  survey area}
\maketitle
%

\section{Introduction}\label{intro}

Cataclysmic variable stars (CVs)  are close binary systems composed by
a  late-type star  which transfers  material to  a more  massive white
dwarf  (WD) companion.   The typical  mass transfer  rates in  CVs are
$10^{-8}  -  10^{-11}$  M$_{\odot}$  yr$^{-1}$,  and  the  progressive
accretion of  hydrogen-rich material from the secondary  star onto the
hot surface  of the WD can  lead to a thermonuclear  runaway, in which
the accreted material is expelled from the system.  This phenomenon is
a            so             called            nova            eruption
\citep[e.g.,][]{2002apa..book.....F,2003cvs..book.....W}.

During  a  nova eruption  the  object  suddenly  rises in  brightness,
becoming typically  8-15 magnitudes brighter than  its progenitor. The
recurrence  between nova  outbursts is  expected to  be $10^3  - 10^6$
years,  and depends  on  the mass  of  the white  dwarf  and the  mass
transfer  rate.  The systems  are classified  as Classical  Novae when
just one nova eruption was  recorded. However, this is strongly biased
by  ancient  records, which  are  limited  to  a few  thousand  years.
Systems with two  or more registered nova eruptions  are classified as
Recurrent Novae, in general  associated with a high-mass transfer rate
system onto  a primary star with  the mass close  to the Chandrasekhar
limit \citep{2003cvs..book.....W}.

Since novae are usually found in eruption, the novae classification is
in general based  on the features seen during this  phase, such as the
fading    (speed    class)    rate    after    the    nova    eruption
\citep[e.g.,][]{1957gano.book.....G}  and   the  spectral  differences
\citep[e.g.,][]{1992AJ....104..725W}.  

More recently,  alternative classifications are  proposed based either
on   the    Galactic   component,   with   disk    and   bulge   novae
\citep[e.g.,][]{1998ApJ...506..818D}, or using automated selection and
classification        based        on        optical        photometry
\citep{2004MNRAS.353..571D}.

The  infrared light  can also  contain  the thermal  emission by  dust
formed  during the  shell ejection.   Models for  thermonuclear runway
predict the presence  of material from the white  dwarf in the ejecta,
thus its  spectroscopic analysis allows to distinguish  between CO and
ONe  white  dwarfs  \citep{2011ApJ...727...50S}.   The  importance  of
observing  Galactic  novae  in   the  IR  spectral  region  was  fully
understood when it was realized  that they can represent a significant
contributor to the interestellar medium  (at least on local scales) of
highly processed material, thus playing an active role in the chemical
evolution  of  the   Galaxy.   Furthermore,  IR  observations  provide
independent  methods  for determining  the  ejected  gas  mass, a  key
parameter      of      every      thermonuclear     runaway      model
\citep{1998ApJ...494..783M}.

IR data can also help in the determination both of the global Galactic
nova rate and the separate Galactic bulge and disk nova rates. Current
Galactic  nova  samples, obtained  mostly  from  optical surveys,  are
biased towards  regions of lower optical extinction  and therefore are
incomplete within both the bulge  and disk towards the Galactic plane.
Previous determinations  of the Galactic  nova rate therefore  vary by
more   than  an  order   of  magnitude,   from  11   to  260~yr$^{-1}$
\citep{1997ApJ...487..226S}.    The  nova  rate   in  M31   is  better
constrained from optical studies \citep[e.g., ][]{2006MNRAS.369..257D}
but  it remains  unclear whether  the rates  within both  galaxies are
consistent  with fixed  bulge and  disk nova  rates per  unit galactic
luminosity.   Again  this uncertainty  is  due  mostly  to the  poorly
constrained   knowledge  of   the  Galactic   rate.   A   large  scale
near-infrared  time domain  survey can  allow novae  to  be identified
throughout the Galactic bulge and disk and therefore facilitate a more
complete  determination   of  the   nova  rate  within   each  stellar
population.

The  VISTA Variables in  the V\'ia  L\'actea is  a ESO  Public near-IR
survey scanning the  Milky Way (MW) bulge and  southern plane, in five
near-IR bands ($ZYJHK_{\rm s}$),  plus a variability campaign of about
100  epochs   in  the  $K_{\rm  s}$-band  spanning   over  many  years
\citep{2010NewA...15..433M}.  VVV is about four magnitudes deeper than
previous IR  surveys, and thanks  to its higher spatial  resolution in
the  near-IR,  enables  deep  observations  in the  most  crowded  and
high-extinction       regions       of       the       Milky       Way
\citep[e.g.,][]{2012A&A...537A.107S}.

Here we present a catalogue with the $JHK_{\rm s}$ colours of 93 novae
in the VVV area. We discuss the colour properties and features seen in
the  near-IR.   We  also  present  the  first  results  from  the  VVV
variability campaign, including light-curves  for some novae.  VVV can
be a  major contributor for  the discovery and  study of novae  in the
innermost regions of the Galaxy,  objects that are beyond detection in
the current novae searches.

\begin{figure*} 
\includegraphics[bb=4cm 0cm 15cm 21cm,angle=-90,scale=0.7]{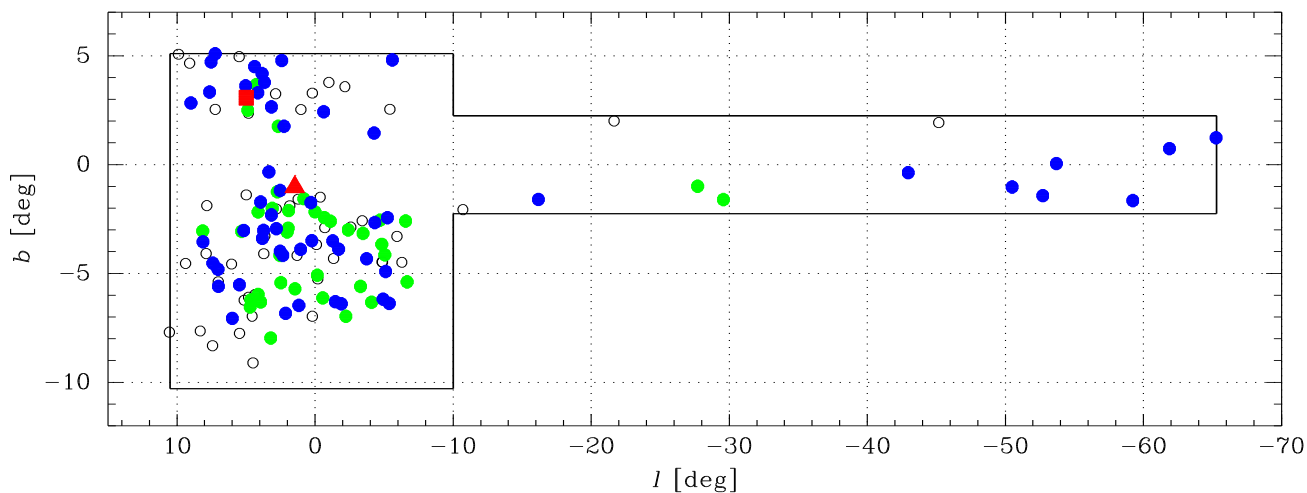}
 \caption{Spatial distribution of all  known Galactic novae in the VVV
   survey area.  A  total of 140 objects are  present.  Filled symbols
   mark the objects  detected in the VVV data  while open symbols mark
   the  novae with no  detection in  our data.   The blue  circles are
   novae  with  the  coordinates   matching  a  single  source  within
   $1\arcsec$ from the position given by the catalogue.  Green circles
   mark the novae with doubtful  photometry (see text). The red square
   is  Nova Sgr  2012 progenitor,  while  the red  triangle marks  the
   position of  Nova Sgr 2010b,  caught during eruption.  We  note the
   presence of  a ``zone  of avoidance'' on  the Galactic  plane, with
   just a few objects belonging  to the high-extinction regions of the
   MW.}
\label{fig:area}
\end{figure*}

\begin{figure} 
\includegraphics[bb=3cm 2cm 18cm 21cm,angle=-90,scale=0.41]{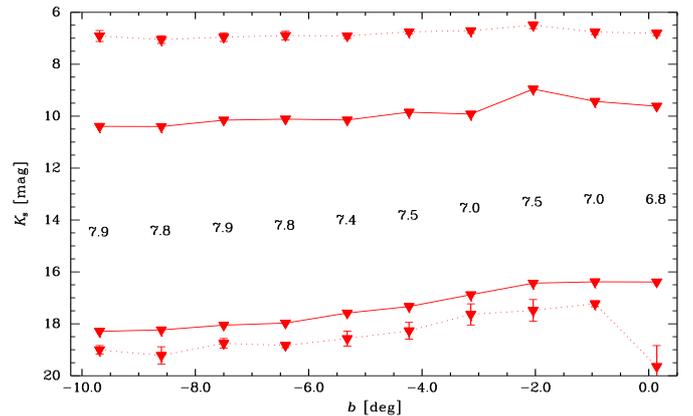}
 \caption{Magnitude  range   covered  by  the   VVV  $K_{\rm  s}$-band
   variability  campaign observations along  the Galactic  minor axis.
   The  solid lines  with triangles  mark  the limits  covered by  the
   stellar sources, with the value  of the total magnitude range shown
   in  each  position.   The  increasing sky  brightness  towards  the
   Galactic  centre is  due  to the  contribution  of the  underlying,
   unresolved faint stars.  The upper dotted line with triangles shows
   the mean value of  the brightest sources, including objects flagged
   as  saturated.  Similarly,  the  lower dotted  line with  triangles
   marks the limiting magnitude  of all sources, including those below
   the 5-$\sigma$ limit.}
\label{fig:maglim}
\end{figure}

\section{Observations}\label{obs}

VVV is an ESO Public Survey  observing the MW bulge and inner southern
plane in the near-IR with the VISTA Telescope located at Cerro Paranal
Observatory        in       Chile        \citep[][and       references
  therein]{2010NewA...15..433M}.   VVV  observes   an  area  of  about
562~deg$^2$, within $-10.0^\circ  \lesssim l \lesssim +10.5^\circ$ and
$-10.3^\circ \lesssim b \lesssim  +5.1^\circ$ in the bulge, and within
$294.7^\circ  \lesssim   l  \lesssim  350.0^\circ$   and  $-2.25^\circ
\lesssim    b    \lesssim    +2.25^\circ$    in   the    plane    (see
Fig.~\ref{fig:area}).   The  whole bulge  and  disk  areas were  fully
observed   in  five   near-IR  passbands   ($ZYJHK_{\rm   s}$)  during
2010--2011, while  a variability campaign  in the $K_{\rm  s}-$band is
ongoing,   with   $\sim$100   pointings   planned  over   many   years
\citep{2012A&A...537A.107S}.

Photometric catalogues  for the VVV  observations are provided  by the
Cambridge     Astronomical     Survey    Unit     (CASU)\footnote{{\tt
    http://casu.ast.cam.ac.uk/vistasp/}}.  The  catalogues contain the
positions, fluxes, and some shape measurements obtained from different
apertures,  with a  flag  indicating the  most probable  morphological
classification. The VVV data are  in the natural VISTA Vegamag system,
with the photometric calibration  in $JHK_{\rm s}$ performed using the
VISTA magnitudes of unsaturated 2MASS stars present in the images. The
single-band  CASU catalogues for  each tile  were matched,  creating a
$JHK_{\rm s}$  catalogue of more than  173 million sources  in the VVV
bulge area \citep{2012A&A...544A.147S},  and about 148 million sources
in the  disk.  A  similar procedure was  adopted to match  the $K_{\rm
  s}$-band catalogues from the variability campaign in order to create
variability tables for selected fields.

Different observing  strategies between the  bulge and disk  area, and
effects  caused by  crowding and  extinction cause  both  the limiting
magnitude   and    saturation   to    vary   along   the    VVV   area
\citep{2012A&A...537A.107S,2012A&A...544A.147S}.   For  instance,  the
$K_{\rm  s}$-band  limiting magnitude  on  the  outer  bulge is  about
18.5~mag,  while   in  the  Galactic  center  the   limit  is  $K_{\rm
  s}\sim16.5$~mag  (see Fig.~\ref{fig:maglim}).  The  saturation limit
varies in a shorter range.  The $JHK_{\rm s}$ observations for a given
field  were  taken in  the  same  ``observational  block'' (OB)  which
guarantees quasi-simultaneous  observations in the  three bands (there
is a  time gap of only  $\sim190$~s between each band).   On the other
hand, due to multiple scheduling  constraints on the side of the VISTA
telescope operations, there is a  little control on the cadence of the
$K_{\rm  s}$-band variability  data within  a given  season,  with the
number of epochs planned in  a given year being observed together with
the other VISTA surveys according with the surveys requirements on the
weather and visibility, for instance.

\begin{figure*} 
\includegraphics[bb=0cm -1.0cm 19cm 21cm,angle=-90,scale=0.65]{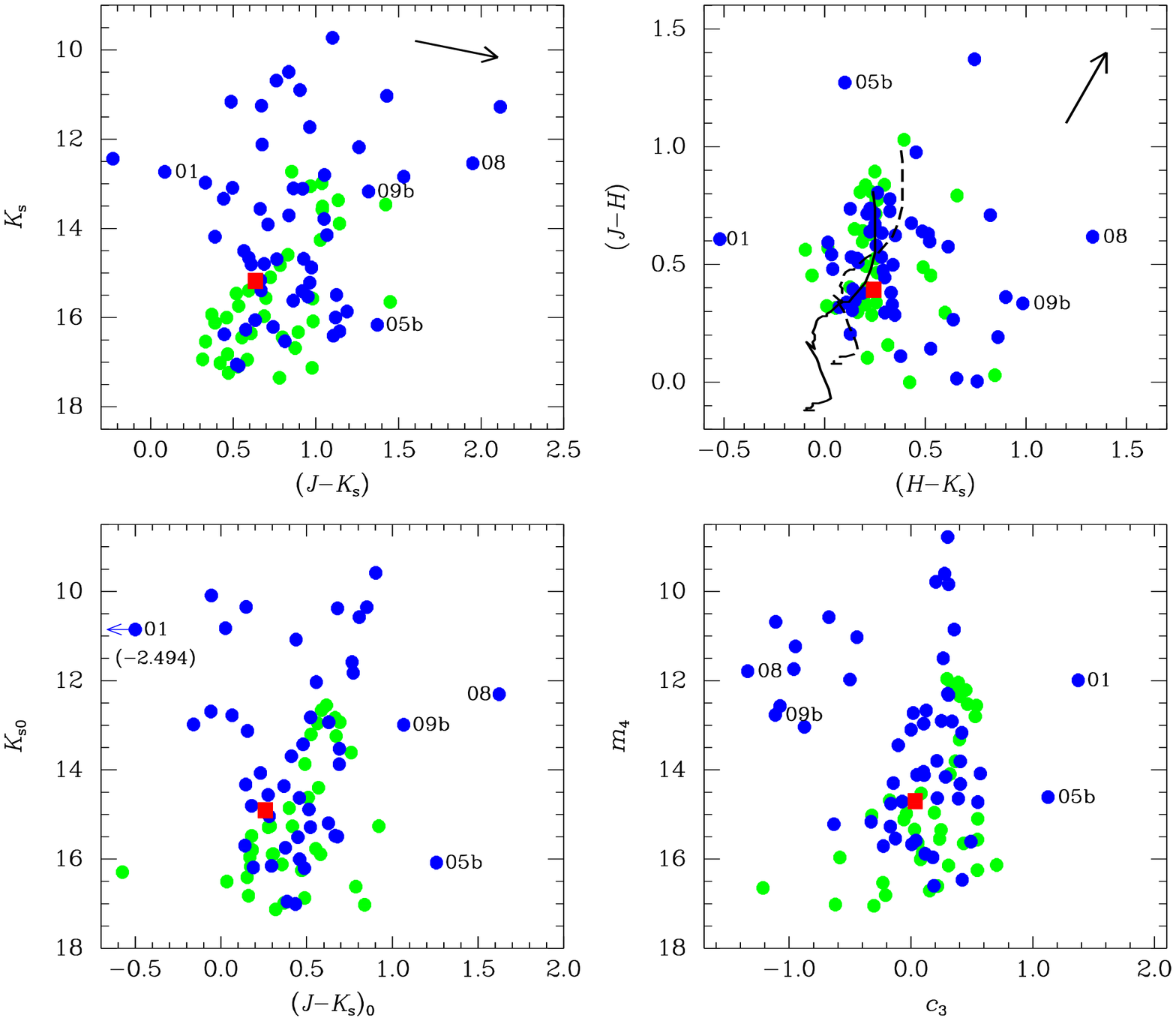}
 \caption{Top-left  panel: $K_{\rm s}~vs.~(J-K_{\rm  s})$ CMD  for all
   novae detected in the VVV area.  The colour pattern and symbols are
   the  same as  used  in Fig.~\ref{fig:area}.   Top-right panel:  the
   corresponding  $  (J-H)~vs.~(H-K_{\rm  s})$ colour-colour  diagram.
   Bottom-left:  dereddened  CMD  for   all  bulge  Novae,  using  the
   reddening  maps  of  \cite{2012A&A...543A..13G}  and  assuming  the
   \cite{1989ApJ...345..245C}  extinction  law   (no  disk  novae  are
   plotted).   Bottom-right: CMD  using the  reddening-free parameters
   provided by \cite{2011rrls.conf..145C}. The colour of main sequence
   stars is shown in the $ (J-H)~vs.~(H-K_{\rm s})$ CCD in the case of
   $A_{Ks}=0$~mag (no  extinction, solid line)  and for $A_{Ks}=0.243$
   \citep[the median value of extinction among our targets, shown as a
     dashed line; adapted from][]{2001ApJ...558..309D}.  The reddening
   vector associated  with an extinction  of $E(B-V)=1$, based  on the
   relative  extinctions  of  the  VISTA  filters,  and  assuming  the
   \cite{1989ApJ...345..245C} extinction law, is also shown in the top
   panels. Four objects are labelled in the plots: Nova Sgr 2001, Nova
   Sgr 2005b, Nova Sgr 2008, and Nova Sgr 2009b.}
\label{fig:cmd}
\end{figure*}

\section{The VVV Novae Catalogue}\label{sec:cat}

Our novae  list was taken  from The International Variable  Star Index
(VSX),  provided   by  the  American  Association   of  Variable  Star
Observers\footnote{{\tt  http://www.aavso.org/vsx/index.php}} (AAVSO).
In this catalogue  there are about 400 known novae  in the Galaxy, 140
of  them lie  within the  VVV area.   Fig.~\ref{fig:area}  shows their
spatial  distribution,  with  most  objects concentrated  towards  the
Galactic bulge.   We note  that the information  about some  novae are
from historical  records or amateur astronomers,  with the coordinates
in general taken from observations while the objects were in eruption.
Thus, in order to certify  the coordinates with enough accuracy during
the  quiescent  state,  we  checked  for all  objects  in  the  SIMBAD
database\footnote{{\tt   simbad.u-strasbg.fr/simbad/}}.   Finally,  we
defined  as our  coordinates  the ICRS  J2000  RA and  DEC taken  from
SIMBAD, which were used as entries in our search for the VVV $JHK_{\rm
  s}$ photometry and composite colour images.

\begin{landscape}
\begin{table}
  \caption[]{Previously known  novae in the VVV area  (completed as of
    December 2012).  We adopted  the novae type designations described
    by VSX  database$^1$: Novae (N),  subdivided into fast  (NA), slow
    (NB),  and  very slow  (NC)  categories.   The coordinates  (epoch
    J2000) were taken from Simbad$^2$. The distance between the Simbad
    coordinates and the VVV source is presented.  Values of extinction
    ($A_{\rm  Ks}$)  are for  a  2  arcmin  region around  the  target
    position, taken from  the maps of \cite{2012A&A...543A..13G}.  The
    abbreviations used  in the last column are  MS (multiple sources),
    ND  (no detection) and  ST (saturated  source).  The  ``:'' symbol
    stands for doubtful sources (see text).}
\label{table:novae}
\centering
\small
\begin{tabular}{lllcccccccccl}
\hline \hline
\noalign{\smallskip}
Nova    & Other        & Nova  &  RA            &  DEC         &   RA (VVV)    &  DEC (VVV)     & Dist.      &$J$     &   $H$  & $K_{\rm s}$ & $A_{\rm Ks}$  & Note \\
(Year)   & Designation  & Type  &  [hh:mm:ss.ss] & [dd:mm:ss.s] & [hh:mm:ss.ss] & [dd:mm:ss.ss] & [$\arcsec$] &  [mag] &  [mag] & [mag]      & [mag]  & \\
\noalign{\smallskip}                                                                                                    
\hline                                                                                                                  
\noalign{\smallskip}                                                                                                    
Sgr 1893   & \object{V5557 Sgr}  & N    & 18:01:43.15 & $-$35:39:27.8 & 18:01:43.14  & $-$35:39:28.95 & 1.155  & 17.528$\pm$0.076 & 16.959$\pm$0.093 & 16.943$\pm$0.154 & 0.070 & MS \\
Sgr 1897   & LQ Sgr              & NA   & 18:28:28.91 & $-$27:55:19.2 & ---	     &  ---   	      &        &                  &                  &                  &       & ST \\
Sgr 1899   & \object{V1016 Sgr}  & NA   & 18:19:57.63 & $-$25:11:14.6 & 18:19:57.63  & $-$25:11:14.52 & 0.089  & 11.329$\pm$0.010 & 10.749$\pm$0.010 & 10.493$\pm$0.010 & 0.113 & \\
Sco 1901   & V0382 Sco           & NA   & 17:51:56.13 & $-$35:25:05.4 & ---          & ---            &        &                  &                  &                  &       & ND \\ 
Sgr 1901   & V1014 Sgr           & NB   & 18:06:45.70 & $-$27:26:15.0 & ---          & ---            &        &                  &                  &                  &       & ND \\ 
Sgr 1905   & \object{V1015 Sgr}  & NA   & 18:09:02.00 & $-$32:28:32.0 & 18:09:02.06  & $-$32:28:30.84 & 1.407  & 17.250$\pm$0.063 & 17.147$\pm$0.102 & 16.935$\pm$0.154 & 0.112 & MS: \\
Cir 1906   & \object{AR Cir}     & NB   & 14:48:09.53 & $-$60:00:27.5 & 14:48:09.52  & $-$60:00:27.69 & 0.204  & 12.211$\pm$0.010 & 12.701$\pm$0.010 & 12.439$\pm$0.010 &  ---  & \\
Sco 1906   & \object{V0711 Sco}  & NB   & 17:54:06.16 & $-$34:21:15.5 & 17:54:06.16  & $-$34:21:15.55 & 0.070  & 13.967$\pm$0.010 & 13.328$\pm$0.010 & 13.103$\pm$0.010 & 0.171 & \\
Sgr 1910   & V0999 Sgr           & NB   & 18:00:05.59 & $-$27:33:14.0 & ---          & ---            &        &                  &                  &                  &       & ND \\ 
Sgr 1914c  & V1012 Sgr           & NA   & 18:06:14.00 & $-$31:44:27.0 & 18:06:14.10  & ---            &        &                  &                  &                  &       & ND \\
Sgr 1917   & \object{BS Sgr}     & NB:  & 18:26:46.72 & $-$27:08:19.9 & 18:26:46.72  & $-$27:08:19.84 & 0.061  & 12.795$\pm$0.010 & 12.286$\pm$0.010 & 12.120$\pm$0.010 & 0.087 & \\
Sgr 1919   & V1017 Sgr           & NA+UG& 18:32:04.17 & $-$29:23:12.6 & ---          & ---            &        &                  &                  &                  &       & ND \\
Sco 1922   & V0707 Sco           & N    & 17:48:26.38 & $-$36:37:54.9 & ---          & ---  	      &        &                  &                  &                  &       & ND \\ 
Sgr 1924   & \object{FL Sgr}     & NA   & 18:00:30.20 & $-$34:36:13.0 & 18:00:30.27  & $-$34:36:11.99 & 1.353  & 17.710$\pm$0.159 & 17.553$\pm$0.242 & 17.239$\pm$0.260 & 0.111 & MS: \\
Sgr 1926   & \object{FM Sgr}     & NA   & 18:17:18.10 & $-$23:38:27.0 & 18:17:18.05  & $-$23:38:27.30 & 0.787  & 16.819$\pm$0.082 & 16.481$\pm$0.115 & 16.373$\pm$0.144 & 0.187 & MS \\
Sgr 1926a  & \object{KY Sgr}     & NA   & 18:01:21.02 & $-$26:24:40.0 & 18:01:21.02  & $-$26:24:39.95 & 0.073  & 11.804$\pm$0.010 & 11.539$\pm$0.010 & 10.900$\pm$0.010 & 0.551 & \\
Sco 1928   & KP Sco              & NA   & 17:44:16.47 & $-$35:43:23.6 & ---          & ---            &        &                  &                  &                  &       & ND \\ 
Sgr 1928   & \object{V1583 Sgr}  & NA   & 18:15:26.30 & $-$23:23:18.0 & 18:15:26.27  & $-$23:23:18.95 & 1.020  & 16.998$\pm$0.094 & 16.637$\pm$0.129 & 16.445$\pm$0.151 & 0.277 & MS: \\ 
Sgr 1930   & V0441 Sgr           & NA   & 18:22:08.09 & $-$25:28:53.7 & ---          & ---     	      &        &                  &                  &                  &       & ND  \\
Cen 1931   & \object{MT Cen}     & NA   & 11:44:00.80 & $-$60:33:39.5 & 11:44:00.80  & $-$60:33:39.55 & 0.050  & 15.607$\pm$0.010 & 14.892$\pm$0.010 & 14.681$\pm$0.010 &  ---  & \\
Sgr 1932   & V1905 Sgr           & NA   & 18:33:42.10 & $-$25:20:42.0 & ---	     & ---     	      &        &                  &                  &                  &       & ND \\ 
Sgr 1933   & \object{V0737 Sgr}  & N    & 18:07:08.66 & $-$28:44:52.3 & 18:07:08.71  & $-$28:44:52.24 & 0.621  & 15.483$\pm$0.036 & 14.959$\pm$0.037 & 14.796$\pm$0.041 & 0.167 & MS \\
Cru 1935   & \object{AP Cru}     & NA/DQ& 12:31:20.45 & $-$64:26:25.2 & 12:31:20.44  & $-$64:26:25.23 & 0.083  & 15.852$\pm$0.011 & 15.229$\pm$0.011 & 14.878$\pm$0.015 &  ---  & \\
Sco 1935   & \object{V0744 Sco}  & N:   & 17:53:18.08 & $-$31:13:35.1 & 17:53:18.17  & $-$31:13:35.37 & 1.210  & 17.098$\pm$0.201 & 16.306$\pm$0.189 & 15.648$\pm$0.135 & 0.385 & MS:\\
Sgr 1936   & \object{V0732 Sgr}  & NA   & 17:56:07.51 & $-$27:22:16.1 & 17:56:07.51  & $-$27:22:16.36 & 0.261  & 13.391$\pm$0.010 & 12.020$\pm$0.010 & 11.275$\pm$0.010 & 0.922 & \\
Sgr 1936b  & \object{V0726 Sgr}  & NA   & 18:19:33.69 & $-$26:53:19.6 & 18:19:33.73  & $-$26:53:19.66 & 0.591  & 16.484$\pm$0.048 & 15.748$\pm$0.045 & 15.621$\pm$0.059 & 0.143 & \\
Sgr 1936c  & \object{V0630 Sgr}  & NA   & 18:08:48.25 & $-$34:20:21.4 & 18:08:48.39  & $-$34:20:21.70 & 1.723  & 17.557$\pm$0.051 & 17.528$\pm$0.109 & 16.682$\pm$0.106 & 0.064 & : \\
Sgr 1936d  & V0990 Sgr           & NA   & 17:57:19.00 & $-$28:19:07.0 & ---          & ---            &        &                  &                  &                  &       & ND \\
Sgr 1937   & \object{V0787 Sgr}  & NA   & 18:00:02.20 & $-$30:30:31.0 & 18:00:02.18  & $-$30:30:30.22 & 0.821  & 15.415$\pm$0.054 & 14.822$\pm$0.052 & 14.806$\pm$0.067 & 0.243 & MS \\
Oph 1940   & \object{V0553 Oph}  & NA   & 17:42:53.50 & $-$24:51:26.2 & 17:42:53.46  & $-$24:51:25.45 & 0.951  & 16.480$\pm$0.093 & 15.756$\pm$0.102 & 15.528$\pm$0.118 & 0.488 & MS \\
Sco 1941   & \object{V0697 Sco}  & NA/DQ& 17:51:21.83 & $-$37:24:55.2 & 17:51:21.91  & $-$37:24:56.91 & 1.934  & 16.870$\pm$0.047 & 16.547$\pm$0.068 & 16.537$\pm$0.109 & 0.131 & : \\
Sgr 1943   & \object{V1148 Sgr}  & N    & 18:09:05.85 & $-$25:59:08.0 & 18:09:05.85  & $-$25:59:08.06 & 0.065  & 13.308$\pm$0.010 & 13.103$\pm$0.010 & 12.976$\pm$0.010 & 0.285 & \\
Sco 1944   & \object{V0696 Sco}  & NA   & 17:53:11.56 & $-$35:50:14.4 & 17:53:11.55  & $-$35:50:14.41 & 0.109  & 14.577$\pm$0.010 & 14.259$\pm$0.010 & 14.187$\pm$0.016 & 0.116 & \\
Sgr 1944   & \object{V0927 Sgr}  & NA   & 18:07:42.70 & $-$33:21:17.0 & 18:07:42.72  & $-$33:21:17.77 & 0.805  & 16.846$\pm$0.042 & 16.304$\pm$0.049 & 16.270$\pm$0.082 & 0.065 & MS \\
Sgr 1945a  & \object{V1149 Sgr}  & N    & 18:18:30.40 & $-$28:17:17.0 & 18:18:30.37  & $-$28:17:18.35 & 1.389  & 17.626$\pm$0.102 & 17.223$\pm$0.127 & 17.098$\pm$0.191 & 0.116 & MS:\\
Sgr 1945b  & \object{V1431 Sgr}  & N:   & 18:03:29.61 & $-$29:59:54.6 & 18:03:29.55  & $-$29:59:54.84 & 0.852  & 12.692$\pm$0.010 & 11.975$\pm$0.010 & 11.729$\pm$0.010 & 0.144 & \\
Sgr 1947a  & V0928 Sgr           & N    & 18:18:59.30 & $-$28:06:00.0 & ---          & ---            &        &                  &                  &                  &       & ND \\
Sgr 1948   & V1150 Sgr           & N    & 18:18:55.40 & $-$24:05:32.0 & ---          & ---            &        &                  &                  &                  &       & ND \\
Sco 1949   & V0902 Sco           & N:   & 17:26:08.30 & $-$39:04:05.0 & ---          & ---            &        &                  &                  &                  &       & ND \\ 
Oph 1950   & \object{V2110 Oph}  & NC   & 17:43:33.32 & $-$22:45:37.0 & 17:43:33.32  & $-$22:45:36.95 & 0.074  & 11.647$\pm$0.010 & 11.537$\pm$0.010 & 11.160$\pm$0.010 & 0.336 & ST \\ 

\noalign{\smallskip}
\hline
\end{tabular}
\tablefoottext{1}{{\tt http://www.aavso.org/}}
\tablefoottext{2}{{\tt http://simbad.u-strasbg.fr/simbad/}}
\end{table}
\end{landscape}

\addtocounter{table}{-1}

\begin{landscape}
\begin{table}
\caption[]{{\bf cont.} Previously known  novae in the VVV area.}
\label{tab:tiles}
\centering
\small
\begin{tabular}{lllcccccccccl}
\hline \hline
\noalign{\smallskip}
Nova    & Other        & Nova  &  RA            &  DEC         &   RA (VVV)    &  DEC (VVV)     & Dist.      &$J$     &   $H$  & $K_{\rm s}$ & $A_{\rm Ks}$  & Note \\
(Year)   & Designation  & Type  &  [hh:mm:ss.ss] & [dd:mm:ss.s] & [hh:mm:ss.ss] & [dd:mm:ss.ss] & [$\arcsec$] &  [mag] &  [mag] & [mag]      & [mag]  & \\
\noalign{\smallskip}                                                                                                    
\hline                                                                                                                  
\noalign{\smallskip}
Sco 1950a  & \object{V0719 Sco} & NA & 17:45:43.90 & $-$34:00:55.0 & 17:45:43.85  & $-$34:00:55.41 & 0.724 & 16.179$\pm$0.070 & 15.442$\pm$0.067 & 15.216$\pm$0.068 & 0.328 & MS: \\
Sco 1950b  & V0720 Sco          & NA & 17:51:58.00 & $-$35:23:22.0 & ---          & ---		   &       &                  &                  &                  &       & ND \\ 
Sco 1950c  & \object{V0721 Sco} & N  & 17:42:29.10 & $-$34:40:41.5 & 17:42:29.10  & $-$34:40:41.60 & 0.103 & 13.587$\pm$0.010 & 13.241$\pm$0.010 & 13.091$\pm$0.012 & 0.314 & \\
Sgr 1951a  & \object{V1172 Sgr} & N  & 17:50:23.66 & $-$20:40:29.9 & 17:50:23.65  & $-$20:40:29.88 & 0.114 & 12.460$\pm$0.010 & 11.484$\pm$0.010 & 11.030$\pm$0.010 & 0.454 & \\
Sgr 1951b  & \object{V2415 Sgr} & N: & 17:53:12.00 & $-$29:34:25.0 & 17:53:11.99  & $-$29:34:24.39 & 0.629 & 17.053$\pm$0.293 & 16.478$\pm$0.329 & 15.864$\pm$0.242 & 0.371 & \\
Sco 1952   & ---                & n  & 17:47:38.00 & $-$33:11:55.0 & ---          & ---            &       &                  &                  &                  &       & ND \\ 
Sco 1952a & \object{V0722 Sco}  & NA & 17:48:37.00 & $-$34:57:50.0 & 17:48:36.88  & $-$34:57:50.78 & 1.679 & 15.607$\pm$0.034 & 15.011$\pm$0.033 & 14.823$\pm$0.037 & 0.201 & MS:\\
Sco 1952b & \object{V0723 Sco}  & NA & 17:50:05.20 & $-$35:23:58.0 & 17:50:05.32  & $-$35:23:56.79 & 1.939 & 15.422$\pm$0.028 & 14.780$\pm$0.029 & 14.591$\pm$0.030 & 0.191 & MS:\\
Sgr 1952a & \object{V1175 Sgr}  & NA & 18:14:17.00 & $-$31:07:08.0 & 18:14:17.01  & $-$31:07:08.42 & 0.432 & 17.614$\pm$0.097 & 17.220$\pm$0.120 & 17.080$\pm$0.178 & 0.072 & \\
Sgr 1952b & \object{V1174 Sgr}  & N  & 18:01:37.00 & $-$28:44:24.0 & 18:01:37.07  & $-$28:44:22.94 & 1.439 & 14.021$\pm$0.015 & 13.307$\pm$0.013 & 13.053$\pm$0.013 & 0.220 & :\\
Sgr 1953  & NSV 10158           & N: & 18:05:12.00 & $-$29:54:24.0 & ---          & ---            &       &                  &                  &                  &       & ND \\ 
Oph 1954  & V0908 Oph           & N  & 17:28:41.00 & $-$27:45:24.0 & ---          & ---            &       &                  &                  &                  &       & ND \\
Sco 1954  & \object{NSV 9808}   & N: & 17:53:40.00 & $-$30:45:32.0 & 17:53:40.08  & $-$30:45:31.48 & 1.177 & 15.288$\pm$0.056 & 14.516$\pm$0.056 & 14.261$\pm$0.057 & 0.391 & :\\
Sgr 1954a & V1274 Sgr           & NB & 17:48:55.00 & $-$17:51:55.0 & ---          & ---            &       &                  &                  &                  &       & ND \\
Sgr 1954b & \object{V1275 Sgr}  & NA & 17:59:06.35 & $-$36:18:40.8 & 17:59:06.34  & $-$36:18:40.80 & 0.169 & 14.225$\pm$0.010 & 13.694$\pm$0.010 & 13.562$\pm$0.010 & 0.134 & \\
Sgr 1955  & \object{V1572 Sgr}  & N  & 18:05:37.00 & $-$31:37:42.0 & 18:05:36.87  & $-$31:37:43.02 & 1.894 & 16.506$\pm$0.053 & 16.053$\pm$0.064 & 16.116$\pm$0.106 & 0.162 & :\\
Oph 1957  & \object{V0972 Oph}  & NB & 17:34:43.81 & $-$28:10:35.8 & 17:34:43.87  & $-$28:10:36.34 & 0.927 & 17.450$\pm$0.229 & 16.821$\pm$0.287 & 16.306$\pm$0.262 & 0.560 & \\
Sgr 1960  & \object{V1944 Sgr}  & N  & 18:00:36.90 & $-$27:17:13.0 & 18:00:36.98  & $-$27:17:13.70 & 1.320 & 14.550$\pm$0.027 & 13.743$\pm$0.025 & 13.510$\pm$0.026 & 0.268 & MS: \\
Oph 1961  & \object{V1012 Oph}  & N  & 17:41:34.38 & $-$23:23:33.1 & 17:41:34.43  & $-$23:23:32.34 & 1.063 & 17.284$\pm$0.149 & 16.722$\pm$0.167 & 16.818$\pm$0.249 & 0.315 & :\\
Sgr 1963  & NSV 9828            & N: & 17:54:43.00 & $-$28:41:41.0 & ---          & ---            &       &                  &                  &                  &       & ND \\
Sco 1964  & \object{V0825 Sco}  & N  & 17:49:53.65 & $-$33:32:13.7 & 17:49:53.73  & $-$33:32:13.13 & 1.194 & 16.259$\pm$0.076 & 15.766$\pm$0.096 & 15.561$\pm$0.101 & 0.298 & :\\
Sgr 1968  & \object{V4027 Sgr}  & NB & 18:02:29.21 & $-$28:45:19.9 & 18:02:29.18  & $-$28:45:21.23 & 1.403 & 13.582$\pm$0.010 & 12.954$\pm$0.010 & 12.728$\pm$0.010 & 0.174 & :\\
Sgr 1974  & V3888 Sgr           & NA & 17:48:41.07 & $-$18:45:37.0 & ---          & ---            &       &                  &                  &                  &       & ND\\
Sgr 1975  & \object{V3889 Sgr}  & NA & 17:58:21.32 & $-$28:21:52.7 & 17:58:21.46  & $-$28:21:53.20 & 1.927 & 15.035$\pm$0.042 & 14.141$\pm$0.036 & 13.892$\pm$0.037 & 0.279 & MS:\\
Sgr 1977  & V4021 Sgr           & NA & 18:38:14.26 & $-$23:22:47.0 & ---	  & ---	           &       &                  & 	         &                  &       & ND\\
Sgr 1978  & \object{V4049 Sgr}  & N  & 18:20:38.06 & $-$27:56:26.0 & 18:20:37.99  & $-$27:56:24.59 & 1.673 & 16.460$\pm$0.036 & 16.163$\pm$0.050 & 16.000$\pm$0.073 & 0.113 & :\\
Sgr 1980  & \object{V4065 Sgr}  & NA & 18:19:38.20 & $-$24:43:55.9 & 18:19:38.22  & $-$24:43:55.87 & 0.260 & 10.829$\pm$0.010 & 10.052$\pm$0.010 &  9.728$\pm$0.010 & 0.144 & ST \\
Sgr 1982  & V4077 Sgr           & N  & 18:34:39.41 & $-$26:26:02.9 & ---          & ---            &       &                  &                  &                  &       & ND \\
Nor 1983  & \object{V0341 Nor}  & NA & 16:13:44.40 & $-$53:19:08.0 & 16:13:44.23  & $-$53:19:07.13 & 1.754 & 18.129$\pm$0.133 & 17.603$\pm$0.153 & 17.349$\pm$0.181 &  ---  & : \\
Oph 1983  & \object{V0794 Oph}  & NB & 17:38:49.25 & $-$22:50:48.9 & 17:38:49.25  & $-$22:50:49.00 & 0.114 & 14.031$\pm$0.010 & 13.361$\pm$0.010 & 13.111$\pm$0.010 & 0.289 & \\
Sgr 1983  & \object{V4121 Sgr}  & N  & 18:07:54.77 & $-$28:49:26.8 & 18:07:54.90  & $-$28:49:26.45 & 1.769 & 15.980$\pm$0.057 & 15.695$\pm$0.072 & 15.461$\pm$0.074 & 0.177 & MS: \\
Sgr 1984  & \object{V4092 Sgr}  & NA & 17:53:41.99 & $-$29:02:08.4 & 17:53:41.91  & $-$29:02:08.73 & 1.099 & 14.618$\pm$0.035 & 13.782$\pm$0.033 & 13.579$\pm$0.034 & 0.374 & MS: \\
Sco 1985  & \object{V0960 Sco}  & N  & 17:56:34.14 & $-$31:49:36.3 & 17:56:34.17  & $-$31:49:35.55 & 0.830 & 15.216$\pm$0.037 & 14.412$\pm$0.033 & 14.149$\pm$0.034 & 0.274 & MS\\
Sgr 1986  & \object{V4579 Sgr}  & NB & 18:03:37.89 & $-$28:00:08.9 & 18:03:37.90  & $-$28:00:08.65 & 0.296 & 13.778$\pm$0.012 & 13.473$\pm$0.015 & 13.336$\pm$0.017 & 0.209 & \\
Sgr 1987  & V4135 Sgr           & N  & 17:59:45.08 & $-$32:16:20.8 & ---          & ---            &       &                  &                  &                  &       & ND \\ 
Sco 1989b & \object{V0977 Sco}  & N  & 17:51:50.25 & $-$32:31:58.0 & 17:51:50.35  & $-$32:31:57.59 & 1.389 & 14.024$\pm$0.012 & 13.250$\pm$0.011 & 12.988$\pm$0.012 & 0.327 & :\\
Cen 1991  & \object{V0868 Cen}  & NA & 13:50:10.60 & $-$63:08:52.0 & 13:50:10.69  & $-$63:08:51.94 & 0.619 & 17.513$\pm$0.032 & 16.838$\pm$0.041 & 16.407$\pm$0.064 &  ---  & \\
Oph 1991b & \object{V2290 Oph}  & NA & 17:43:05.50 & $-$20:07:00.0 & 17:43:05.47  & $-$20:07:00.59 & 0.734 & 16.951$\pm$0.057 & 16.507$\pm$0.082 & 16.209$\pm$0.100 & 0.206 & \\
Sgr 1991  & V4160 Sgr           & NA & 18:14:13.83 & $-$32:12:28.5 & ---	  & ---	           &       &                  & 	         &                  &       & ND \\
Sco 1992  & \object{V0992 Sco}  & NA & 17:07:17.44 & $-$43:15:22.0 & 17:07:17.45  & $-$43:15:21.86 & 0.199 & 15.455$\pm$0.010 & 14.981$\pm$0.014 & 14.691$\pm$0.019 &  ---  & \\
\noalign{\smallskip}
\hline
\end{tabular}
\end{table}
\end{landscape}

\addtocounter{table}{-1}

\begin{landscape}
\begin{table}
\caption[]{{\bf cont.} Previously known  novae in the VVV area.}
\label{tab:tiles}
\centering
\small
\begin{tabular}{lllcccccccccl}
\hline \hline
\noalign{\smallskip}
Nova    & Other        & Nova  &  RA            &  DEC         &   RA (VVV)    &  DEC (VVV)     & Dist.      &$J$     &   $H$  & $K_{\rm s}$ & $A_{\rm Ks}$  & Note \\
(Year)   & Designation  & Type  &  [hh:mm:ss.ss] & [dd:mm:ss.s] & [hh:mm:ss.ss] & [dd:mm:ss.ss] & [$\arcsec$] &  [mag] &  [mag] & [mag]      & [mag]  & \\
\noalign{\smallskip}                                                                                                    
\hline                                                                                                                  
\noalign{\smallskip} 
Sgr 1992a & \object{V4157 Sgr} & NA & 18:09:34.90 & $-$25:51:58.0 & 18:09:34.81  & $-$25:51:56.55 & 1.900 & 15.820$\pm$0.045 & 15.355$\pm$0.053 & 15.095$\pm$0.057 & 0.238 & MS: \\
Sgr 1992b & V4169 Sgr          & NA & 18:23:26.94 & $-$28:21:59.7 & ---		 & ---      	  &       &                  &                  &                  &       & ND \\
Sgr 1992c & V4171 Sgr          & NA & 18:23:41.34 & $-$22:59:28.7 & ---          & ---            &       &                  &                  &                  &       & ND \\
Sgr 1993  & \object{V4327 Sgr} & NA & 18:12:49.82 & $-$29:29:04.9 & 18:12:49.70  & $-$29:29:04.38 & 1.613 & 16.960$\pm$0.085 & 16.560$\pm$0.112 & 16.352$\pm$0.136 & 0.100 & : \\
Cru 1996  & \object{CP Cru}    & N+E& 12:10:31.39 & $-$61:45:09.5 & 12:10:31.36  & $-$61:45:09.75 & 0.315 & 16.617$\pm$0.016 & 15.977$\pm$0.016 & 15.492$\pm$0.021 &  ---  & \\
Sco 1997  & \object{V1141 Sco} & NA & 17:54:11.33 & $-$30:02:53.0 & 17:54:11.37  & $-$30:02:51.51 & 1.600 & 16.272$\pm$0.136 & 15.943$\pm$0.192 & 15.739$\pm$0.204 & 0.258 & MS: \\
Sco 1998  & V1142 Sco          & NA & 17:55:24.99 & $-$31:01:41.5 & ---		 & ---      	  &       &                  & 	                &                  &       & ND \\
Sgr 1998  & V4633 Sgr          & NA & 18:21:40.49 & $-$27:31:37.3 & ---		 & ---      	  &       &                  & 	                &                  &       & ND \\
Sgr 1999  & \object{V4444 Sgr} & NA & 18:07:36.20 & $-$27:20:13.2 & 18:07:36.18  & $-$27:20:13.92 & 0.787 & 16.323$\pm$0.094 & 15.690$\pm$0.091 & 15.406$\pm$0.089 & 0.211 & \\
Sgr 2000  & \object{V4642 Sgr} & NA & 17:55:09.84 & $-$19:46:01.0 & 17:55:09.89  & $-$19:46:00.34 & 0.976 & 17.343$\pm$0.117 & 16.812$\pm$0.145 & 16.530$\pm$0.170 & 0.378 & \\
Sco 2001  & \object{V1178 Sco} & NA & 17:57:06.98 & $-$32:23:05.0 & 17:57:06.92  & $-$32:23:05.32 & 0.801 & 14.838$\pm$0.019 & 14.112$\pm$0.016 & 13.788$\pm$0.017 & 0.261 & MS\\
Sgr 2001  & \object{V4643 Sgr} & NA & 17:54:40.46 & $-$26:14:15.2 & 17:54:40.47  & $-$26:14:14.30 & 0.499 & 12.819$\pm$0.001 & 12.212$\pm$0.010 & 12.733$\pm$0.010 & 1.880 & \\
Sgr 2001b & \object{V4739 Sgr} & NA & 18:24:46.04 & $-$30:00:41.1 & 18:24:46.11  & $-$30:00:39.45 & 1.929 & 16.654$\pm$0.029 & 16.145$\pm$0.038 & 15.967$\pm$0.065 & 0.077 & MS: \\
Sgr 2001c & \object{V4740 Sgr} & NA & 18:11:45.82 & $-$30:30:49.9 & 18:11:45.71  & $-$30:30:50.14 & 1.377 & 18.103$\pm$0.239 & 17.615$\pm$0.257 & 17.126$\pm$0.249 & 0.101 & MS: \\
Sgr 2002  & V4741 Sgr          & NA & 17:59:59.63 & $-$30:53:20.5 & ---          & ---            &       &                  &                  &                  &       & ND \\
Sgr 2002b & V4742 Sgr          & NA & 18:02:21.86 & $-$25:20:32.2 & ---          & ---            &       &                  &                  &                  &       & ND \\
Sgr 2002d & \object{V4744 Sgr} & NA & 17:47:21.74 & $-$23:28:23.1 & 17:47:21.79  & $-$23:28:24.49 & 1.544 & 17.066$\pm$0.140 & 16.259$\pm$0.129 & 16.083$\pm$0.150 & 0.313 & MS:\\
Sgr 2003b & V5113 Sgr          & NA & 18:10:10.42 & $-$27:45:35.2 & ---          & ---            &       &                  &                  &                  &       & ND \\
Oph 2004  & \object{V2574 Oph} & NA & 17:38:45.49 & $-$23:28:18.5 & 17:38:45.52  & $-$23:28:18.55 & 0.415 & 15.825$\pm$0.028 & 15.496$\pm$0.040 & 15.160$\pm$0.046 & 0.354 & MS\\
Sco 2004  & \object{V1186 Sco} & NA & 17:12:51.21 & $-$30:56:37.2 & 17:12:51.24  & $-$30:56:37.56 & 0.509 & 14.622$\pm$0.010 & 14.242$\pm$0.013 & 13.912$\pm$0.015 & 0.218 & \\
Sco 2004b & \object{V1187 Sco} & NA & 17:29:18.81 & $-$31:46:01.5 & 17:29:18.84  & $-$31:46:01.35 & 0.363 & 14.545$\pm$0.010 & 14.047$\pm$0.013 & 13.708$\pm$0.018 & 0.727 & \\
Sgr 2004  & \object{V5114 Sgr} & NA & 18:19:32.29 & $-$28:36:35.7 & 18:19:32.38  & $-$28:36:36.57 & 1.502 & 16.302$\pm$0.031 & 15.987$\pm$0.043 & 15.932$\pm$0.069 & 0.136 & : \\
Cen 2005  & \object{V1047 Cen} & N  & 13:20:49.74 & $-$62:37:50.5 & 13:20:49.79  & $-$62:37:50.63 & 0.387 & 14.373$\pm$0.010 & 13.664$\pm$0.010 & 12.841$\pm$0.010 &  ---  & \\
Nor 2005  & \object{V0382 Nor} & NA & 16:19:44.74 & $-$51:34:53.1 & 16:19:44.79  & $-$51:34:52.09 & 1.124 & 16.554$\pm$0.028 & 16.101$\pm$0.039 & 15.574$\pm$0.049 &  ---  & \\
Sco 2005  & \object{V1188 Sco} & NA & 17:44:21.59 & $-$34:16:35.7 & 17:44:21.58  & $-$34:16:33.87 & 1.836 & 17.239$\pm$0.174 & 16.589$\pm$0.180 & 16.441$\pm$0.193 & 0.322 & MS: \\
Sgr 2005a & V5115 Sgr          & NA & 18:16:59.04 & $-$25:56:38.8 & ---		 & ---      	  &       &                  & 	                &                  &       & ND \\
Sgr 2005b & \object{V5116 Sgr} & NA & 18:17:50.77 & $-$30:26:31.2 & 18:17:50.75  & $-$30:26:31.86 & 0.708 & 17.535$\pm$0.088 & 16.263$\pm$0.050 & 16.163$\pm$0.081 & 0.084 & \\
Oph 2006  & \object{V2575 Oph} & N  & 17:33:13.06 & $-$24:21:07.1 & 17:33:13.05  & $-$24:21:07.05 & 0.132 & 17.117$\pm$0.072 & 16.520$\pm$0.091 & 15.998$\pm$0.098 & 0.488 & \\
Sgr 2006  & \object{V5117 Sgr} & NA & 17:58:52.61 & $-$36:47:35.1 & 17:58:52.57  & $-$36:47:34.33 & 0.883 & 17.572$\pm$0.068 & 17.092$\pm$0.090 & 17.052$\pm$0.154 & 0.098 & MS: \\
Oph 2007  & \object{V2615 Oph} & NA & 17:42:44.00 & $-$23:40:35.1 & 17:42:44.00  & $-$23:40:35.16 & 0.059 & 16.689$\pm$0.088 & 16.404$\pm$0.127 & 16.056$\pm$0.131 & 0.359 & \\
Nor 2007  & V0390 Nor          & NA & 16:32:11.51 & $-$45:09:13.4 & ---		 & ---      	  &       &                  & 	                &                  &       & ND \\
Oph 2008a & \object{V2670 Oph} & NA & 17:39:50.94 & $-$23:50:01.0 & 17:39:50.92  & $-$23:50:00.88 & 0.274 & 13.442$\pm$0.010 & 13.081$\pm$0.010 & 12.181$\pm$0.010 & 0.356 & \\
Oph 2008b & V2671 Oph          & N  & 17:33:29.67 & $-$27:01:16.4 & ---		 & ---      	  &       &                  & 	                &                  &       & ND \\
Sgr 2008  & \object{V5579 Sgr} & NA & 18:05:58.88 & $-$27:13:56.0 & 18:05:58.89  & $-$27:13:55.91 & 0.129 & 14.490$\pm$0.018 & 13.873$\pm$0.018 & 12.541$\pm$0.010 & 0.237 & \\
Sgr 2008b & \object{V5580 Sgr} & N  & 18:22:01.39 & $-$28:02:39.8 & 18:22:01.48  & $-$28:02:39.73 & 1.153 & 15.988$\pm$0.024 & 15.650$\pm$0.033 & 15.394$\pm$0.048 & 0.129 & :\\
Oph 2009  & V2672 Oph          & NA & 17:38:19.72 & $-$26:44:13.7 & ---		 & ---      	  &       &                  & 	                &                  &       & ND \\
Sgr 2009a & \object{V5581 Sgr} & N: & 17:44:08.44 & $-$26:05:48.7 & 17:44:08.46  & $-$26:05:47.78 & 0.948 & 11.448$\pm$0.010 & 11.445$\pm$0.010 & 10.687$\pm$0.010 & 0.595 & ST\\
Sgr 2009b & \object{V5582 Sgr} & N  & 17:45:05.40 & $-$20:03:21.5 & 17:45:05.42  & $-$20:03:21.43 & 0.263 & 14.493$\pm$0.010 & 14.159$\pm$0.010 & 13.174$\pm$0.010 & 0.184 & \\
Sgr 2009c & \object{V5583 Sgr} & NA & 18:07:07.67 & $-$33:46:33.9 & 18:07:07.68  & $-$33:46:34.48 & 0.605 & 16.060$\pm$0.021 & 15.918$\pm$0.035 & 15.391$\pm$0.037 & 0.107 & \\
Oph 2010  & V2673 Oph          & NA & 17:39:40.90 & $-$21:39:50.5 & ---          & ---            &       &                  &                  &                  &       & ND\\
\noalign{\smallskip}
\hline
\end{tabular}
\end{table}
\end{landscape}

\addtocounter{table}{-1}

\begin{landscape}
\begin{table}
\caption[]{{\bf cont.} Previously known  novae in the VVV area.}
\label{tab:tiles}
\centering
\small
\begin{tabular}{lllcccccccccl}
\hline \hline
\noalign{\smallskip}
Nova    & Other        & Nova  &  RA            &  DEC         &   RA (VVV)    &  DEC (VVV)     & Dist.      &$J$     &   $H$  & $K_{\rm s}$ & $A_{\rm Ks}$  & Note \\
(Year)   & Designation  & Type  &  [hh:mm:ss.ss] & [dd:mm:ss.s] & [hh:mm:ss.ss] & [dd:mm:ss.ss] & [$\arcsec$] &  [mag] &  [mag] & [mag]      & [mag]  & \\
\noalign{\smallskip}                                                                                                    
\hline                                                                                                                  
\noalign{\smallskip}
Oph 2010b & V2674 Oph               & NA &  17:26:32.19 & $-$28:49:36.3 & ---          & ---            &                  &                  &                  &        &  & ND\\
Sgr 2010  & \object{V5585 Sgr}      & NA &  18:07:26.95 & $-$29:00:43.6 & 18:07:26.95  & $-$29:00:43.59 & 0.016  & 11.921$\pm$0.010 & 11.906$\pm$0.010 & 11.250$\pm$0.010 & 0.170 & \\
Sgr 2010b$^{*}$ & \object{V5586 Sgr} & N &  17:53:03.00 & $-$28:12:19.0 & 17:53:02.98  & $-$28:12:18.84 & 0.339 & 7.764$\pm$0.010 &      ---         & 10.609$\pm$0.010 & 0.724 & ST\\
Sgr 2011  & V5587 Sgr               & NA &  17:47:46.33 & $-$23:35:13.1 & ---          & ---            &  &                &                  &                  &        & ND\\
Sgr 2011b & V5588 Sgr               & NA &  18:10:21.35 & $-$23:05:30.6 & ---		 & ---   	     & &                 & 	      &                  &        & ND \\
Cen 2012b$^{**}$ & {\tiny TCP J14250600-5845360}  & N: & 14:25:06.00 & $-$58:45:36.0 &  ---         &  ---             &  &                &                  &                  &   & ND\\
Oph 2012b & {\tiny PNV J17395600-2447420}  & N  & 17:39:56.00 & $-$24:47:42.0 & ---          &  ---          &                  &  &                &                  &      &  ND \\
Sco 2012  & {\tiny MOA 2012 BLG-320     }  & NB:& 17:50:53.90 & $-$32:37:20.5 & ---          &  ---          &                  &  &                &                  &      &  ND \\
Sgr 2012$^{**}$  & {\tiny \object{PNV J17452791-2305213}}  & N: & 17:45:28.03 & $-$23:05:22.8 & 17:45:28.02  & $-$23:05:22.72 & 0.125 & 15.813$\pm$0.035 & 15.421$\pm$0.047 & 15.178$\pm$0.056 & 0.274 & \\
Sgr 2012c & {\tiny PNV J17522579-2126215} & NA & 17:52:25.79 & $-$21:26:21.5 & ---          & ---     	     &        &                  & 	           &                   &        & ND \\
Sgr 2012d & {\tiny PNV J18202726-2744263} & N  & 18:20:27.26 & $-$27:44:26.3 & ---          & ---    	     &        &                  & 	           &                   &        & ND \\
          & {\tiny \object{OGLE-2012-NOVA-01}}     & N  & 17:56:49.39 & $-$27:13:28.2 & 17:56:49.50  & $-$27:13:28.59 & 1.518  & 17.439$\pm$0.12  & 17.440$\pm$0.366 & 17.018$\pm$0.455 & 0.726 & :\\
          & \object{V1213 Cen}            & NA & 13:31:15.80 & $-$63:57:39.0 & 13:31:15.74  & $-$63:57:38.41 & 0.717  & 13.855$\pm$0.01  & 13.664$\pm$0.010 & 12.803$\pm$0.010 &  ---  & MS\\
          & \object{PN G002.6+01.7}       & n: & 17:45:08.60 & $-$25:44:01.0 & 17:45:08.54  & $-$25:44:01.99 & 1.288  & 14.888$\pm$0.02  & 13.859$\pm$0.022 & 13.465$\pm$0.024 & 0.531 & :\\
          & \object{V0733 Sco}            & N: & 17:39:42.88 & $-$35:52:38.4 & 17:39:43.00  & $-$35:52:38.62 & 1.523  & 14.505$\pm$0.01  & 13.667$\pm$0.010 & 13.370$\pm$0.011 & 0.416 & :\\
          & \object{GR Sgr}               & NA:& 18:22:58.50 & $-$25:34:47.3 & 18:22:58.52  & $-$25:34:47.28 & 0.217  & 15.071$\pm$0.01  & 14.693$\pm$0.017 & 14.506$\pm$0.023 & 0.143 & \\
          & \object{AT Sgr}               & NA & 18:03:30.87 & $-$26:28:28.5 & 18:03:31.00  & $-$26:28:28.74 & 1.815  & 17.218$\pm$0.17  & 16.923$\pm$0.271 & 16.325$\pm$0.225 & 0.432 & MS:\\
          & V2859 Sco                     & NA & 17:50:36.09 & $-$30:01:46.6 & ---          &   ---          &        &                  &                  &                  &       & ND \\
          & \object{[KW2003] 105}         & n: & 18:01:56.24 & $-$27:22:55.6 & 18:01:56.25  & $-$27:22:56.16 & 0.590  & 15.250$\pm$0.030 & 14.955$\pm$0.050 & 14.656$\pm$0.054 & 0.327 & \\
          & V0729 Sco                     & N: & 17:22:02.66 & $-$32:05:48.8 &  ---         &  ---           &        &                  &                  &                  &       & ND\\
\noalign{\smallskip}
\hline
\end{tabular}
\tablefoot{$^{*}$\,during eruption, $^{**}$\,nova progenitor}
\end{table}
\end{landscape}


We defined as a valid match  all cases where a source was found within
1~arcsec  from the  position given  by the  catalogue.  This  value is
agreement with the median values of the VVV image quality in the first
season  ($0\farcs89-0\farcs87$, respectively  to $J$  and  $K_{\rm s}$
observations) and the typical  astrometric accuracy of the photometric
catalogues   \citep[$\lesssim0$\farcs$5$;][]{2012A&A...537A.107S}.   A
complementary visual  inspection of the  images was performed  for all
novae candidates, in  order to certify the presence  (or absence) of a
target in  the given position. In  some cases the  source was rejected
because of the presence of a  faint object in a closer position to the
coordinate,  seen during  the  visual inspection  of  the images,  but
beyond the  detection limit of  the CASU aperture photometry,  used to
build the VVV photometric catalogues (see Section~\ref{obs}). In other
cases multiple sources  are seen in a similar  close distance from the
entry coordinates.   In this procedure we secured  the information for
55 novae  within $\leq1~$arcsec from the catalogue  position, being 27
of them within the VIRCAM  pixel scale of 0$\farcs$34 pixel$^{-1}$.  A
complementary search included also objects matching the position given
by the catalogue within 2~arcsec.   By this procedure we retrieved the
photometry of another 37 objects.   These are all fainter than $K_{\rm
  s}\sim13$~mag and were classified as doubtful sources.  We note that
due to  the source confusion in  the most crowded  regions of Galactic
plane  and bulge,  a search  based only  on the  coordinates  does not
guarantee that in  all cases the source found in  a closer position to
the coordinates  is in  fact the nova  remnant.  Thus these  should be
seen as candidate near-IR counterparts for the novae remnant.

Therefore, from  the 140  novae within  the VVV area,  we are  able to
provide  the $JHK_{\rm  s}$ colours  of 93  objects: 55  classified as
valid matches (27 within the pixel scale), and 37 as doubtful sources.
The  other objects  are  beyond our  detection  limit (progenitors  in
quiescence  with $K_{\rm  s}\gtrsim18$~mag), or  with  coordinates not
sufficiently accurate to allow us to identify the target in the field.
Table~\ref{table:novae} presents  the $JHK_{\rm s}$  photometry of all
novae detected  in the VVV area  while in the Appendix  we provide the
composite $JHK_{\rm s}$ finding charts for all objects.

Since  the  VVV  $JHK_{\rm  s}$  observations  were  performed  during
2010--2011,  we are  also able  to provide  the information  about the
progenitors  of the most  recent novae,  as well  as the  fading curve
after the  nova eruption (see  Section~\ref{sec:var}).  Interestingly,
the observations of  Nova Sgr 2010b were secured  while the object was
in  eruption \citep{2012ATel.4353....1S}.   Complementary information,
including  the timings  of  the VVV  colour  and variability  campaign
observations for  all novae erupted  in this century, is  presented in
Table~\ref{table:time}.

\begin{table*}
\begin{center}
\caption{Novae  eruptions in  the VVV  area from  January 2001  to the
  December  2012.  More  detailed  information are  provided  for  the
  2010-2012 novae. }
\begin{tabular}{l l l l l l}
\hline
\hline
\noalign{\smallskip}
Nova &  Other        & Eruption & VVV $JHK_{\rm s}$    & First $K_{\rm s}$-band epoch & Note \\
     &  Designation  & Date     & Observation         & (variability campaign)      &      \\
\noalign{\smallskip}
\hline
\noalign{\smallskip}
Sco 2001  & V1178 Sco & 2001~~~~~~ & 2010~08~10 & 2010~09~12 &  Remnant\\
Sgr 2001  & V4643 Sgr & 2001~~~~~~ & 2010~04~09 & 2011~08~06 &  Remnant\\
Sgr 2001b & V4739 Sgr & 2001~~~~~~ & 2010~08~14 & 2010~10~15 &  Remnant\\
Sgr 2001c & V4740 Sgr & 2001~~~~~~ & 2010~08~15 & 2011~08~05 &  Remnant\\
Sgr 2002  & V4741 Sgr & 2002~~~~~~ & 2010~08~10 & 2010~09~12 &  Remnant\\
Sgr 2002b & V4742 Sgr & 2002~~~~~~ & 2010~03~28 & 2010~03~30 &  Remnant\\
Sgr 2002d & V4744 Sgr & 2002~~~~~~ & 2010~03~18 & 2010~09~29 &  Remnant\\
Sgr 2003b & V5113 Sgr & 2003~~~~~~ & 2010~03~28 & 2010~08~26 &  Remnant\\
Oph 2004  & V2574 Oph & 2004~~~~~~ & 2010~03~30 & 2010~04~10 &  Remnant\\
Sco 2004  & V1186 Sco & 2004~~~~~~ & 2010~08~03 & 2010~08~26 &  Remnant\\
Sco 2004b & V1187 Sco & 2004~~~~~~ & 2010~08~03 & 2010~08~26 &  Remnant\\
Sgr 2004  & V5114 Sgr & 2004~~~~~~ & 2010~08~15 & 2010~10~24 &  Remnant\\
Cen 2005  & V1047 Cen & 2005~~~~~~ & 2010~03~07 & 2010~03~29 &  Remnant\\
Nor 2005  & V0382 Nor & 2005~~~~~~ & 2010~03~06 & 2010~03~06 &  Remnant\\
Sco 2005  & V1188 Sco & 2005~~~~~~ & 2010~10~01 & 2011~07~27 &  Remnant\\
Sgr 2005a & V5115 Sgr & 2005~~~~~~ & 2010~03~28 & 2010~03~30 &  Remnant\\
Sgr 2005b & V5116 Sgr & 2005~~~~~~ & 2010~08~15 & 2010~10~15 &  Remnant\\
Oph 2006  & V2575 Oph & 2006~~~~~~ & 2010~03~30 & 2010~04~09 &  Remnant\\
Sgr 2006  & V5117 Sgr & 2006~~~~~~ & 2010~04~23 & 2010~10~25 &  Remnant\\
Oph 2007  & V2615 Oph & 2007~~~~~~ & 2010~03~17 & 2011~06~12 &  Remnant\\
Nor 2007  & V0390 Nor & 2007~~~~~~ & 2010~03~16 & 2010~08~15 &  Remnant\\
Oph 2008a & V2670 Oph & 2008~~~~~~ & 2010~03~17 & 2011~06~12 &  Remnant\\
Oph 2008b & V2671 Oph & 2008~~~~~~ & 2011~05~07 & 2011~06~13 &  Remnant\\
Sgr 2008  & V5579 Sgr & 2008~~~~~~ & 2010~04~12 & 2011~08~05 &  Remnant\\
Sgr 2008b & V5580 Sgr & 2008~~~~~~ & 2010~04~08 & 2010~08~27 &  Remnant\\
Oph 2009  & V2672 Oph & 2009~~~~~~ & 2010~04~13 & 2011~06~13 &  Remnant\\
Sgr 2009a & V5581 Sgr & 2009~~~~~~ & 2010~03~30 & 2010~03~30 &  Remnant\\
Sgr 2009b & V5582 Sgr & 2009~~~~~~ & 2010~04~10 & 2011~06~14 &  Remnant\\
Sgr 2009c & V5583 Sgr & 2009~~~~~~ & 2010~04~23 & 2010~08~18 &  Remnant\\
Oph 2010  & V2673 Oph             & 2010~01~15~$^{a}$ & 2010~04~10     & 2011~06~14 & Remnant\\
Sgr 2010  & V5585 Sgr             & 2010~01~20~$^{b}$ & 2010~04~23     & 2010~08~26 & Remnant\\
Oph 2010b & V2674 Oph             & 2010~02~18~$^{c}$ & 2011~05~09     & 2011~07~26 & Remnant\\
Sgr 2010b & V5586 Sgr             & 2010~04~23.782~$^{d}$ & 2010~04~23.2457 ($H$)       & 2010~09~12 & During eruption \\ 
          &                       &                      & 2010~04~23.2479 ($K_{\rm s}$) &         &                 \\ 
          &                       &                      & 2010~04~23.2501 ($J$)        &         &                \\ 
Sgr 2011  & V5587 Sgr             & 2011~01~25~$^{e}$ & 2010~03~18     & 2010~09~29 &  Progenitor\\
Sgr 2011b & V5588 Sgr             & 2011~04~07~$^{f}$ & 2010~04~21     & 2010~10~06 &  Progenitor\\
Cen 2012b & TCP J14250600-5845360 & 2012~04~05~$^{g}$ & 2010~03~27     & 2010~03~09 &  Progenitor\\
Sgr 2012  & PNV J17452791-2305213 & 2012~04~21~$^{h}$ & 2010~03~18     & 2010~09~29 &  Progenitor\\
Oph 2012b & PNV J17395600-2447420 & 2012~05~19~$^{i}$ & 2010~04~13     & 2011~06~12 &  Progenitor\\
Sco 2012  & MOA 2012 BLG-320      & 2012~05~22~$^{j}$ & 2010~08~29     & 2010~09~25 &  Progenitor\\
Sgr 2012c & PNV J17522579-2126215 & 2012~06~26~$^{k}$ & 2010~03~25     & 2010~09~29 &  Progenitor\\   
Sgr 2012d & PNV J18202726-2744263 & 2012~07~12~$^{l}$ & 2010~04~08     & 2010~08~27 &  Progenitor\\
          & OGLE-2012-NOVA-01     & 2012~05~02~$^{m}$ & 2010~03~28     & 2010~08~26 &  Progenitor\\ 
\noalign{\smallskip}
\hline
\end{tabular}
\tablebib{$^{a}$~\cite{2010IAUC.9111....2N},
  $^{b}$~\cite{2010IAUC.9140....1N},
  $^{c}$~\cite{2010IAUC.9119....1N},
  $^{d}$~\cite{2010IAUC.9112....1K},
  $^{e}$~\cite{2011IAUC.9196....1N},
  $^{g}$~\cite{2011IAUC.9203....1N},
  $^{g}$~\cite{2012CBET.3073....1G},
  $^{h}$~\cite{2012CBET.3089....1G},
  $^{i}$~\cite{2012CBET.3124....1W}, 
  $^{j}$~\cite{2012AAN...461....1W},
  $^{k}$~\cite{2012CBET.3156....1Y},
  $^{l}$~\cite{2012CBET.3166....1N},
  and $^{m}$~\cite{2012ATel.4323....1K}.}
\end{center}
\label{table:time}
\end{table*}

\section{The near-IR Colours}\label{sec:col}

Figure~\ref{fig:cmd}   shows   in   the   top   panels   the   $K_{\rm
  s}~vs.~(J-K_{\rm  s})$   colour-magnitude  diagram  (CMD)   and  the
$(J-H)~vs.~(H-K_{\rm s})$ colour-colour diagram (CCD) for all novae in
the  VVV  area.  We  used  different  symbols  to denote  the  objects
classified as valid matches and the doubtful ones, as described in the
previous Section.  The reddening vector associated  with an extinction
of $E(B-V)=1$, based on the relative extinctions of the VISTA filters,
and  assuming the \cite{1989ApJ...345..245C}  extinction law,  is also
shown.

The Galactic  novae range  in magnitude and  colour, with  all objects
appearing red, with $(J-K_{\rm s})>0$  (the only exception is Nova Cir
1906, with  $(J-K_{\rm s})=-0.228$).  Interestingly,  a large fraction
of novae are beyond detection in previous near-IR surveys \citep[e.g.,
  2MASS, DENIS;  ][]{2006AJ....131.1163S,1994Ap&SS.217....3E}, thus we
are  now  reporting  their  IR-colours   for  the  first  time  in  an
homogeneous data  set.  We note that  the novae are  spread across the
whole VVV  area (see Fig.~\ref{fig:area}), and  therefore are affected
by different extinction.   For instance, in the bulge  area where most
novae  are  located,   reddening  spans  from  $E(B-V)\lesssim0.2$  at
$b\sim-10^{\circ}$  to $E(B-V)\sim10$  closer to  the  Galactic centre
\citep[e.g.,][]{2012A&A...543A..13G}.

\begin{figure} 
\includegraphics[bb=1cm 4.5cm 15cm 20.3cm,scale=0.48]{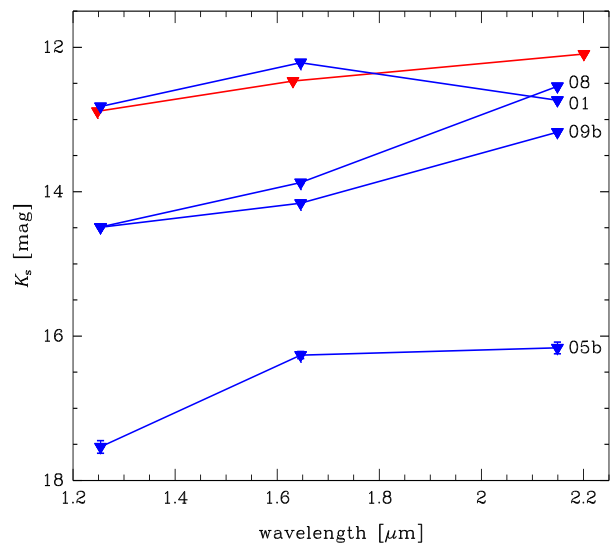}
 \caption{Spectral  energy  distribution  for  the novae  showing  the
   extreme  colours in  Fig.~\ref{fig:cmd}:  Nova Sgr  2001, Nova  Sgr
   2005b, Nova  Sgr 2008, and Nova  Sgr 2009b. Data for  Nova Sgr 2001
   taken   with    UKIRT   are    also   shown   (red    points;   see
   Section~\ref{sec:col}).}
\label{fig:sed}
\end{figure}

In  the  $(J-H)~vs.~(H-K_{\rm  s})$  CCD  the objects  are  seen  more
concentrated  at  low values  of  $(H-K_{\rm  s})$,  but spread  along
$(J-H)$  by more  than  1 mag.   The  colours of  main sequence  stars
\citep[with    spectral    type     from    B0    to    M4,    adapted
  from][]{2001ApJ...558..309D} are presented in the CMD in the case of
$A_{Ks}=0$~mag  (no  extinction)  an  applying  the  median  value  of
extinction among  our targets of  $A_{Ks}=0.243$~mag (corresponding to
$E(B-V)\sim0.67$~mag). The comparison reveals that part of our sources
coincide in colour with the main sequence stars and therefore suggests
that some  of our  novae candidates can  be actually old  field stars,
which match a position close to the novae coordinates, and suffer from
reddening at  different levels. Individual values of  $A_{Ks}$ for all
bulge    novae    are   listed    in    Table    1.    Although    the
\cite{2001ApJ...558..309D}  data  are  in  the  $JHK$  system,  it  is
sufficiently  close  to  the  VISTA  system that  is  useful  for  the
comparison.  Four objects are seen with the most extreme colours: Nova
Sgr  2001  and  Nova Sgr  2005b,  with  the  largest $(J-H)$  but  low
$(H-K_{\rm s})$;  and Nova  Sgr 2008 and  Nova Sgr 2009b,  showing the
largest $(H-K_{\rm  s})$ values at  relatively low values  of $(J-H)$.
The  complementary $JHK_{\rm s}$  spectral energy  distributions (SED)
for  these selected  objects are  shown in  Fig.~\ref{fig:sed}.  While
Nova Sgr 2001 and Nova Sgr 2005b show a maximum close to the $H$-band,
the  SED of  Nova  Sgr 2008  and  Nova Sgr  2009b  reveal a  monotonic
increase in flux towards longer wavelengths.

In  order  to  minimize   the  reddening  effects  we  calculated  the
dereddened $J_{\rm  0}$, $H_{\rm 0}$, and $K_{\rm  s\,0}$, colours for
the  bulge novae,  using the  maps of  \cite{2012A&A...543A..13G}, the
relative  extinctions   of  the   VISTA  filters,  and   assuming  the
\cite{1989ApJ...345..245C}  extinction   law.   A  dereddened  $K_{\rm
  s\,0}~vs.~(J-K_{\rm s})_0$ CMD is presented in the bottom-left panel
of Fig.~\ref{fig:cmd}.   Since the \cite{2012A&A...543A..13G}  maps do
not cover  the disk area  we prefer to  exclude from the plot  the few
novae  located   in  this  region,   instead  of  using   a  different
prescription for the extinction in  the disk, which does not guarantee
a  good  agreement   with  the  procedure  used  in   the  bulge  area
\citep[e.g.,][]{2012A&A...537A.107S}. We also  note that the reddening
law   is  known   to  change   close   to  in   the  Galactic   center
\citep[e.g.,][]{2009ApJ...696.1407N},  but  in  our  sample  just  one
object (Nova Sgr 2001) belongs  to this region. 

A      set     of      reddening-free     indices      provided     by
\cite{2011rrls.conf..145C},   based   on   the   extinction   law   of
\cite{1985ApJ...288..618R}  for the VISTA  filters was  also computed.
The $m_4$ pseudo-magnitude and the $c_3$ pseudo-colour are defined as
 
\begin{equation}
m_4\equiv K_{\rm s}-1.22(J-H), 
\end{equation}

\begin{equation}
c_3\equiv (J-H)-1.47(H-K_{\rm s}).
\end{equation}

The  $m_4~vs.~c_3$   CMD  is  shown  in  the   bottom-right  panel  of
Fig.~\ref{fig:cmd}. We note the  indices shown in the equations depend
on  the effective wavelengths  of each  filter, assumed  to be  a flat
distribution  here,  which is  certainly  different  from the  near-IR
spectra  of novae.  Refined  indices based  on actual  near-IR spectra
covering different luminosity classes  and spectral types are currently
being computed, and  quantitative tests in the case  of actual near-IR
spectra of novae will also be carried out in the future.

One important aspect  seen in all panels of  Fig.~\ref{fig:cmd} is the
absence of  a clear-cut correlations or trends,  even using dereddened
colours and reddening-free indices, the resulting distributions do not
appear  to have straightforward  interpretations.  The  differences in
colour arise  mostly because ($i$)  the objects are seen  at different
distances  and positions  in  the  Galaxy, and  thus  can suffer  from
extinction  and reddening  at  different levels,  compared  to the  2D
reddening maps of  \citep{2012A&A...543A..13G} and ($ii$) the spectral
energy distribution in  the near-IR can be affected  by emission lines
and thermal  emission by  a dust shell  produced during  eruption. The
latter are time-dependent and valid mostly for recent novae. Depending
on  the evolution  of  the dust  shells  created in  the outbursts,  a
variety  of emission  lines are  seen in  early  post-outburst near-IR
spectra.  Moreover,  for older novae  the peak emission from  the dust
shells       is        shifted       to       longer       wavelengths
\citep[e.g.,][]{2002AJ....124.3009V,2012BASI...40..243B}.

Four  objects  seen   with  the  extreme  colours  on   the  plots  of
Fig.~\ref{fig:cmd}  are discussed  in  more detail  in the  following,
namely  Nova Sgr  2001, Nova  Sgr 2005b,  Nova Sgr  2008 and  Nova Sgr
2009b.

\cite{2006MNRAS.368..592A} classified  Nova Sgr 2001 as  a ``NA'' fast
novae, with absence of dust formation after the eruption.  Its near-IR
spectrum  evolves significantly  from the  early decline  stage (March
2001)     to     the      coronal     phase     \cite[August     2001,
][]{2006MNRAS.368..592A}. Nova Sgr 2001 is the innermost known nova in
the  Galaxy,  with coordinates  $(l,b)=(3.345,  -0.337)$~deg, with  an
extinction  of  $A_{Ks}=1.88$~mag  according to  the  maps  of
\cite{2012A&A...543A..13G}      using     \cite{2009ApJ...696.1407N}
extinction law, equivalent to $A_{V}\simeq15.9$~mag.

Our data  taken during  the quiescent phase  (on April 2010)  show the
remnant of Nova Sgr 2001 as a relatively blue object, with the maximum
intensity in the  near-IR around the$H$-band (see Fig.~\ref{fig:sed}).
However, data taken with UKIRT on March 2012 show Nova Sgr 2001 with a
flatter SED in the near-IR,  with the maximum intensity towards longer
wavelengths \citep{2012ATel.4405....1V}.

\cite{2006MNRAS.368..592A}  estimated  a  distance  to the  object  of
$d=3.3$~kpc.       Thus,     since     our      calculations     using
\cite{2012A&A...543A..13G}  maps assume  the total  extinction  in the
line  of sight,  it  could not  be  a good  approach  to estimate  the
reddening for nearby objects,  overestimating the corrections. In this
case,  three-dimensional extinction  maps are  necessary, in  order to
estimate  the  relative  extinction  at  given  distance  \citep[e.g.,
][]{2013A&A...550A..42C}. On the other hand, the $c_3$ colour seems to
be dominated by  the $H$-band (by construction, see  Equation 2), thus
the presence of features such as emission lines or simply the slope of
the SED contribute to shift the value of the $c_3$ pseudo-color from a
relatively small to a relatively large value, depending on whether the
emission peaks around the $H$-band  or not. This explains why Nova Sgr
2001 and  Nova Sgr 2005b, whose  spectra do peak around  $H$ (Fig. 4),
have large  {\it positive} $c_3$ values, whereas  the remaining novae,
whose  spectra  monotonically  increase  towards  longer  wavelengths,
have~-- on the contrary~-- large {\it negative} $c_3$ values.

Nova Sgr 2005b is similar in color with Nova Sgr 2001. Its remnant is
a Supersoft X-ray variable source, presenting an orbital period of
$P_{orb}=2.97$~hr
\citep{2008A&A...478..815D,2008ApJ...675L..93S,2010AN....331..201S},
which puts the object on the edge of the $2-3$~hr CVs period gap
\citep[e.g.,][]{2003cvs..book.....W}.  The object is located beyond
the bulge, with $(l,b)=(2.136, -6.832)$~deg and a distance of
$d=11\pm3$~kpc estimated by \cite{2008ApJ...675L..93S}.  Nova Sgr
2005b lies in the lower part of the CMDs, with $K_{\rm s}=16.163$ and
$K_{\rm s,0}=16.079$.  The presence of such distant objects in our
data demonstrates our capability to monitor even the most distant
novae in the Galaxy.

In the other  extreme are Nova Sgr 2008 and 2009b.   Nova Sgr 2008 was
discovered in  eruption on 2008  April 18 \citep{2008CBET.1342....1N}.
It is a nearby  fast Fe~II nova, located at $(l,b)=(3.734,-3.022)$~deg
at              distance              of             $d=4.4\pm0.2$~kpc
\citep{2008IAUC.8948....1R,2011MNRAS.415.3455R}.  The  very red colors
even in $(J-K_{\rm s})_0$ are related to the presence of dust. Spectra
taken in  the late stages  after the eruption indicate  dust formation
during  the  nova  remnant's development  \citep{2011MNRAS.415.3455R}.
Nova   Sgr   2009b  \citep{2009IAUC.9049....1S}   is   located  in   a
low-extinction     region    with     $(l,b)=(7.531,4.719)$~deg    and
$A_{K_s}=0.1845$~mag.   There   are  no  entries   in  the  literature
estimating its  distance or classifying the nova  remnant. However, by
comparing their colours, one can infer that Nova Sgr 2009b was similar
to the  Nova Sgr 2008 remnant. In  this case, Nova Sgr  2009b could be
surrounded by dust, expelled from the system after the eruption.

\begin{figure*} 
\includegraphics[bb=6cm -2.5cm 18cm 21cm,angle=-90,scale=0.55]{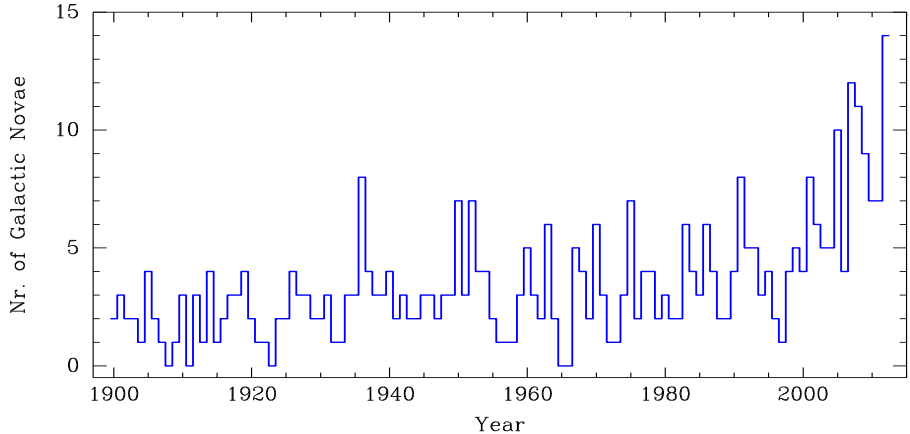}
 \caption{Historical record  of all classical novae  discovered in the
   MW since year 1900. Despite the  increment in the nova rate seen in
   the last  two decades, the  nova rate is  still below that  seen in
   nearby galaxies.  The data are completed as of December 2012.}
\label{fig:year}
\end{figure*}

\begin{figure} 
\includegraphics[bb=3cm 2cm 18cm 15cm,angle=-90,scale=0.41]{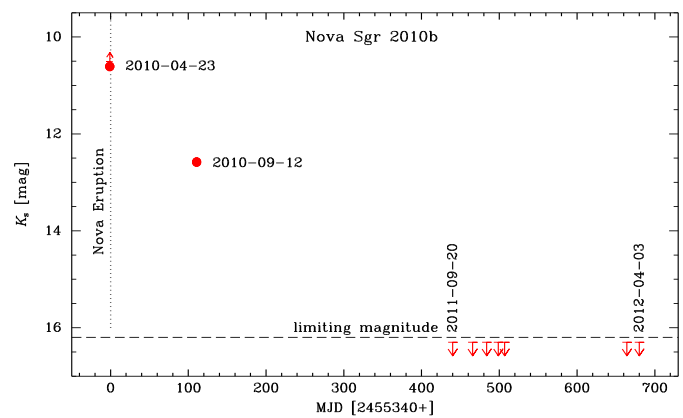}
 \caption{VVV $K_{\rm s}$-band light curve  of Nova Sgr 2010b. The VVV
   data caught the object about 0.5 days before its first detection in
   eruption. The object is seen as a saturated source on 23 March 2010
   and recorded about 2 mags fainter on 12 Sep 2010. Sgr 2010b is
   beyond detection ($K_{\rm  s} \gtrsim 16.2$ mag in  the VVV $K_{\rm
     s}$-band  aperture catalogues for  this field)  in a  sequence of
   observations from Sep 2011 to Apr 2012.}
\label{fig:n2010}
\end{figure}

\begin{figure} 
\includegraphics[bb=3cm 2cm 18cm 15cm,angle=-90,scale=0.41]{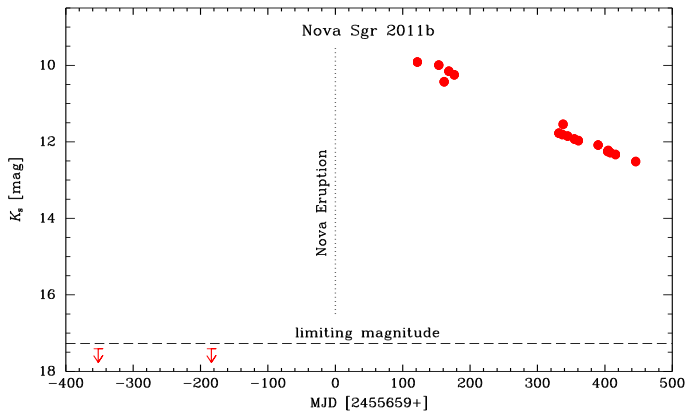}
 \caption{VVV $K_{\rm s}$-band light  curve of Nova Sgr 2011b. Several
   epochs starting on  21 Apr 2011 registered the  object fading after
   its  eruption on 07  Apr 2011.  The Nova  progenitor is  beyond the
   limiting magnitude  in two  observations taken during  2010 ($K_{\rm
     s} \gtrsim 17.3$ mag).}
\label{fig:n2011}
\end{figure}

\begin{figure} 
\includegraphics[bb=3cm 2cm 18cm 15cm,angle=-90,scale=0.41]{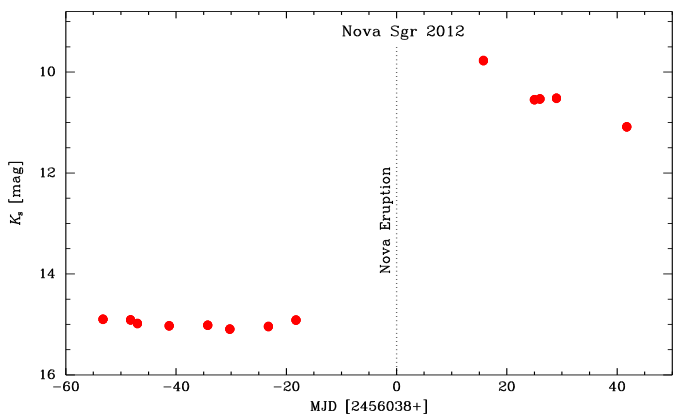}
 \caption{VVV  $K_{\rm s}$-band  light curve  of Nova  Sgr  2012b. Its
   eruption was detected  on 21 Apr 2012. The  Nova progenitor is seen
   at $K_{\rm s} \sim 15.0$  mag in five observations earlier in 2012,
   while a  few observations starting  about a hundred days  after the
   eruption registered its fading curve.}
\label{fig:n2012}
\end{figure}

\section{The VVV variability campaign on Novae}\label{sec:var}

The IR  photometric monitoring of several novae  outbursts has allowed
already  in  the  1990's  to  distinguish  between  two  fundamentally
different  types  of  objects,  usually  referred to  as  CO  and  ONe
novae. While the first seems  to arise from thermonuclear runaway on a
relatively low-mass WD and tends  to form considerable amounts of dust
at later  stages, the  latter turn out  to be associated  with massive
($\geq$1.2 $M_{\sun}$) WD that  have outbursts which exhibit a coronal
line emission phase with little or no dust production.

Particularly useful  in identifying extreme  CO and ONe novae  are the
$K_{\rm s}$-band  light curves: because  of the thermal  emission from
dust during  the condensation phase,  CO Novae reach an  emission peak
within a few months  from the optical maximum, declining exponentially
thereafter as consequence of the decreasing density of the shell owing
to   expansion   \citep[emblematic   is    the   case   of   NQ   Vul;
][]{1978PASJ...30..419S}. ONe novae,  on the contrary, decline rapidly
shortly after the  optical maximum and then have  a positive change in
the  slope  rate   \citep[e.g.,  V1974  Cyg;  ][]{1997ApJ...477..817W}
because  of  the  increasing  contribution  of  several  IR  forbidden
emission lines  (e.g., [Ca VIII]  2.32 $\mu$m, [Si VII]  2.47 $\mu$m),
whose intensity  strengthens during the  coronal phase acting  thus as
effective  coolant   of  the  ejecta   \citep[e.g.,  as  in   QU  Vul;
][]{1988AJ.....95..172G}.

The VVV  Survey, for the first  time, will provide  well-sampled light
curves in  the $K_{\rm  s}$-band for about  $100$ Galactic  novae that
exploded in the last $\sim$150 years  in the Galactic bulge and in the
southern part of  the Galactic disk.  At the time  of writing, the VVV
variability campaign is  ongoing, and in the next  few years until the
completion of the survey (2016 or  later), we would be able to finally
build a coherent  picture of nova IR behavior,  at quiescence, as well
as to discover  new novae in highly-extincted regions  that are missed
in the optical.

There are about  400 known novae in the Milky  Way, but the comparison
with the nova rate in nearby galaxies shows that a number of novae are
lost every  year in  the Galaxy. While  the historical records  show a
rate  below  a  dozen  nova   discovered  per  year  in  the  MW  (see
Fig.~\ref{fig:year}), several  results point  to a nova  rate spanning
between   $\sim30-40$~yr$^{-1}$,  or  even   higher  reaching   up  to
$\sim260$~yr$^{-1}$
\citep{1987PASP...99..606L,1993RMxAA..26...41L,1994A&A...286..786D,
  1997ApJ...487..226S,2001ApJ...563..749S}.

Not coincidentally, the spatial distribution of novae in the bulge and
southern plane shows a ``zone of avoidance''(see Fig.~\ref{fig:area}),
with just  a few objects  belonging to the high-extinction  regions of
the Galaxy.   This arises because  the current searches for  novae are
performed  with small-aperture  telescopes  in the  optical.  The  VVV
Survey combines  deep and high-spatial resolution  observations in the
near-infrared,  allowing one  to search  for  novae even  in the  most
crowded and high extinction  regions of our Galaxy.  Very illustrative
is the recent discovery of  VVV-NOV-001, the first Nova candidate from
the VVV data,  located in a high extinction region  in the inner bulge
with      $(l,b)=(8.897,-0.158)$~deg     and      $A_V\sim     11$~mag
\citep{2012ATel.4426....1S}.

Fig.~\ref{fig:maglim}  shows the  magnitude  range covered  by VVV  at
different  levels  of  crowding  in  the  bulge.   Even  in  the  most
problematic  case which  is  close  to the  Galactic  center, the  VVV
observations allow us to monitor over a range in magnitudes spanning 
 from $\Delta K_{\rm s}>7$~mag, reaching even
$\Delta K_{\rm  s}\sim10$~mag when  using data taken  below 5-$\sigma$
accuracy and saturated sources. As of this writing, we are able to
provide  light  curves with  a  few  points  for some  objects,  which
demonstrates our  capability to monitor novae at  all phases depending
on the magnitude  in each case. Figs.~\ref{fig:n2010}, \ref{fig:n2011}
and  \ref{fig:n2012}   show  VVV  light  curves  of   Nova  Sgr  2010b
\citep{2012ATel.4353....1S},           2011          and          2012
\citep{2012ATel.4372....1S}, respectively.

These $K_{\rm  s}$-band light  curves at quiescence  will allow  us to
study  any kind  of  orbital effects  which  could be  related to  the
influence on the secondary star by the accreting WD, and when combined
with  the  IR  colors  will   also  help  quantify  (via  modeling  of
ellipsoidal variations of the secondary) fundamental parameters of the
nova  system, such  as spectral  type of  the secondary,  mass ratios,
orbital periods and inclinations \citep{1998ApJ...494..783M}.

\section{Perspectives on the search for novae using the VVV images}
\label{sec:ima}

There is a well-known discrepancy on the actual amount of mass ejected
in a CN  outburst as predicted by theory  and numerical simulations on
one side, and  as derived from infrared and  radio observations on the
other side - being the latter  systematically higher by up to a factor
$\sim$10 \citep{2002AIPC..637..259S,2012MmSAI..83..792S}.

Even if there  are strong caveats on the  methods and assumptions used
for      determining       the      mass      of       the      ejecta
\citep[e.g.,][]{2010ApJ...712L.143S}, the analysis  of the most recent
observational  material seems  in fact  to suggest  for the  ejecta of
classical    novae    a    lower    mass   and    higher    velocities
\citep[$10^{-4}-10^{-5}$M$_{\sun}$   and   up   to   6000~km~s$^{-1}$,
  respectively,    as    in    the    extreme    case    of    LMC1991
  --][]{2001MNRAS.320..103S}  than thought before.   Identifying where
such a discrepancy stems from  would have dramatic consequences on our
understanding of the nova phenomenon,  e.g., by clarifying its role in
the chemical evolution of the galaxy.

By  supposing expansion velocities  of the  nova remnants  larger than
1000~km~s$^{-1}$,  it  is  reasonable  to  expect that  at  least  the
closest/fastest novae produce nebular  remnants that eventually may be
spatially  resolved even from  ground-based telescopes.   Imaging such
remnants allows the  investigation of fundamental correlations between
the properties  of the central  binary, the evolution of  the outburst
and the ejecta  shaping mechanism, and also provides  a direct measure
of the  distance to the nova  by combining the  angular expansion rate
with  spectroscopically   derived  expansion  velocities,   i.e.,  the
so-called          \textit{expansion          parallax         method}
\citep{1940PASP...52..386B,1989clno.conf...73M}.

While the infrared imagery has several advantages over the optical one
(e.g., the extinction  -- which is often poorly known  for novae -- is
conveniently reduced), up to now  only a handful of nova remnants have
been  searched (and even  fewer spatially  resolved) in  this spectral
region, compared to more than 40 nova remnants resolved in the optical
and $\sim$10 in the radio.

Thanks  to  the large  sky  area covered  and  to  the higher  spatial
resolution than previous IR surveys  (such as 2MASS and DENIS), VVV is
in unique position  to perform for the first  time a systematic search
for  remnants of  classical novae  that exploded  in the  last decades
within the inner, highly-extincted regions  of the MW. Even assuming a
representative  distance to  a nova  of 4~kpc  (about half-way  to the
Galactic  center)  and  a  very  conservative  expansion  velocity  of
1000~km~s$^{-1}$, VVV is in fact potentially able to spatially resolve
the  corresponding  nebular remnants  after  $\sim$40  years from  the
outburst,  taking into  account  the VIRCAM  pixel scale  (0$\farcs$34
pixel$^{-1}$)  and a  nebula angular  size  of 2".  No resolved  novae
shells  have  yet  been  detected  in  the VVV  imaging  data.   In  a
forthcoming phase of the survey  a search for novae using imaging will
be conducted on the deep,  stacked $K_{\rm s}$-band images, using more
focused image analysis such as point spread function (PSF) subtraction
or         difference        image         analysis        \citep[DIA,
  e.g.,][]{2012A&A...537A.107S}.

\section{Conclusions}\label{conclu}

We presented a near-IR catalogue of novae within the VVV area covering
the MW  bulge and southern Galactic  plane area. From  the 140 objects
found  in  the VSX/AAVSO  catalogue,  we  reported  the $JHK_{\rm  s}$
colours  of 93  objects.  The  rest were  beyond detection  or heavily
blended  sources.  Colour-magnitude  and  colour-colour diagrams  were
presented, as well as CMDs using dereddened colours and reddening-free
indices. These should be used with caution since in the case of nearby
objects the  dereddened colours  can overestimate the  corrections, by
assuming  the  total  extinction  in  the line  of  sight.   Likewise,
reddening-free near-IR  indices specifically devised for  the study of
novae would  also be of  considerable interest, given  their different
spectral  shapes   compared  with  those  typically   assumed  in  the
computation of these indices.
 
Thanks  to its  higher spatial  resolution in  the near-IR,  and large
$K_{\rm s}$-range covered by the observations, the VVV survey can be a
major contributor  for the  search and study  of novae in  the Galaxy,
mainly  in the  most crowded  and high-extinction  regions of  the MW,
beyond the capabilities  of the current searches for  novae in optical
wavelengths.

VVV  can produce  well-sampled light  curves covering  many  years for
Galactic novae  belonging both to the  bulge and the  southern part of
the  disk, even  for objects  during  or fading  after eruption.   The
recent report  of VVV-NOV-001, the  first nova candidate from  the VVV
data   discovered   in   the   inner   bulge,   is   an   illustrative
example. Surely,  the first of many  more to come.  The possibility to
search for novae using the VVV imaging is accordingly a very promising
path to unveiling the heretofore hidden population of heavily obscured
novae.

\begin{acknowledgements}

We  gratefully acknowledge  use of  data  from the  ESO Public  Survey
programme ID 179.B-2002 taken  with the VISTA telescope, data products
from  the Cambridge  Astronomical Survey  Unit, and  funding  from the
FONDAP  Center for Astrophysics  15010003, the  BASAL CATA  Center for
Astrophysics  and Associated  Technologies PFB-06,  the  FONDECYT from
CONICYT, and the Ministry  for the Economy, Development, and Tourism's
Programa  Iniciativa Cient\'{i}fica  Milenio through  grant P07-021-F,
awarded to  The Milky  Way Millennium Nucleus.   Support for  R.A.  is
provided by  Proyecto GEMINI CONICYT  32100022 and via  a Postdoctoral
Fellowship  by the  School  of Engineering  at Pontificia  Universidad
Cat\'olica de Chile.  M.C.  and I.D. acknowledge funding from Proyecto
FONDECYT  Regular 1110326.   J.B. acknowledges  funding  from Proyecto
FONDECYT  Regular  1120601.  R.K.  acknowledges partial  support  from
FONDECYT through grant  n. 1130140. This research has  made use of the
International Variable  Star Index (VSX) database,  operated at AAVSO,
Cambridge, Massachusetts, USA.

\end{acknowledgements}


\begin{appendix}
\section{VVV Novae finding charts}
\label{charts}

In this section we present the finding charts for all Novae in the VVV
Survey  area, listed  in Table~\ref{table:novae}.   In most  cases the
charts  are $JHK_{\rm  s}$  composite images,  unless when  explicitly
marked as a $K_{\rm s}$-band  image.  All charts are $30\arcsec \times
30\arcsec$  in  size,  oriented  in  Galactic  coordinates,  with  the
positive  Galactic   longitude  pointing   up.   A  cross   marks  the
coordinates given by the catalogue and helps one to check the notes we
adopted in Table~\ref{table:novae}.

 \begin{figure*} 
\includegraphics[bb= -1.2cm   1.6cm   2cm  8cm, scale=0.65]{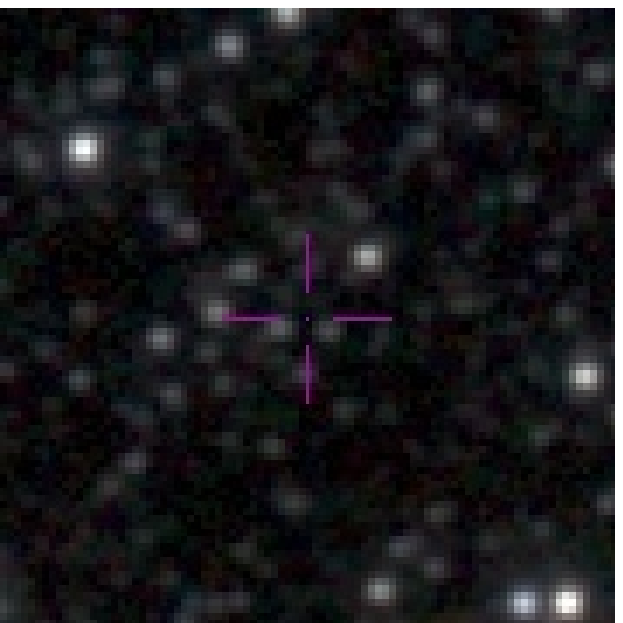}
\includegraphics[bb= -4.4cm   1.6cm   0cm  8cm, scale=0.65]{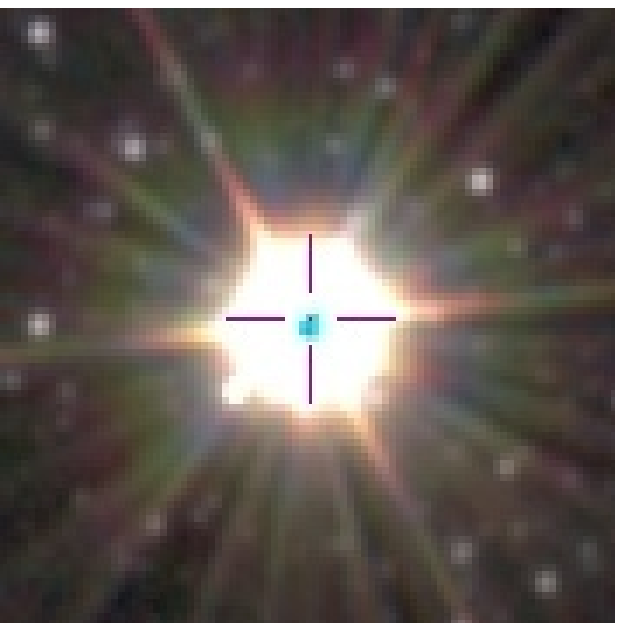}
\includegraphics[bb= -6.4cm   1.6cm   0cm  8cm, scale=0.65]{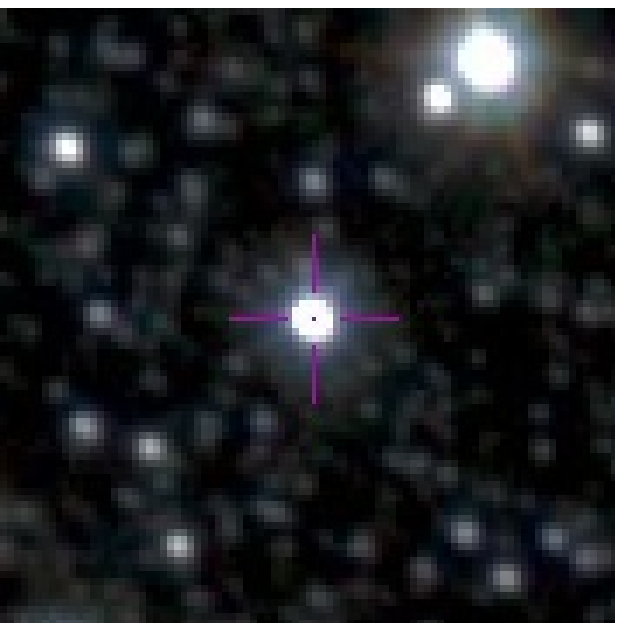}
\includegraphics[bb= -6.4cm   1.6cm   0cm  8cm, scale=0.65]{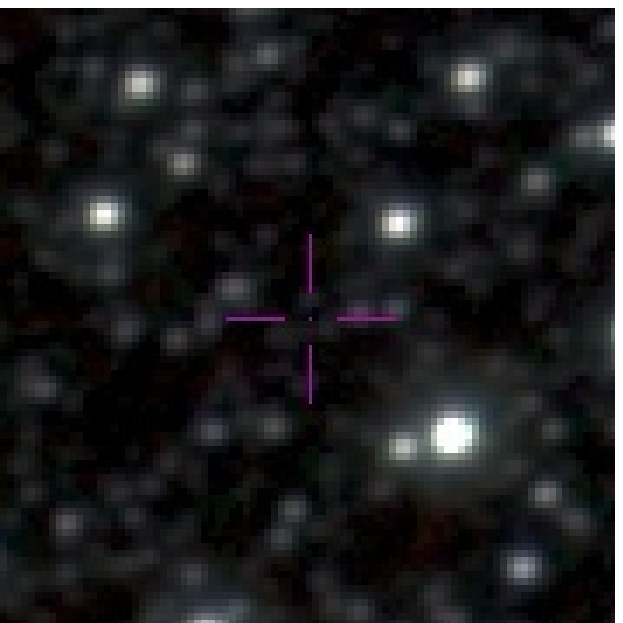}
                              
\includegraphics[bb= -1.2cm   1.6cm   2cm  8cm, scale=0.65]{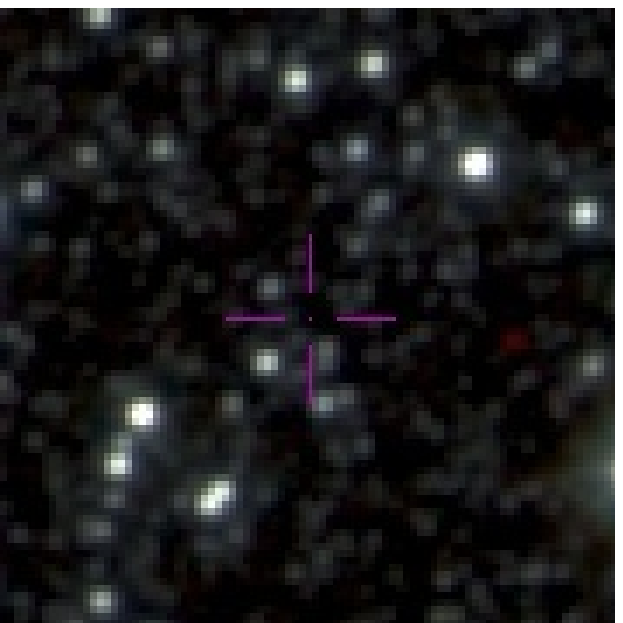} 
\includegraphics[bb= -4.4cm   1.6cm   0cm  8cm, scale=0.65]{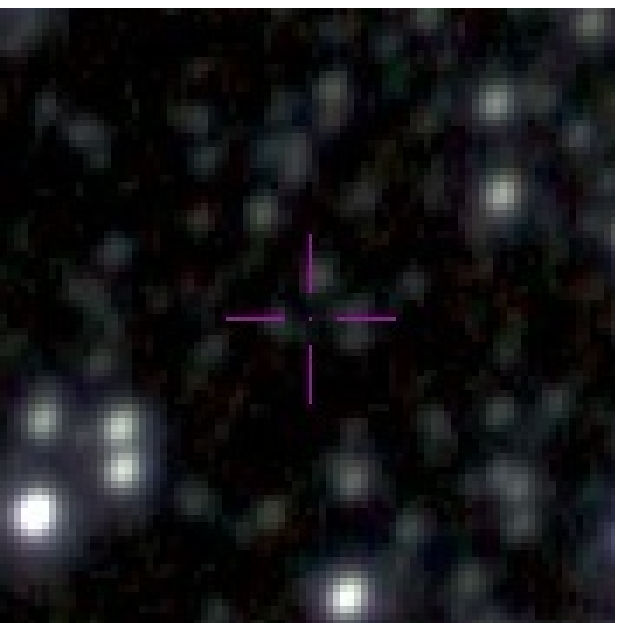} 
\includegraphics[bb= -6.4cm   1.6cm   0cm  8cm, scale=0.65]{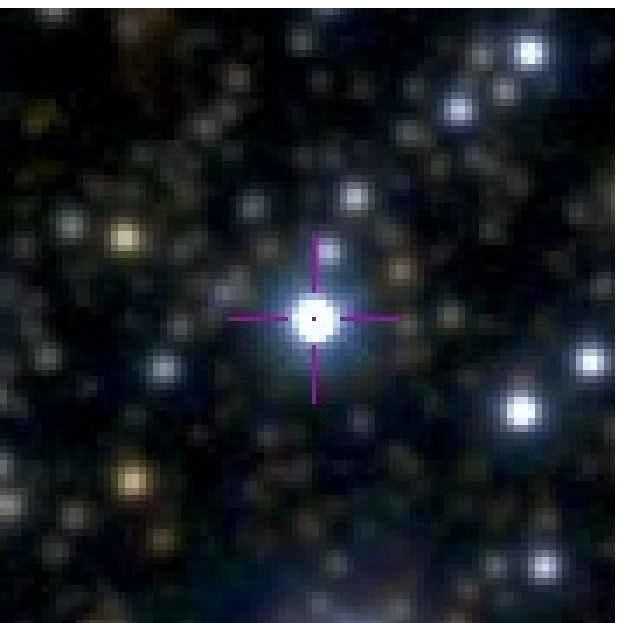} 
\includegraphics[bb= -6.4cm   1.6cm   0cm  8cm, scale=0.65]{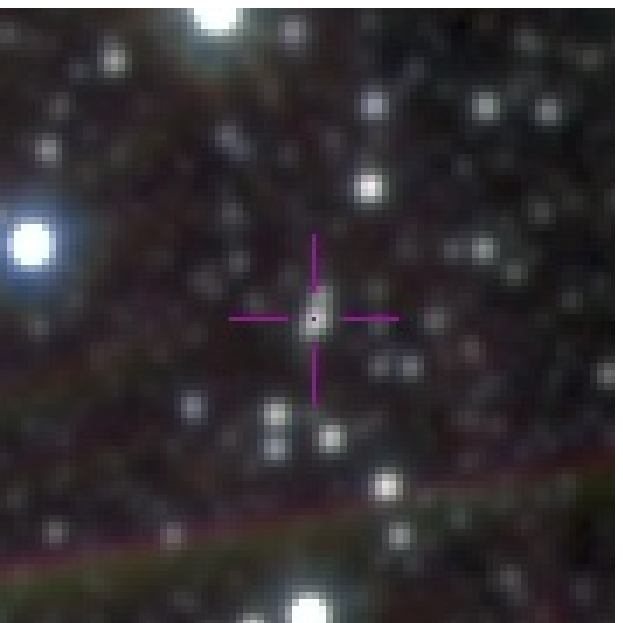} 
                              
\includegraphics[bb= -1.2cm   1.6cm   2cm  8cm, scale=0.65]{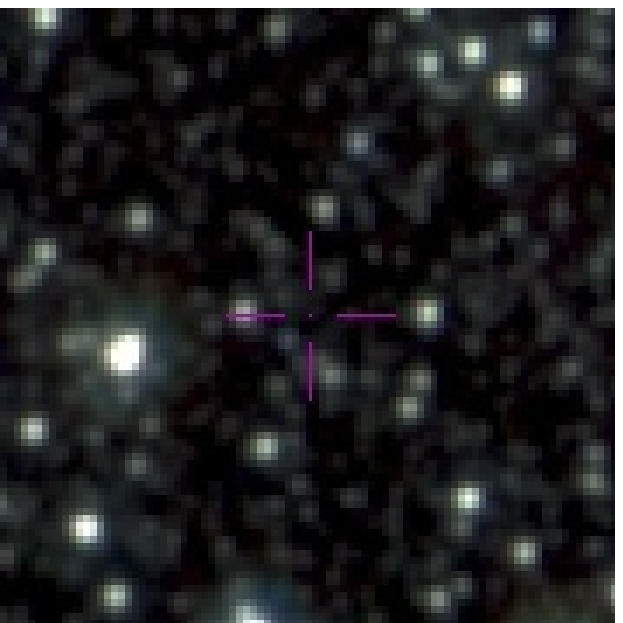} 
\includegraphics[bb= -4.4cm   1.6cm   0cm  8cm, scale=0.65]{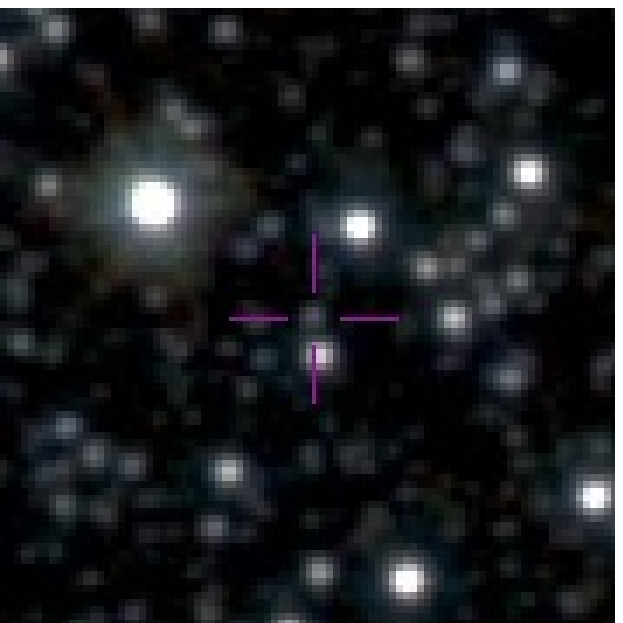}
\includegraphics[bb= -6.4cm   1.6cm   0cm  8cm, scale=0.65]{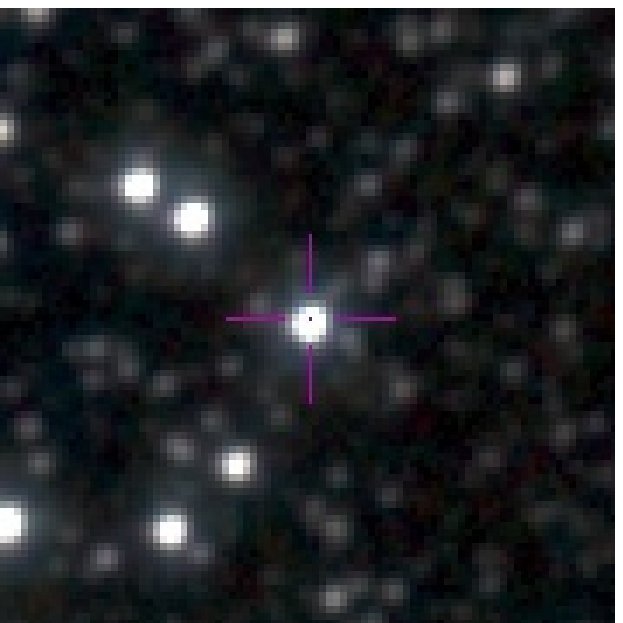} 
\includegraphics[bb= -6.4cm   1.6cm   0cm  8cm, scale=0.65]{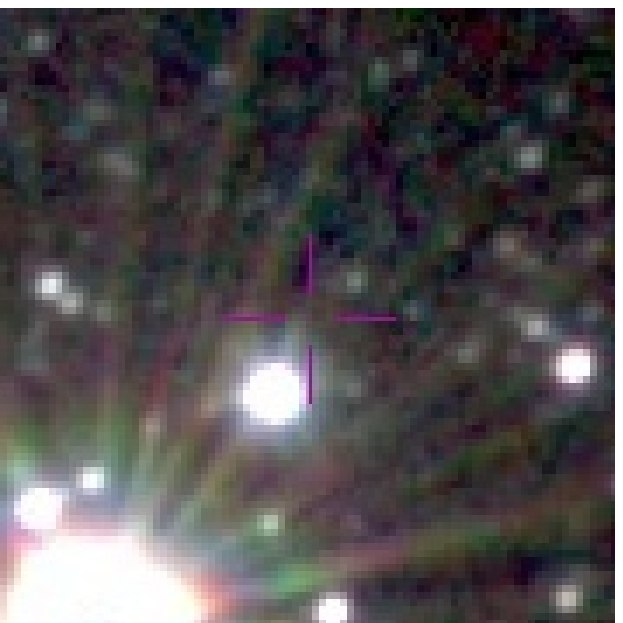} 
                              
\includegraphics[bb= -1.2cm   1.6cm   2cm  8cm, scale=0.65]{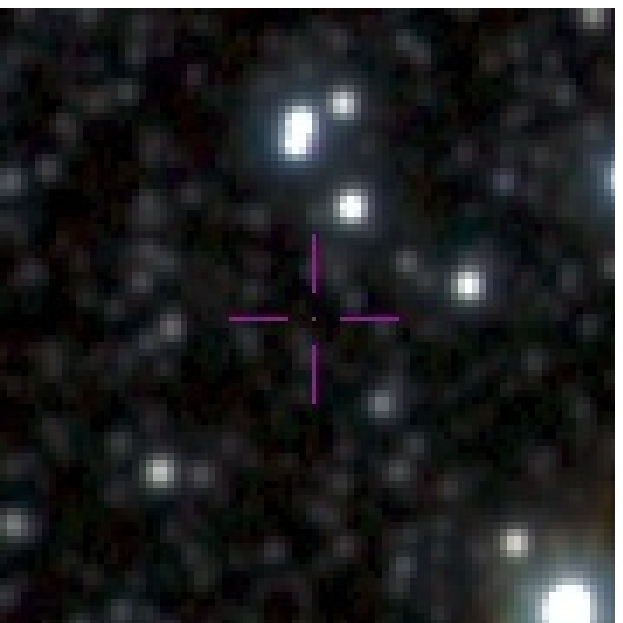} 
\includegraphics[bb= -4.4cm   1.6cm   0cm  8cm, scale=0.65]{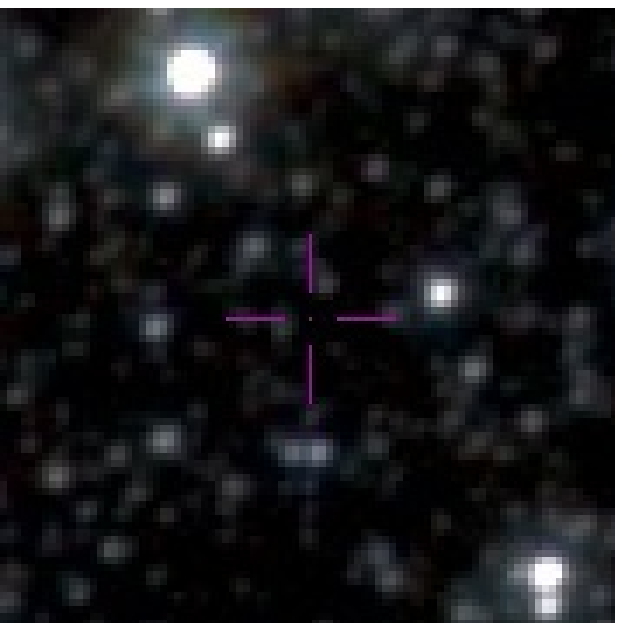} 
\includegraphics[bb= -6.4cm   1.6cm   0cm  8cm, scale=0.65]{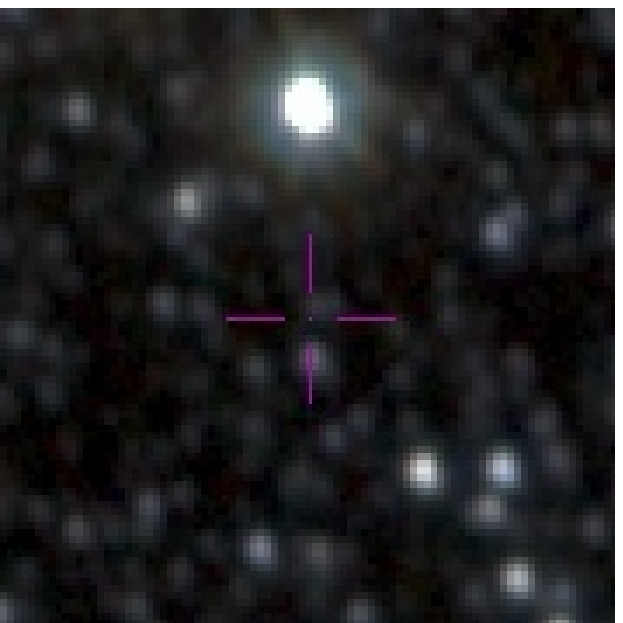}
\includegraphics[bb= -6.4cm   1.6cm   0cm  8cm, scale=0.65]{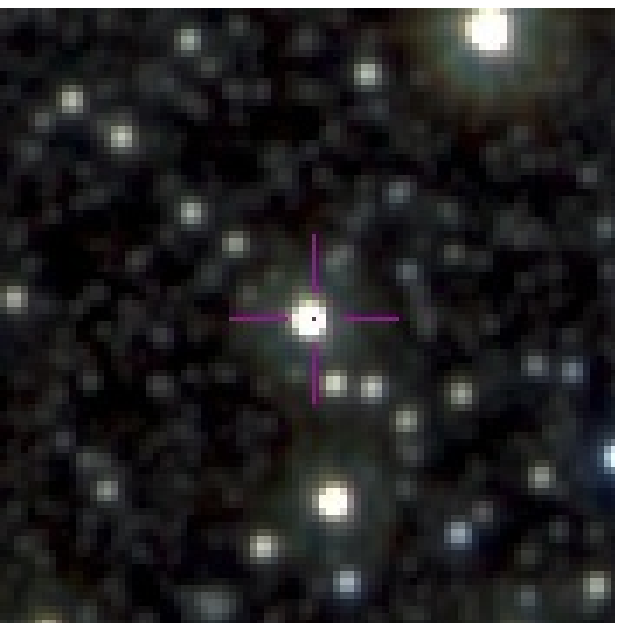}

\includegraphics[bb= -1.2cm  -0.5cm   2cm  8cm, scale=0.65]{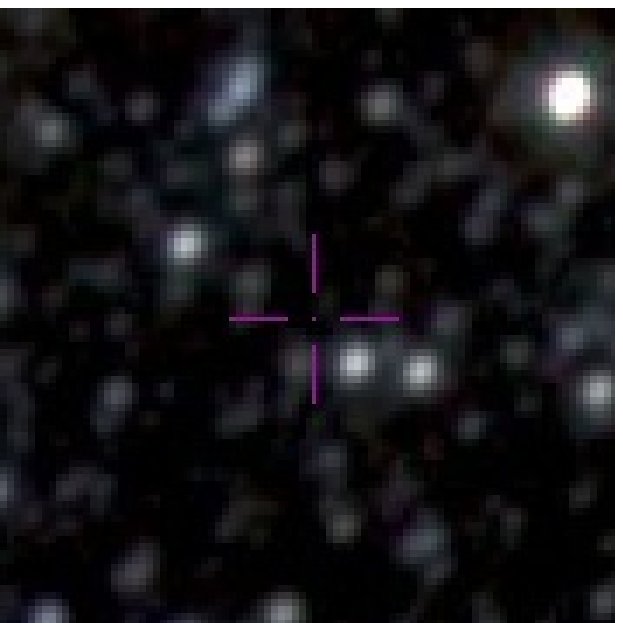} 
\includegraphics[bb= -4.4cm  -0.5cm   0cm  8cm, scale=0.65]{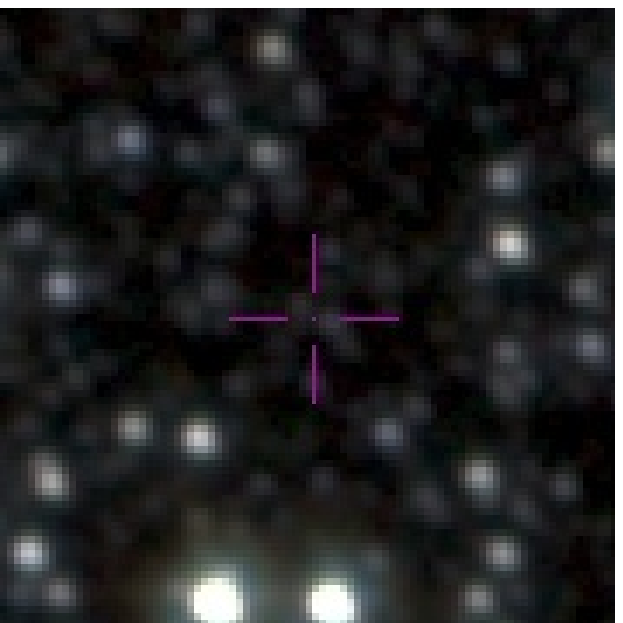} 
\includegraphics[bb= -6.4cm  -0.5cm   0cm  8cm, scale=0.65]{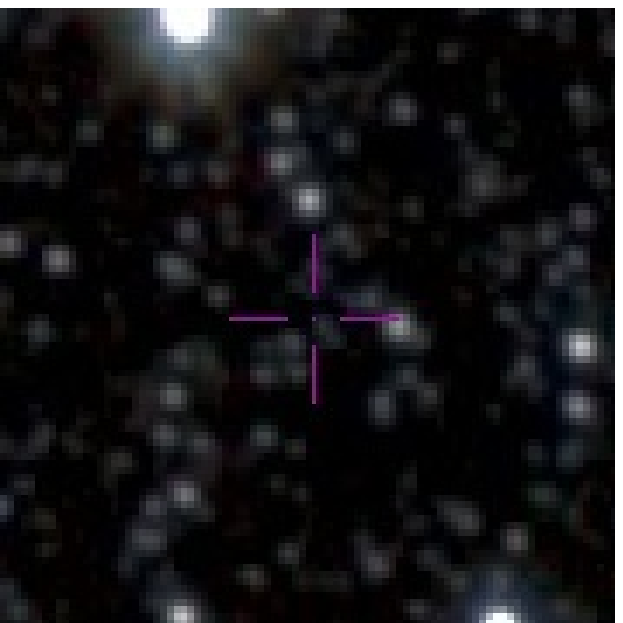} 
\includegraphics[bb= -6.4cm  -0.5cm   0cm  8cm, scale=0.65]{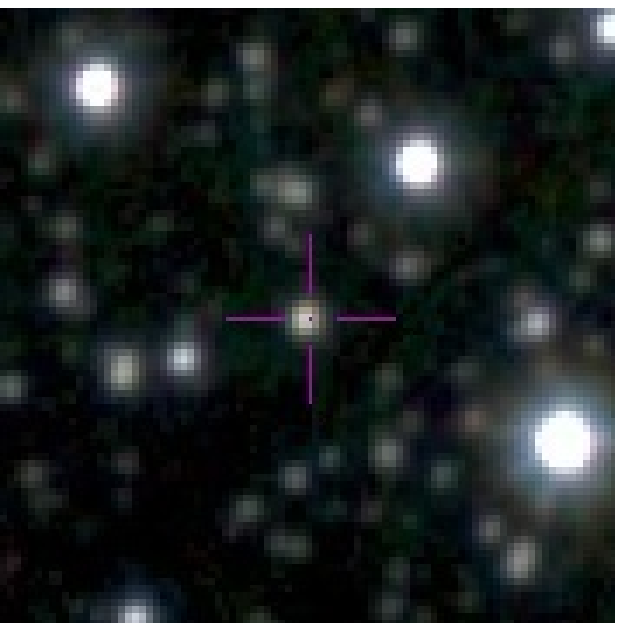}
 
 \caption{First row:  Nova Sgr 1893, Nova  Sgr 1897, Nova  Sgr 1899 and
   Nova Sco  1901. Second row: Nova  Sgr 1901, Nova Sgr  1905, Nova Cir
   1906 and  Nova Sco 1906. Third  row: Nova Sgr 1910,  Nova Sgr 1914c,
   Nova Sgr 1917  and Nova Sgr 1919.  Fourth row: Nova  Sco 1922 , Nova
   Sgr 1924,  Nova Sgr 1926, Nova  Sgr 1926a. Last row:  Nova Sco 1928,
   Nova Sgr 1928, Nova Sgr 1930 and Nova Cen 1931}
 \label{nova_01}
 \end{figure*}
 
\addtocounter{figure}{-1}

\begin{figure*} 
  \includegraphics[bb= -1.2cm   1.6cm   2cm  8cm, scale=0.65]{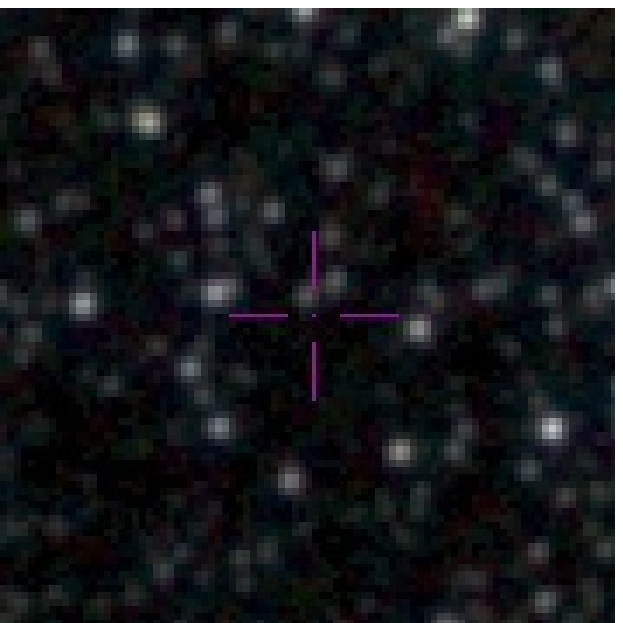}
  \includegraphics[bb= -4.4cm   1.6cm   0cm  8cm, scale=0.65]{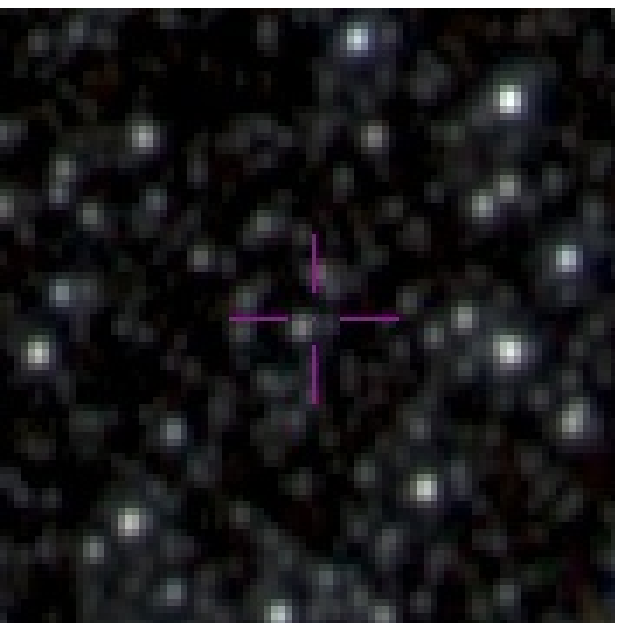}
  \includegraphics[bb= -6.4cm   1.6cm   0cm  8cm, scale=0.65]{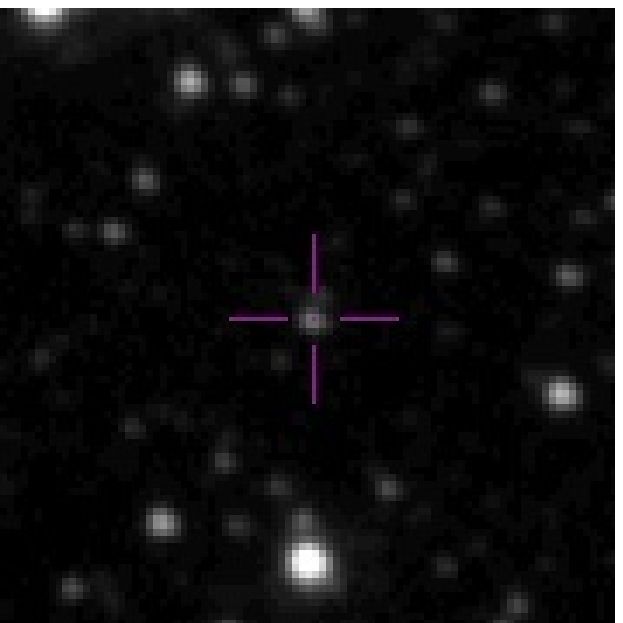}
  \includegraphics[bb= -6.4cm   1.6cm   0cm  8cm, scale=0.65]{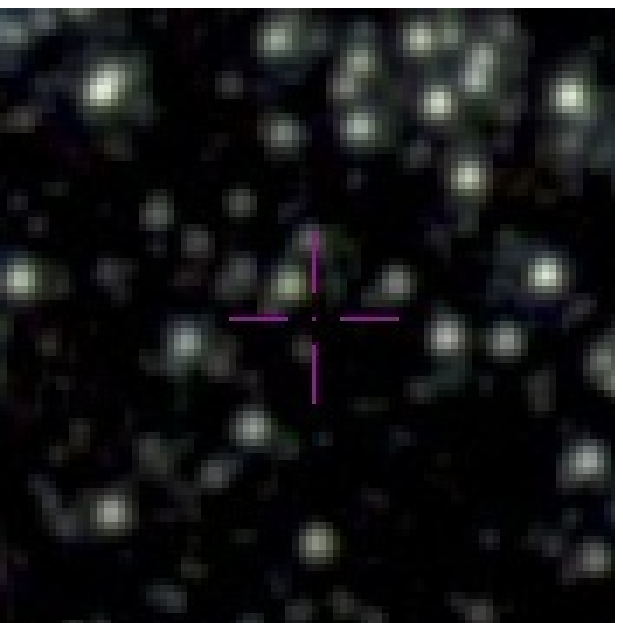}
                                                            
  \includegraphics[bb= -1.2cm   1.6cm   2cm  8cm, scale=0.65]{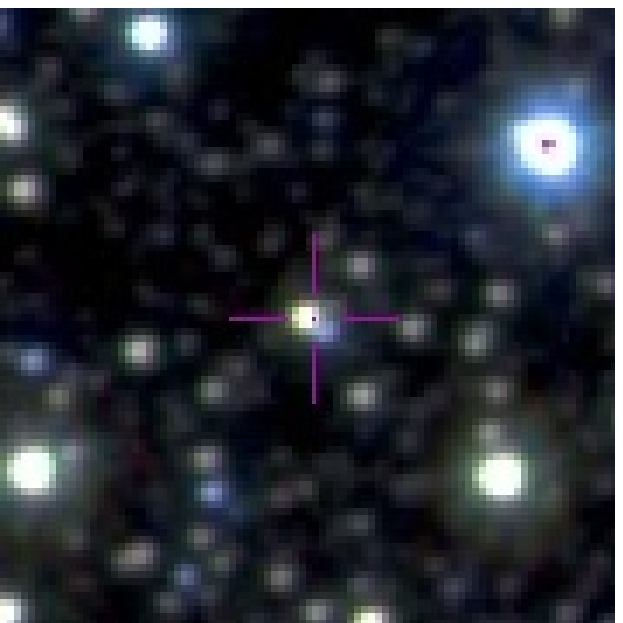}
  \includegraphics[bb= -4.4cm   1.6cm   0cm  8cm, scale=0.65]{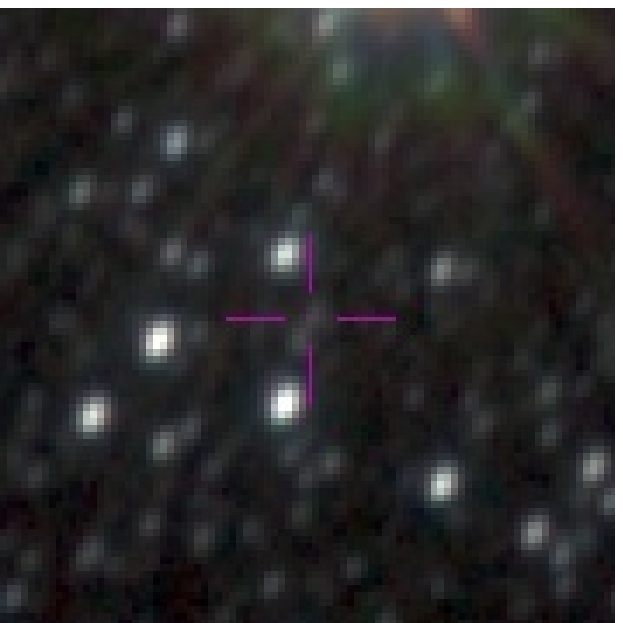} 
  \includegraphics[bb= -6.4cm   1.6cm   0cm  8cm, scale=0.65]{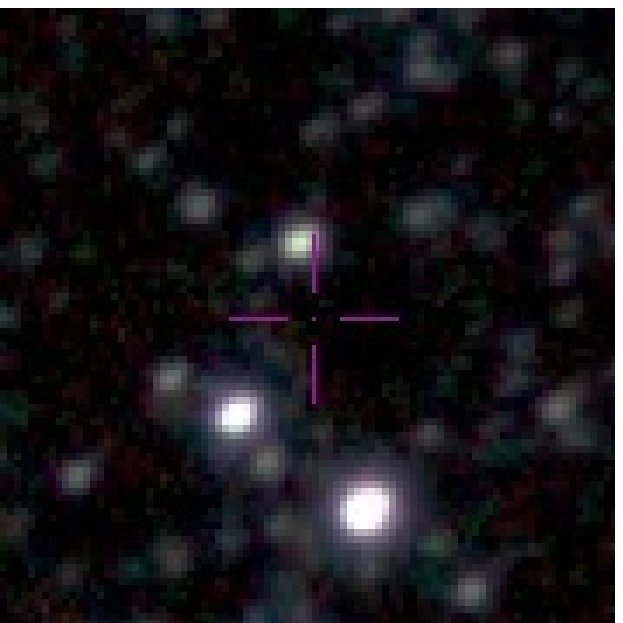} 
  \includegraphics[bb= -6.4cm   1.6cm   0cm  8cm, scale=0.65]{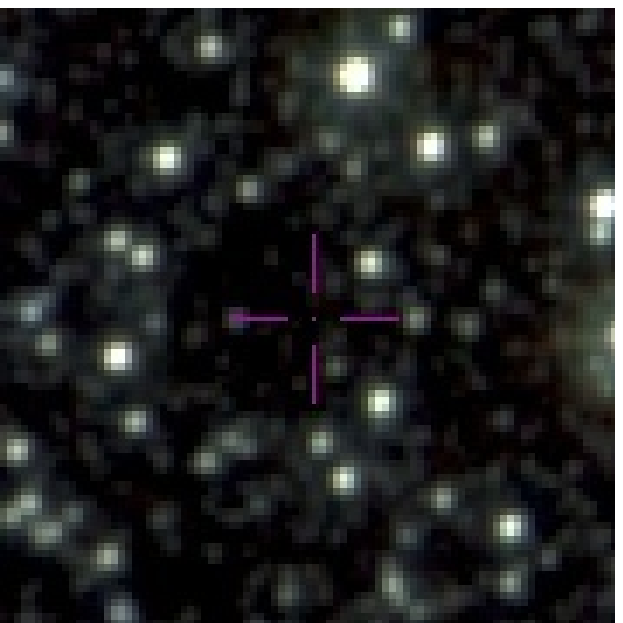} 
                                                            
  \includegraphics[bb= -1.2cm   1.6cm   2cm  8cm, scale=0.65]{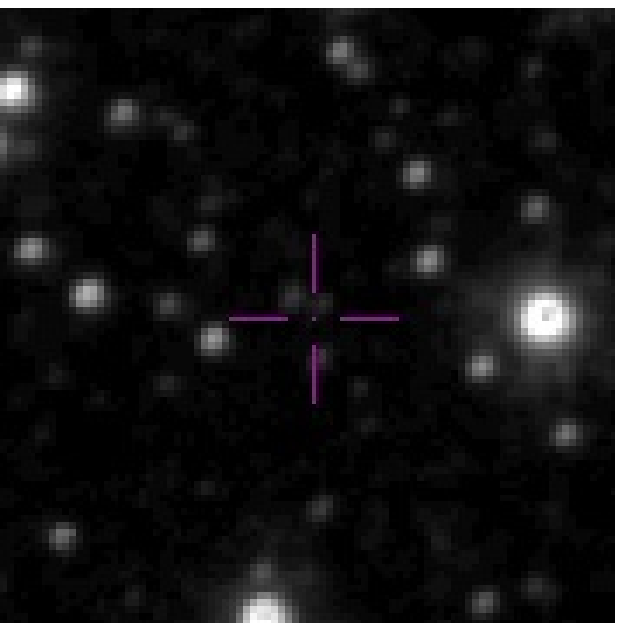}
  \includegraphics[bb= -4.4cm   1.6cm   0cm  8cm, scale=0.65]{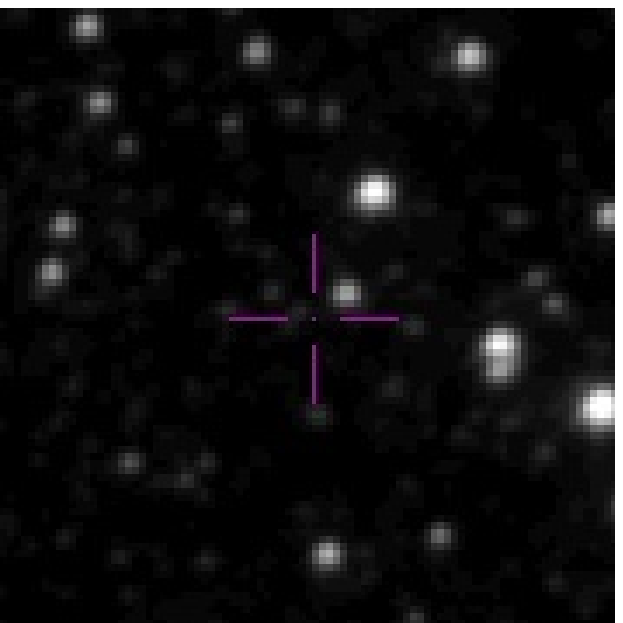}
  \includegraphics[bb= -6.4cm   1.6cm   0cm  8cm, scale=0.65]{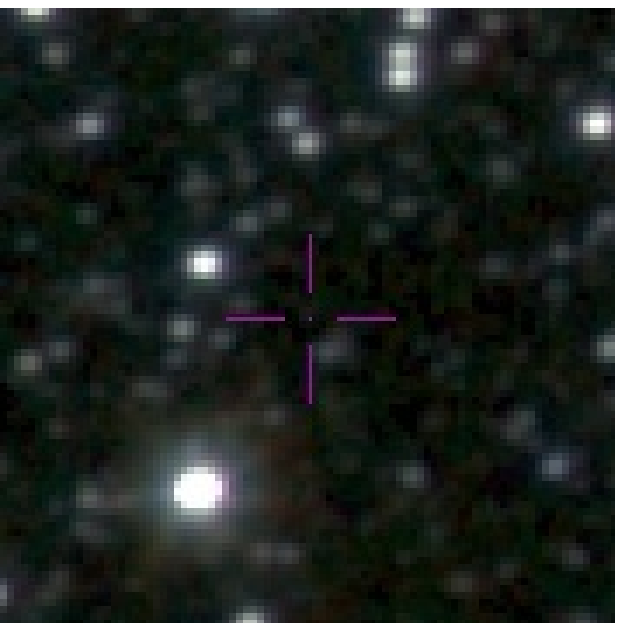}
  \includegraphics[bb= -6.4cm   1.6cm   0cm  8cm, scale=0.65]{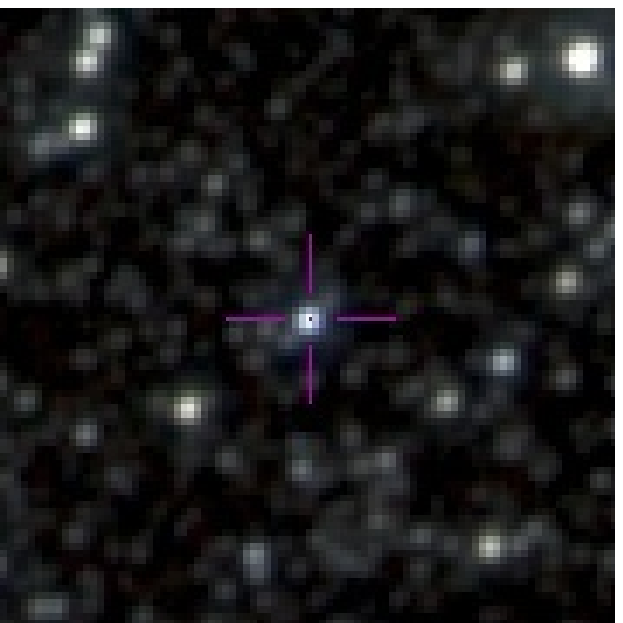}  
                                                            
  \includegraphics[bb= -1.2cm   1.6cm   2cm  8cm, scale=0.65]{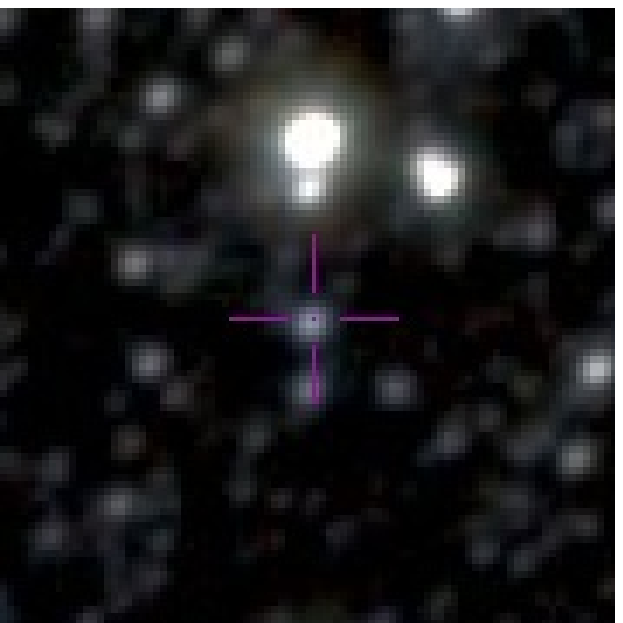}  
  \includegraphics[bb= -4.4cm   1.6cm   0cm  8cm, scale=0.65]{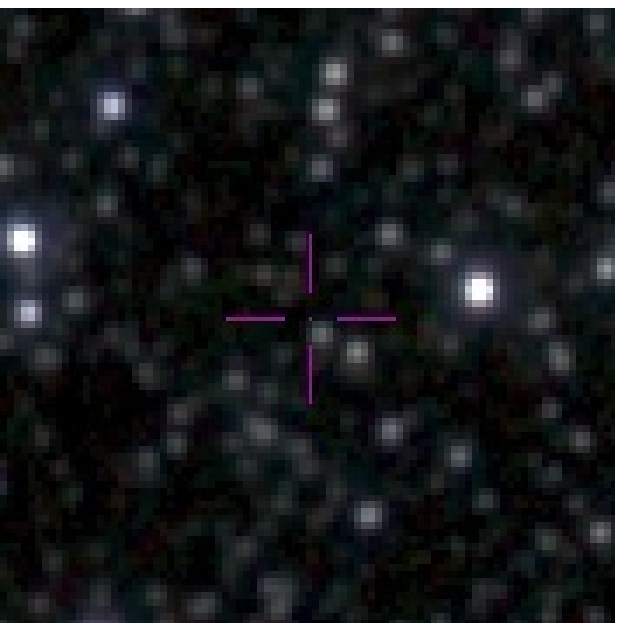}  
  \includegraphics[bb= -6.4cm   1.6cm   0cm  8cm, scale=0.65]{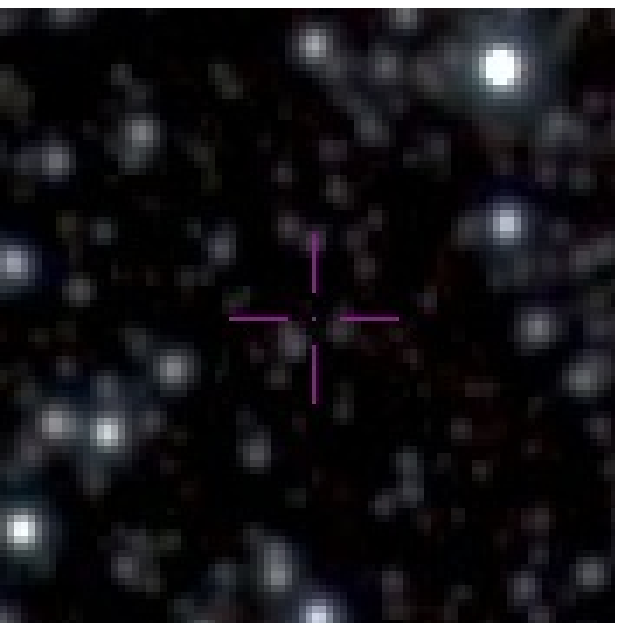} 
  \includegraphics[bb= -6.4cm   1.6cm   0cm  8cm, scale=0.65]{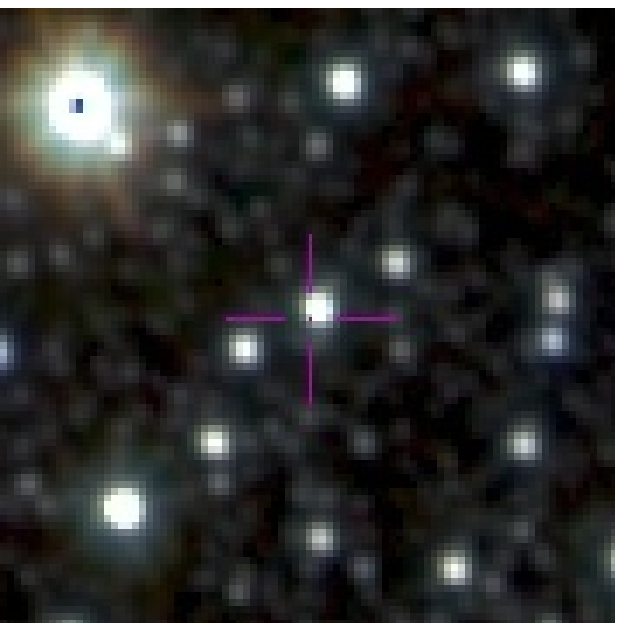} 
                                                            
  \includegraphics[bb= -1.2cm  -0.5cm   2cm  8cm, scale=0.65]{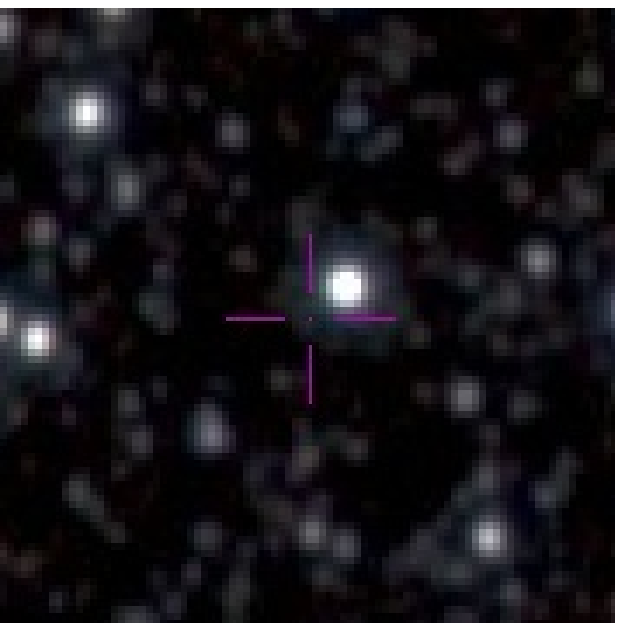} 
  \includegraphics[bb= -4.4cm  -0.5cm   0cm  8cm, scale=0.65]{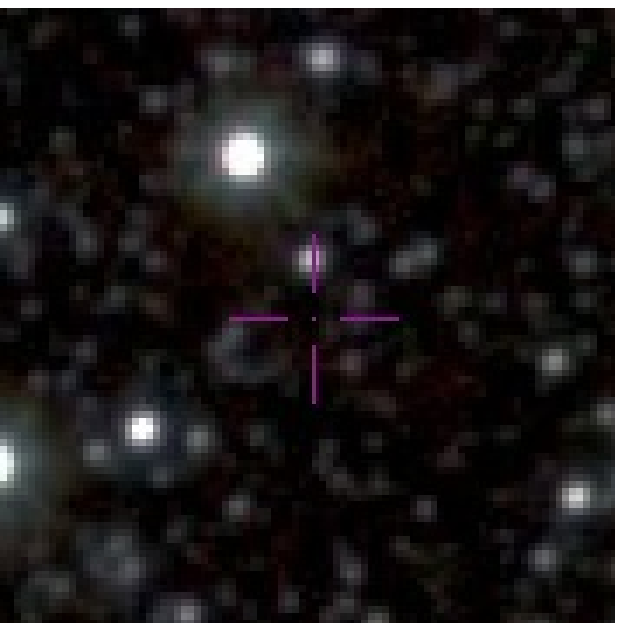}  
  \includegraphics[bb= -6.4cm  -0.5cm   0cm  8cm, scale=0.65]{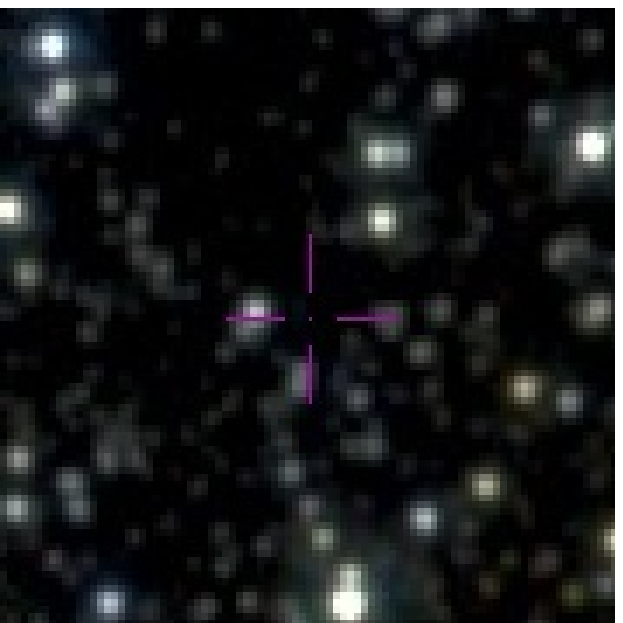}  
  \includegraphics[bb= -6.4cm  -0.5cm   0cm  8cm, scale=0.65]{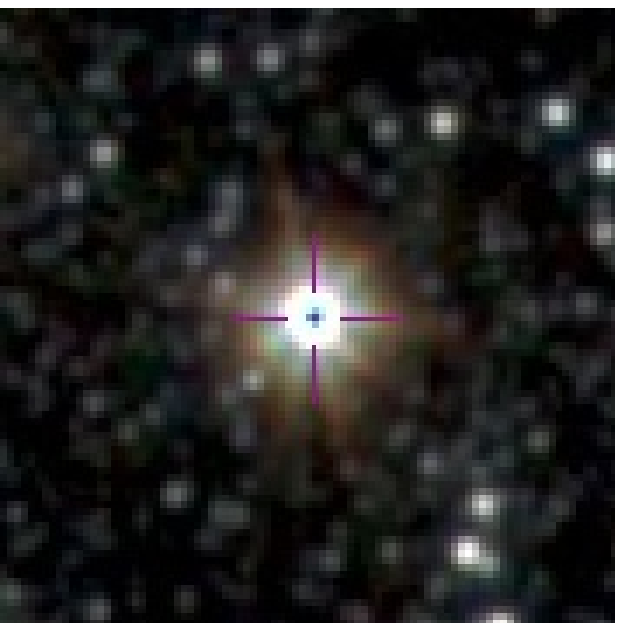}  
\caption{{\bf cont.}   First row: Nova  Sgr 1932, Nova Sgr  1933, Nova
  Cru 1935  ($K_{\rm s}$-band  image) and Nova  Sco 1935.  Second row:
  Nova  Sgr  1936,  Nova  Sgr  1936b,  Nova Sgr  1936c  and  Nova  Sgr
  1936d. Third row:  Nova Sgr 1937 ($K_{\rm s}$-band  image), Nova Oph
  1940  ($K_{\rm s}$-band  image), Nova  Sco 1941  and Nova  Sgr 1943.
  Fourth row:  Nova Sco 1944, Nova  Sgr 1944, Nova Sgr  1945a and Nova
  Sgr 1945b.  Last row: Nova Sgr  1947a, Nova Sgr 1948,  Nova Sco 1949
  and Nova Oph 1950.}
\label{novae_02}
\end{figure*}

\addtocounter{figure}{-1}

\begin{figure*} 
  \includegraphics[bb= -1.2cm   1.6cm   2cm  8cm, scale=0.65]{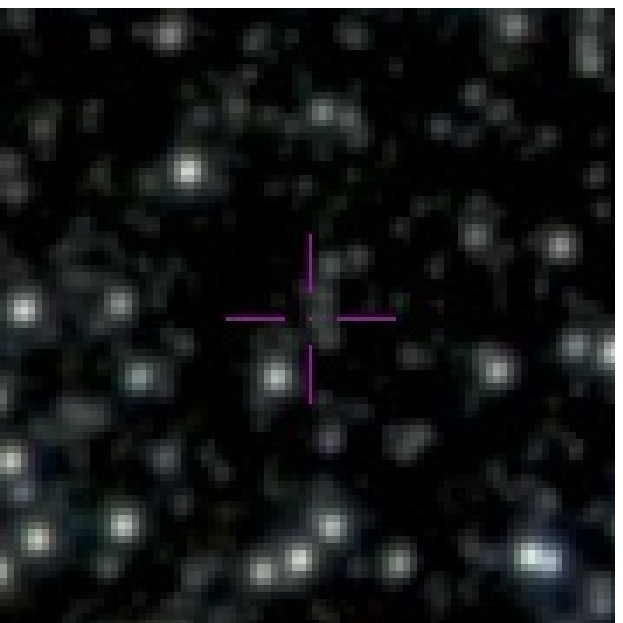} 
  \includegraphics[bb= -4.4cm   1.6cm   0cm  8cm, scale=0.65]{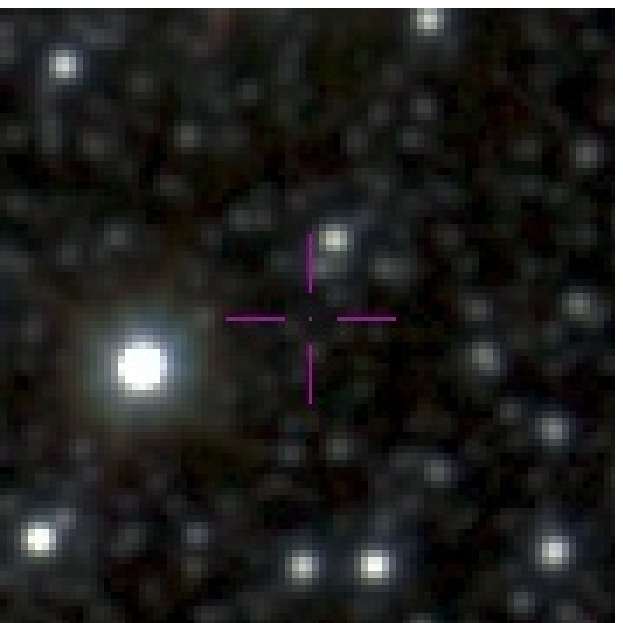} 
  \includegraphics[bb= -6.4cm   1.6cm   0cm  8cm, scale=0.65]{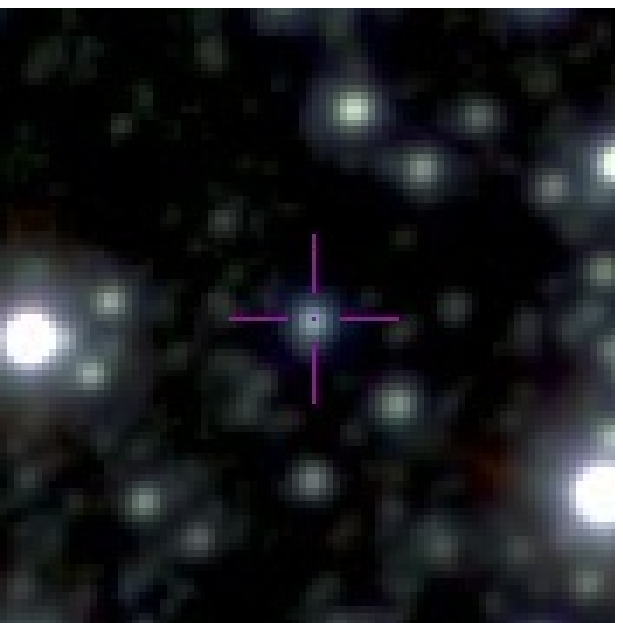} 
  \includegraphics[bb= -6.4cm   1.6cm   0cm  8cm, scale=0.65]{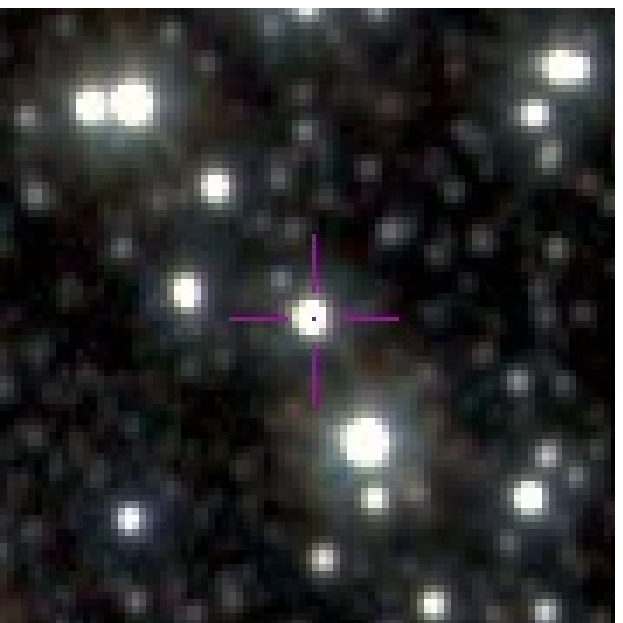} 
                                                            
  \includegraphics[bb= -1.2cm   1.6cm   2cm  8cm, scale=0.65]{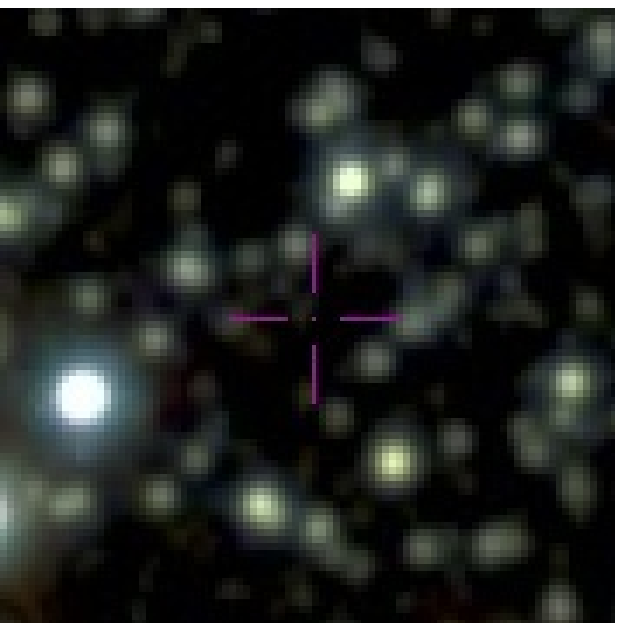} 
  \includegraphics[bb= -4.4cm   1.6cm   0cm  8cm, scale=0.65]{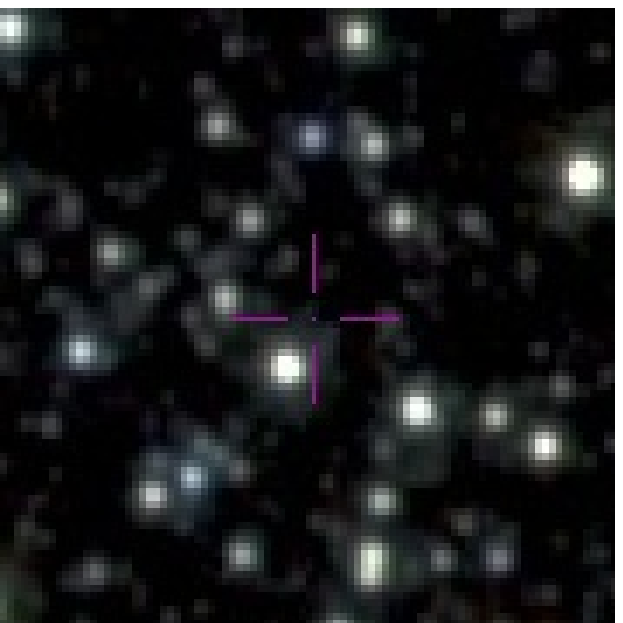}    
  \includegraphics[bb= -6.4cm   1.6cm   0cm  8cm, scale=0.65]{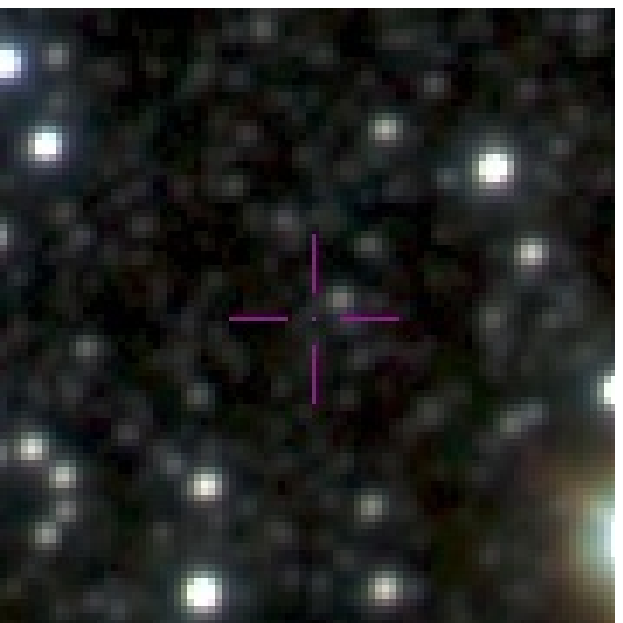}
  \includegraphics[bb= -6.4cm   1.6cm   0cm  8cm, scale=0.65]{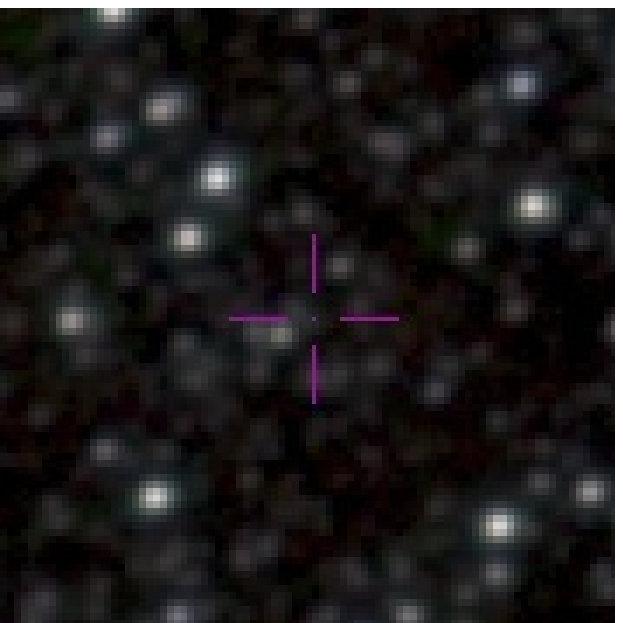}
                                                            
  \includegraphics[bb= -1.2cm   1.6cm   2cm  8cm, scale=0.65]{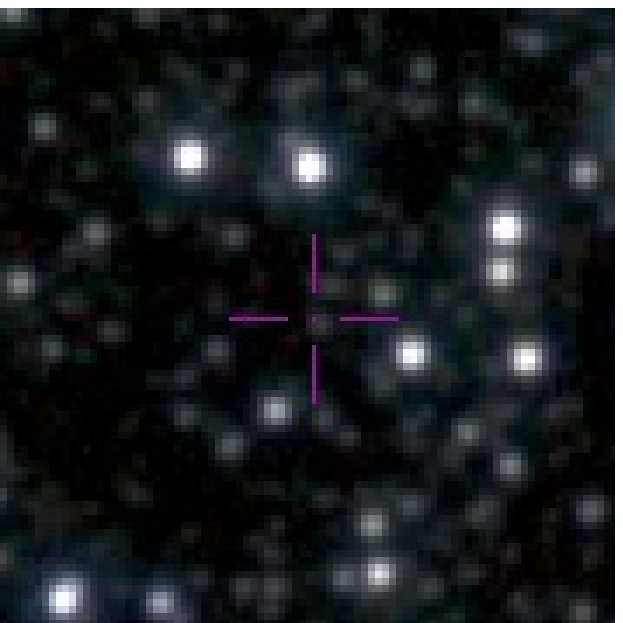}
  \includegraphics[bb= -4.4cm   1.6cm   0cm  8cm, scale=0.65]{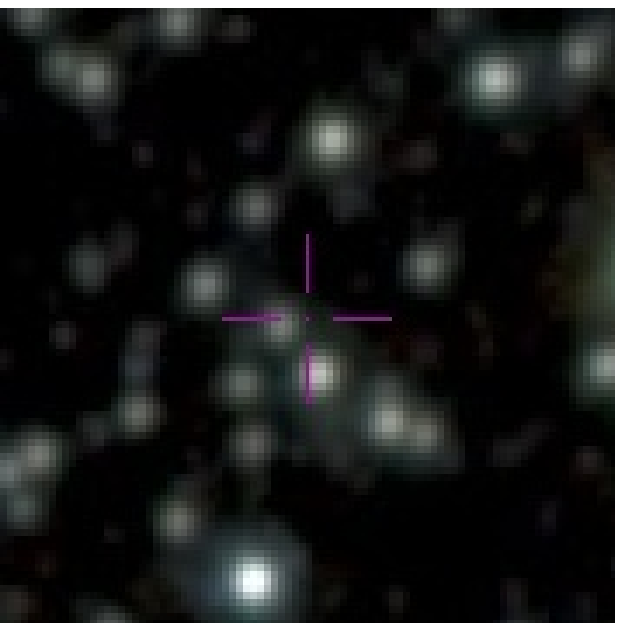}
  \includegraphics[bb= -6.4cm   1.6cm   0cm  8cm, scale=0.65]{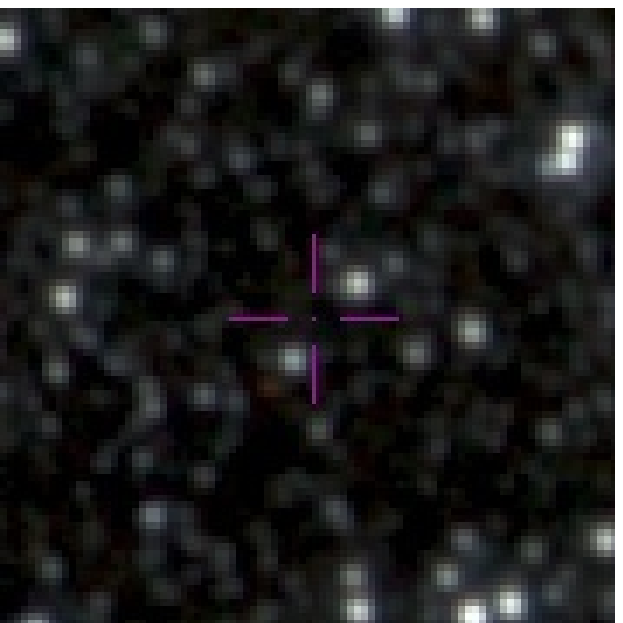}
  \includegraphics[bb= -6.4cm   1.6cm   0cm  8cm, scale=0.65]{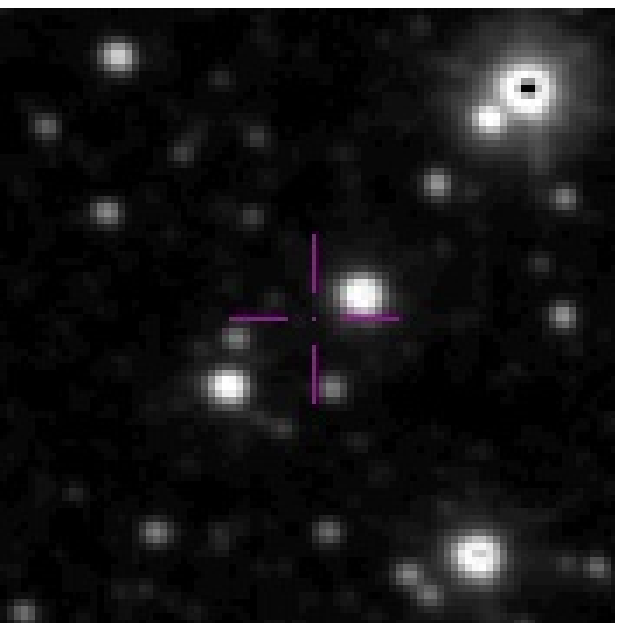}
                                                            
  \includegraphics[bb= -1.2cm   1.6cm   2cm  8cm, scale=0.65]{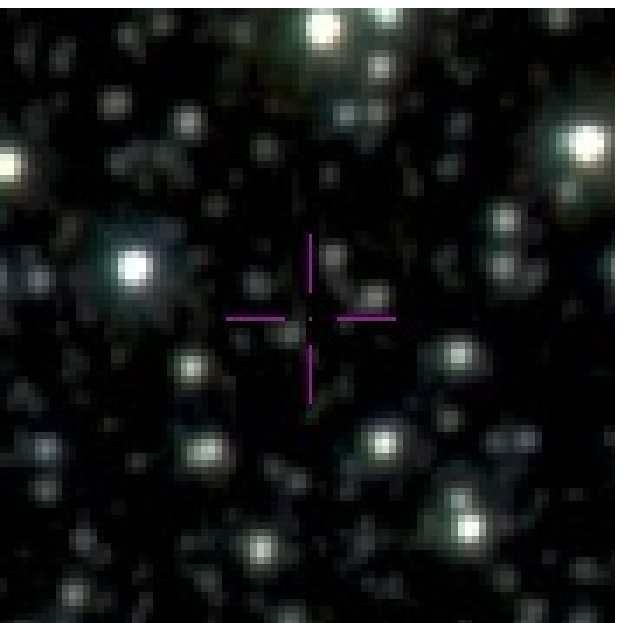}
  \includegraphics[bb= -4.4cm   1.6cm   0cm  8cm, scale=0.65]{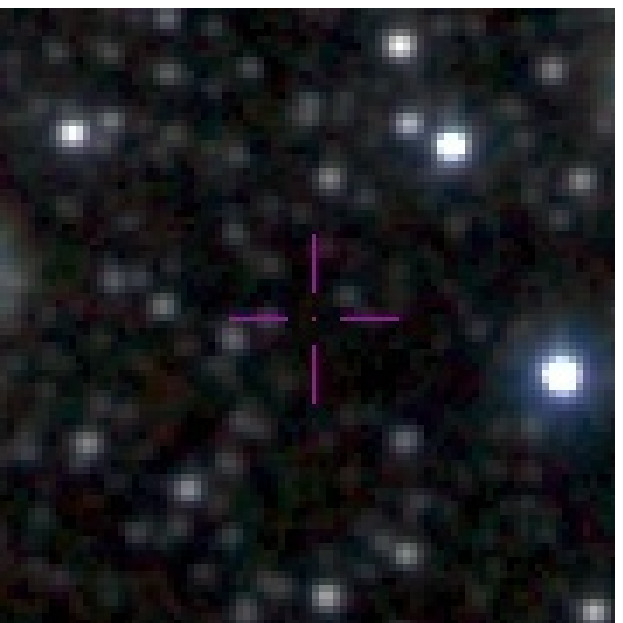} 
  \includegraphics[bb= -6.4cm   1.6cm   0cm  8cm, scale=0.65]{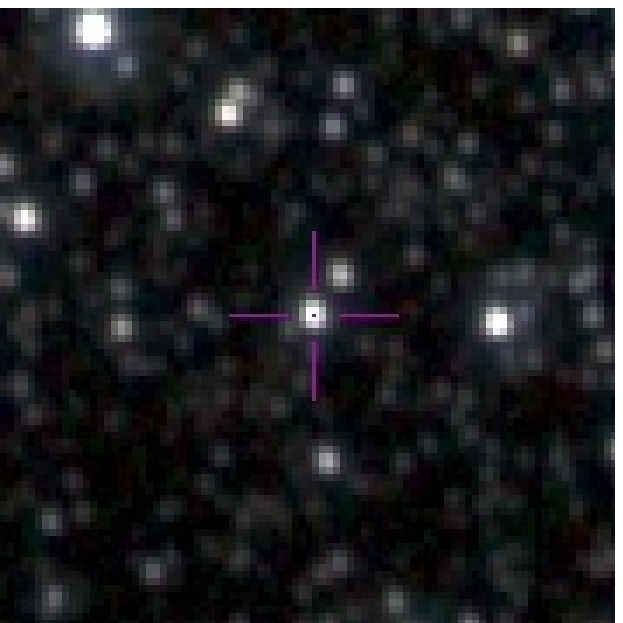} 
  \includegraphics[bb= -6.4cm   1.6cm   0cm  8cm, scale=0.65]{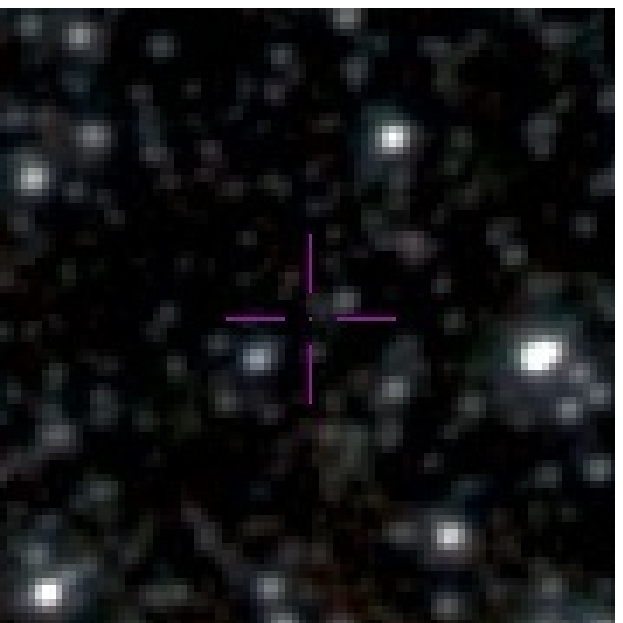}  
                                                            
  \includegraphics[bb= -1.2cm  -0.5cm   2cm  8cm, scale=0.65]{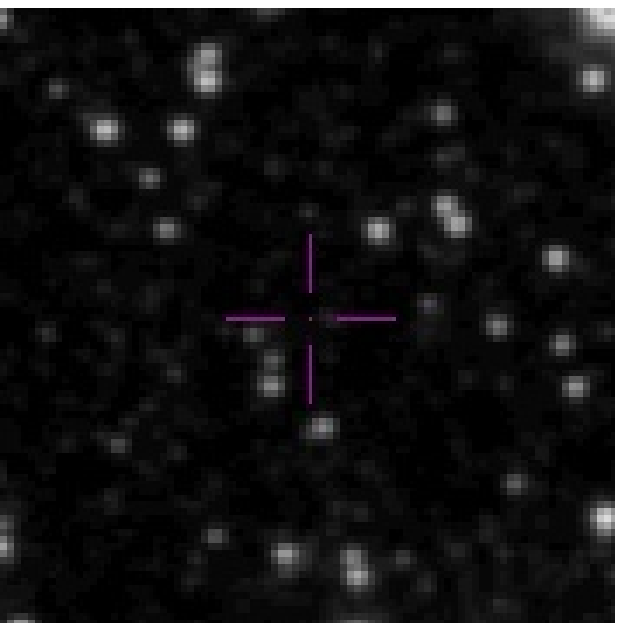}
  \includegraphics[bb= -4.4cm  -0.5cm   0cm  8cm, scale=0.65]{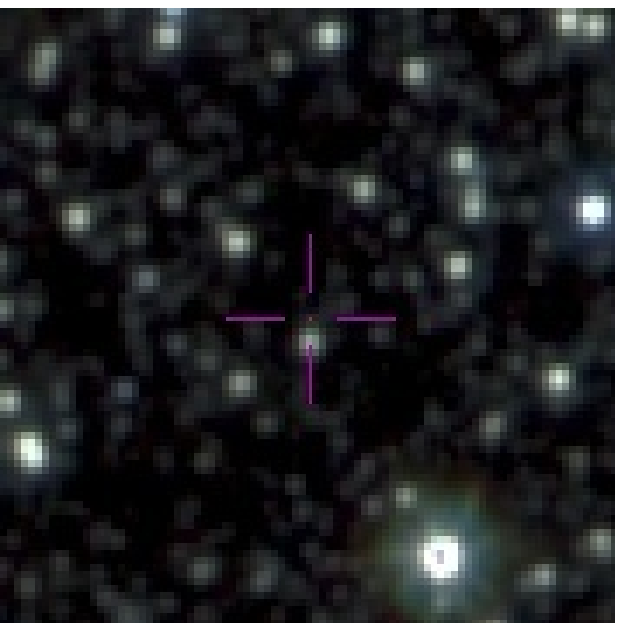}  
  \includegraphics[bb= -6.4cm  -0.5cm   0cm  8cm, scale=0.65]{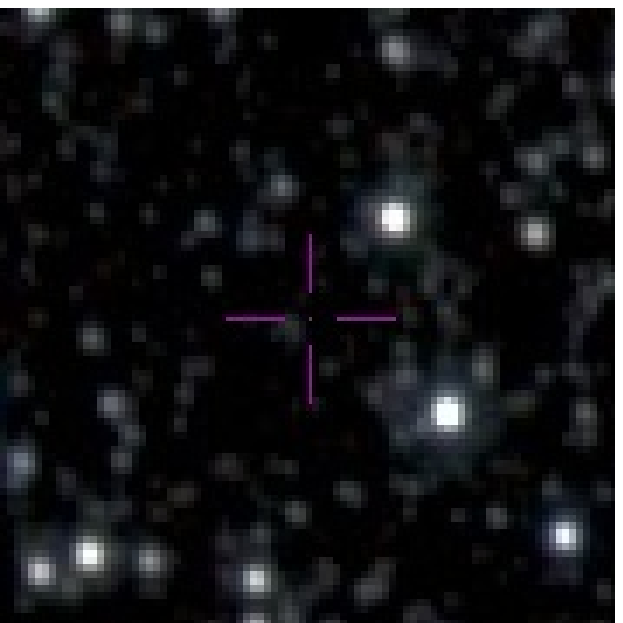}  
  \includegraphics[bb= -6.4cm  -0.5cm   0cm  8cm, scale=0.65]{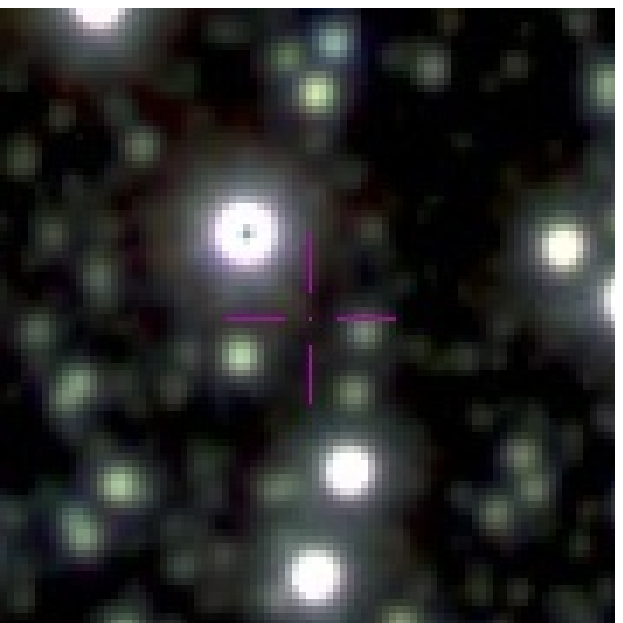}  
\caption{{\bf cont.} First  row: Nova Sco 1950a, Nova  Sco 1950b, Nova
  Sco 1950c and  Nova Sgr 1951a. Second row: Nova  Sgr 1951b, Nova Sco
  1952, Nova Sco 1952a and Nova  Sco 1952b. Third row: Nova Sgr 1952a,
  Nova Sgr  1952b, Nova Sgr 1953  and Nova Oph  1954 ($K_{\rm s}$-band
  image). Fourth  row: Nova Sco 1954,  Nova Sgr 1954a,  Nova Sgr 1954b
  and  Nova Sgr  1955.   Last  row: Nova  Oph  1957 ($K_{\rm  s}$-band
  image), Nova Sgr 1960, Nova Oph 1961 and Nova Sgr 1963.}
\label{novae_03}
\end{figure*}

\addtocounter{figure}{-1}

\begin{figure*} 
  \includegraphics[bb= -1.2cm   1.6cm   2cm  8cm, scale=0.65]{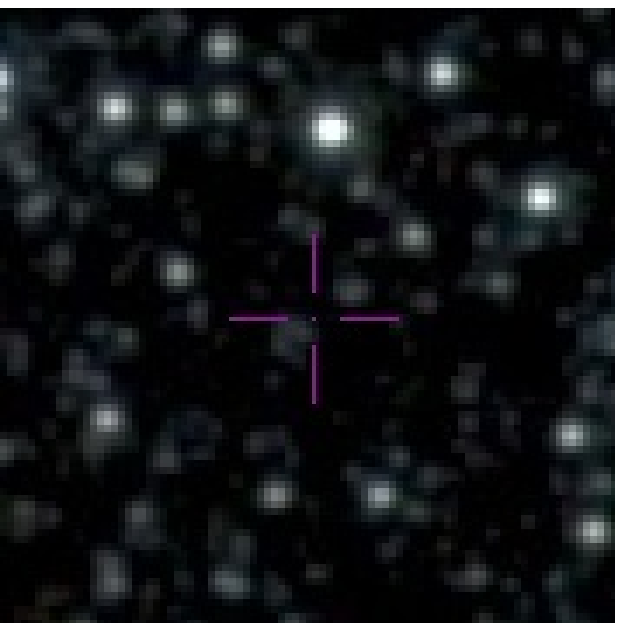}  
  \includegraphics[bb= -4.4cm   1.6cm   0cm  8cm, scale=0.65]{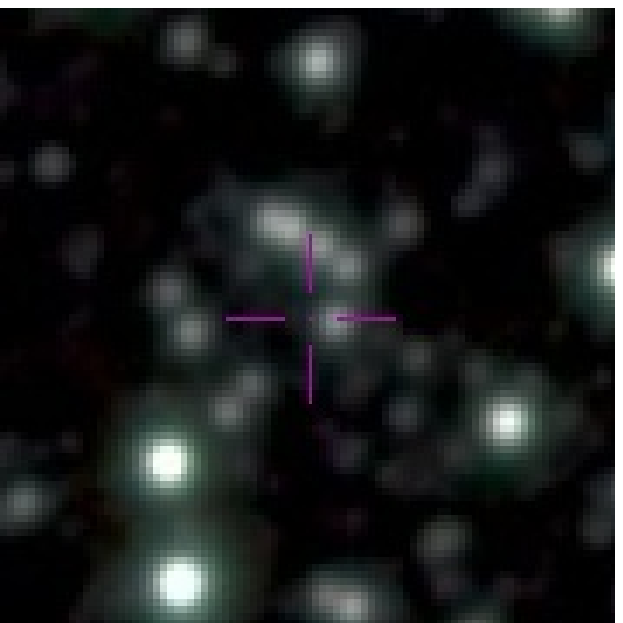}  
  \includegraphics[bb= -6.4cm   1.6cm   0cm  8cm, scale=0.65]{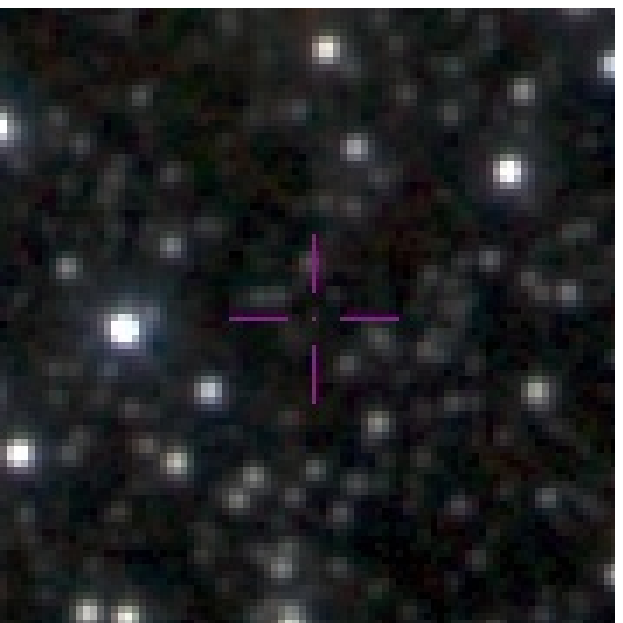}  
  \includegraphics[bb= -6.4cm   1.6cm   0cm  8cm, scale=0.65]{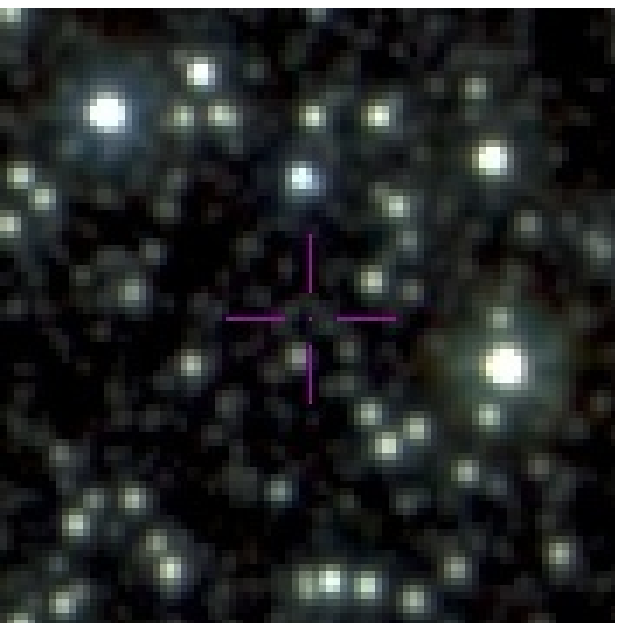}  
                                                            
  \includegraphics[bb= -1.2cm   1.6cm   2cm  8cm, scale=0.65]{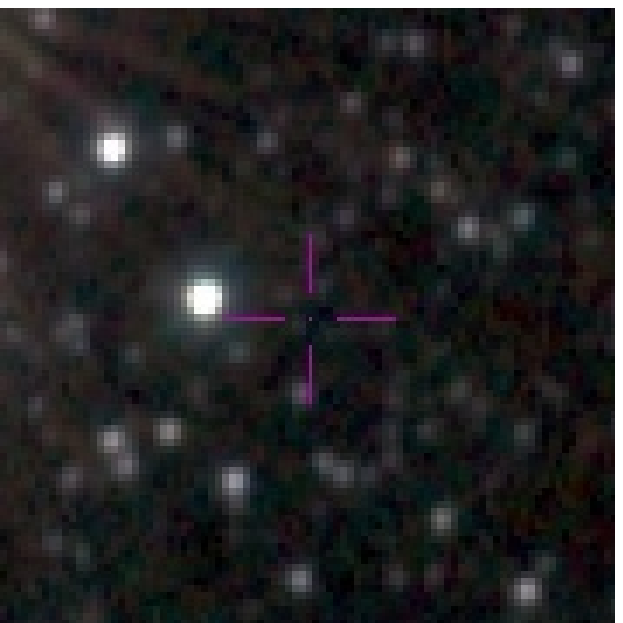}  
  \includegraphics[bb= -4.4cm   1.6cm   0cm  8cm, scale=0.65]{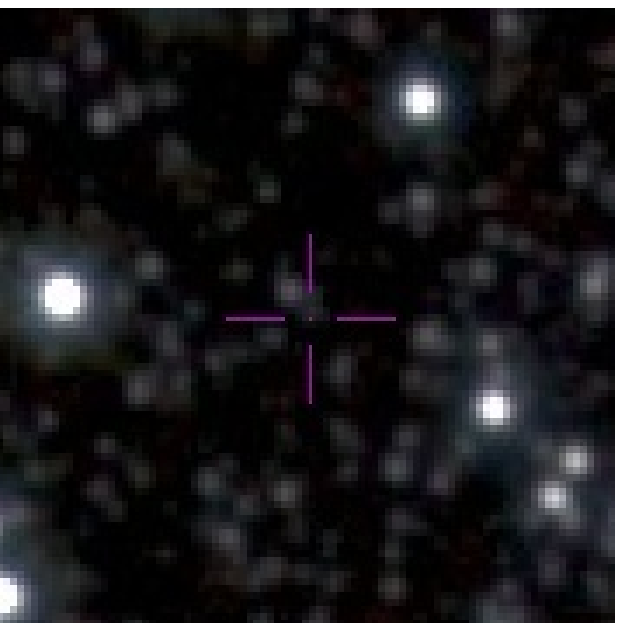} 
  \includegraphics[bb= -6.4cm   1.6cm   0cm  8cm, scale=0.65]{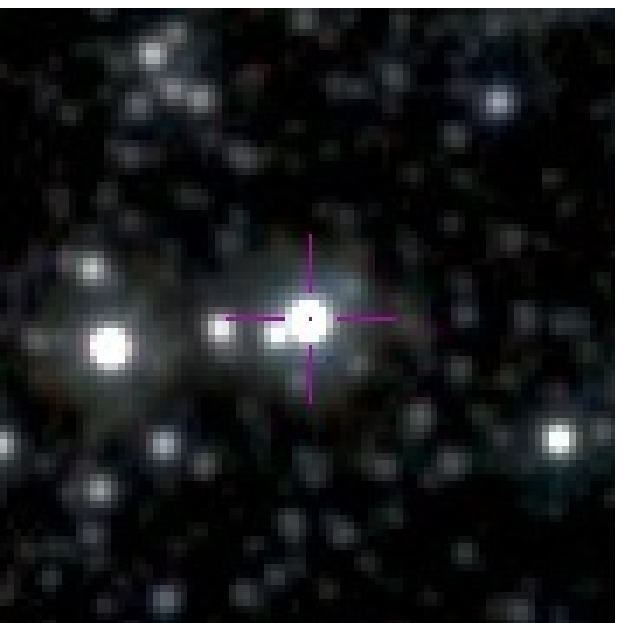}
  \includegraphics[bb= -6.4cm   1.6cm   0cm  8cm, scale=0.65]{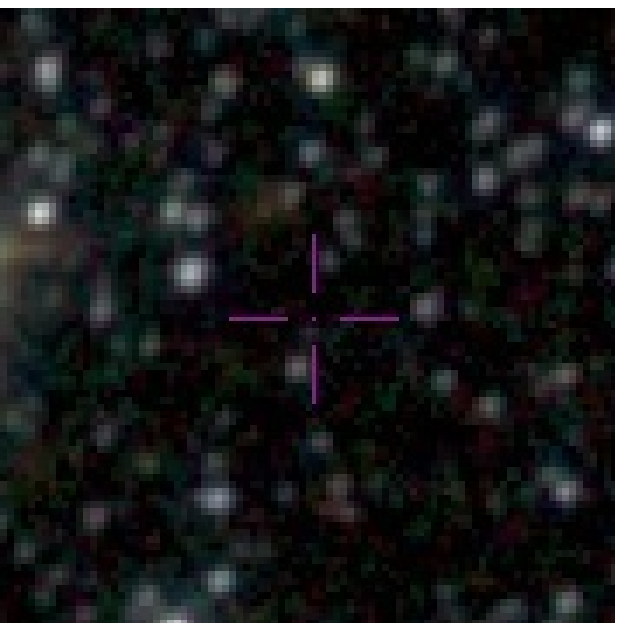}  
                                                            
  \includegraphics[bb= -1.2cm   1.6cm   2cm  8cm, scale=0.65]{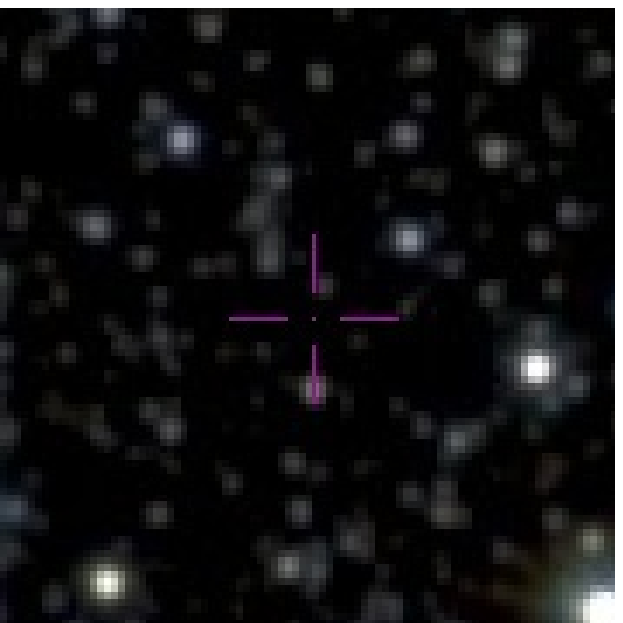} 
  \includegraphics[bb= -4.4cm   1.6cm   0cm  8cm, scale=0.65]{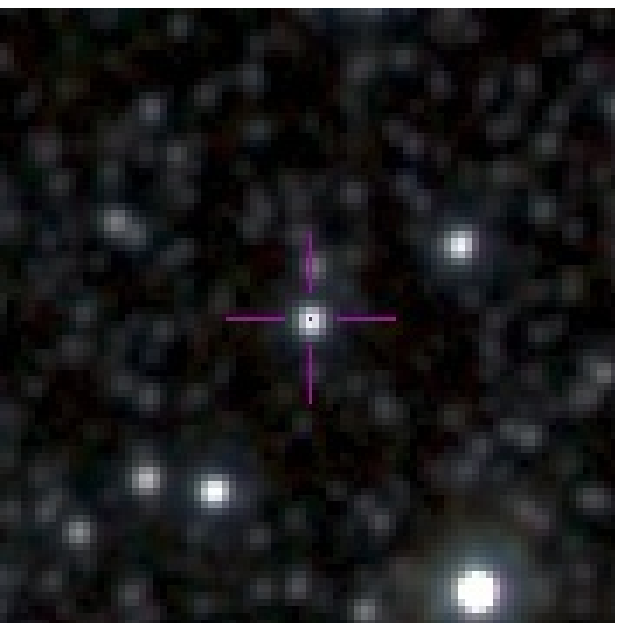}  
  \includegraphics[bb= -6.4cm   1.6cm   0cm  8cm, scale=0.65]{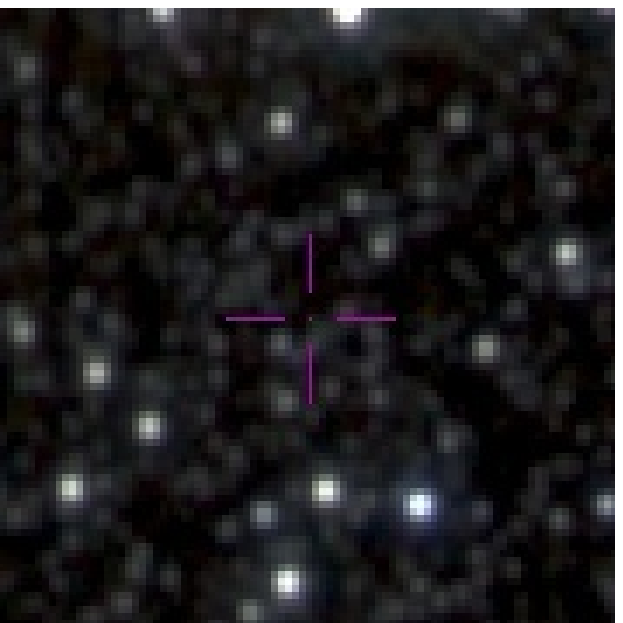}
  \includegraphics[bb= -6.4cm   1.6cm   0cm  8cm, scale=0.65]{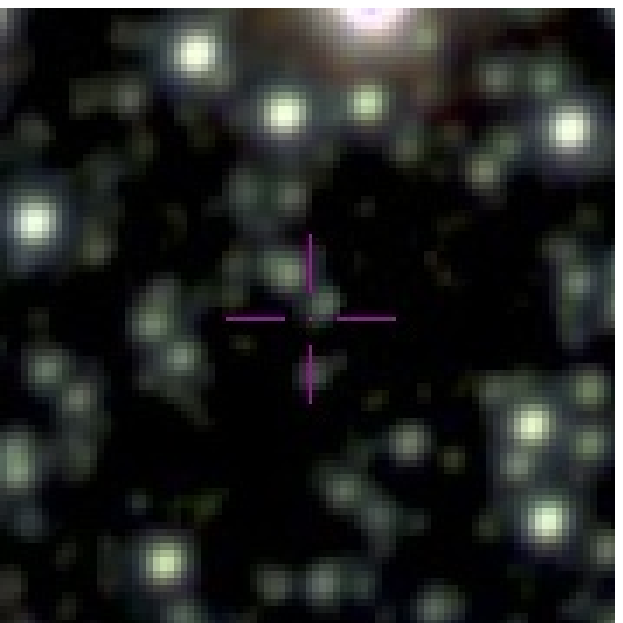}  
                                                            
  \includegraphics[bb= -1.2cm   1.6cm   2cm  8cm, scale=0.65]{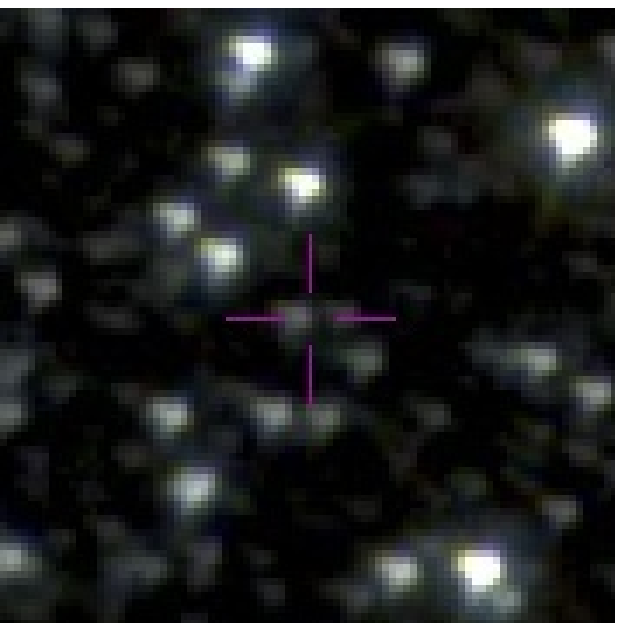}
  \includegraphics[bb= -4.4cm   1.6cm   0cm  8cm, scale=0.65]{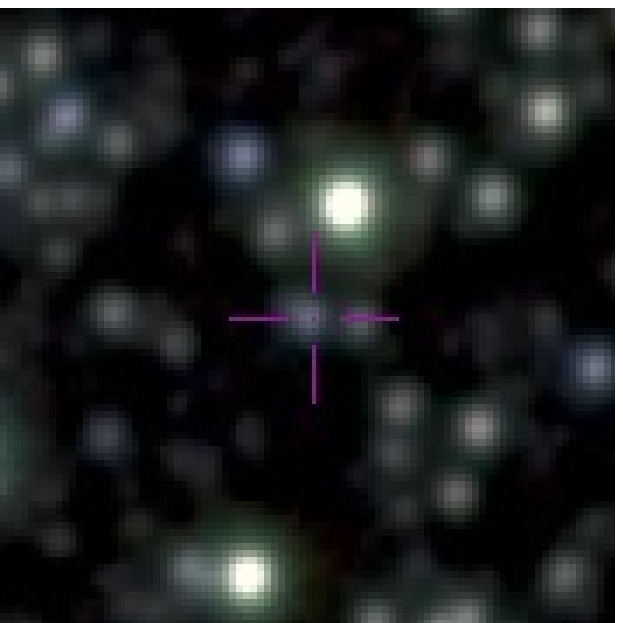}  
  \includegraphics[bb= -6.4cm   1.6cm   0cm  8cm, scale=0.65]{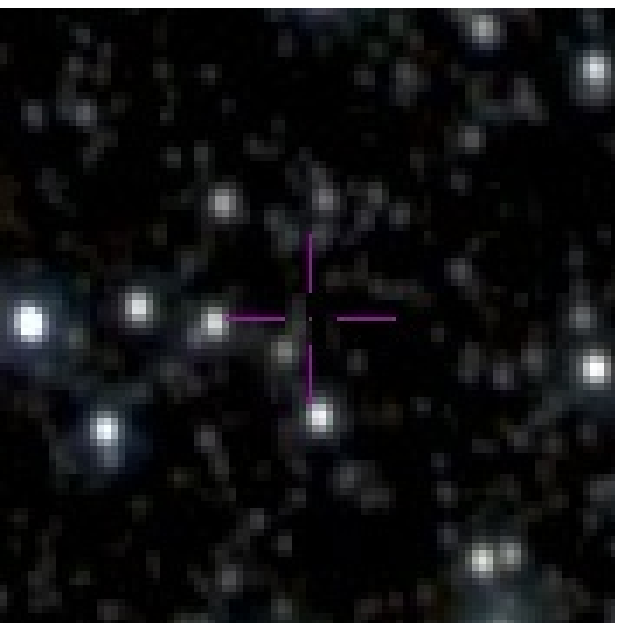}  
  \includegraphics[bb= -6.4cm   1.6cm   0cm  8cm, scale=0.65]{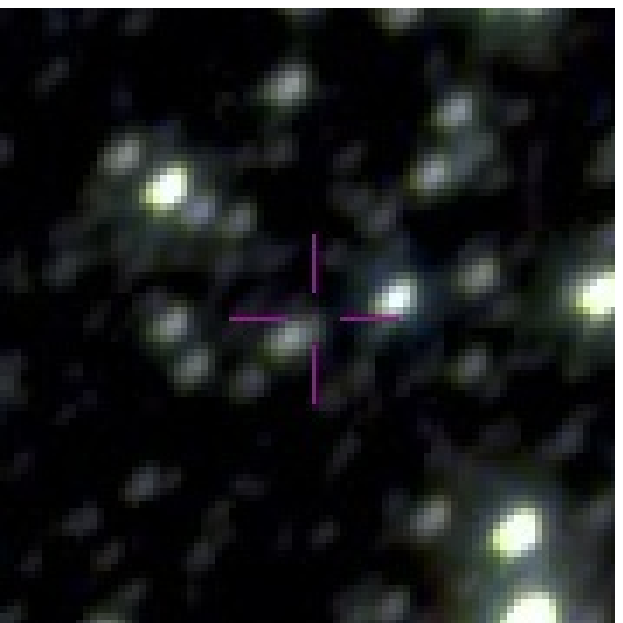} 
                                                            
  \includegraphics[bb= -1.2cm  -0.5cm   2cm  8cm, scale=0.65]{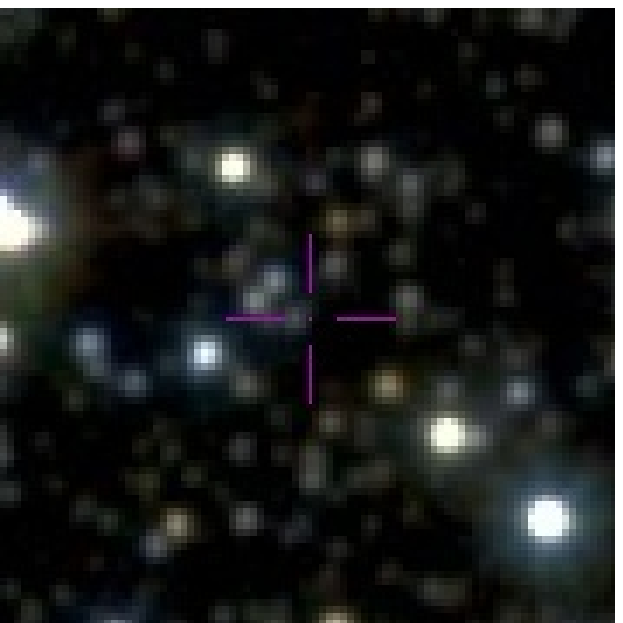}
  \includegraphics[bb= -4.4cm  -0.5cm   0cm  8cm, scale=0.65]{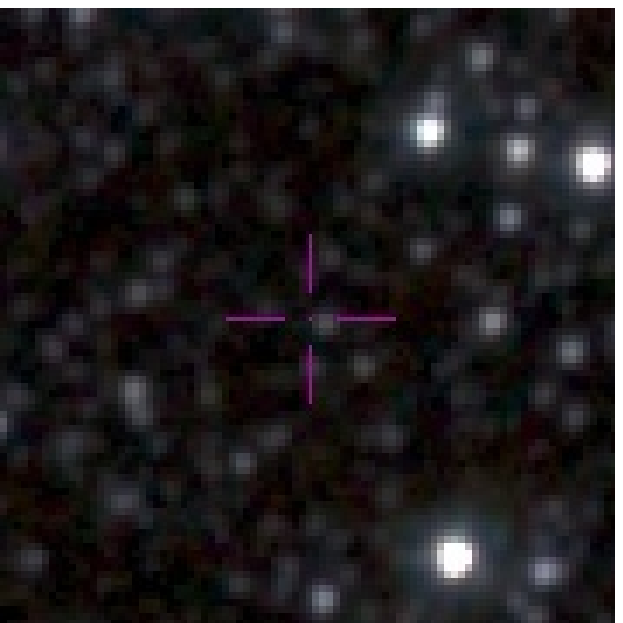}
  \includegraphics[bb= -6.4cm  -0.5cm   0cm  8cm, scale=0.65]{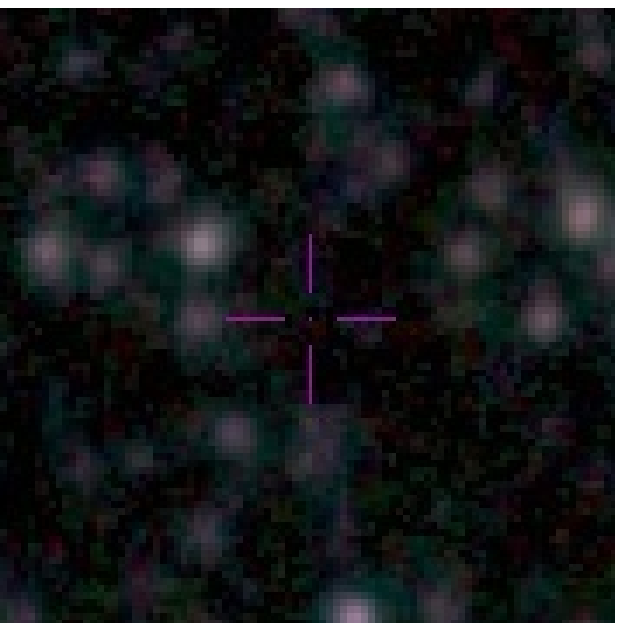}
  \includegraphics[bb= -6.4cm  -0.5cm   0cm  8cm, scale=0.65]{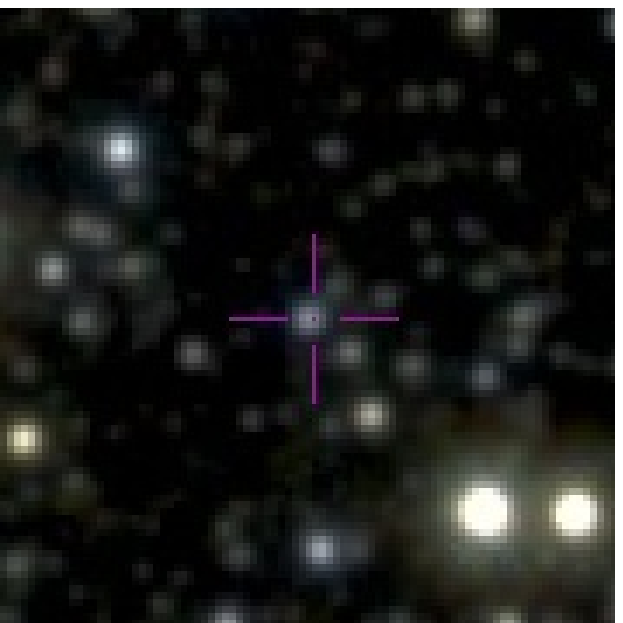} 
\caption{{\bf cont.} First row:  Nova Sco 1964, Nova  Sgr 1968, Nova  Sgr 1974 and
  Nova Sgr  1975. Second row: Nova  Sgr 1977, Nova Sgr  1978, Nova Sgr
  1980 and  Nova Sgr 1982.  Third row: Nova  Nor 1983, Nova  Oph 1983,
  Nova Sgr 1983 and Nova Sgr 1984. Fourth row: Nova Sco 1985, Nova Sgr
  1986, Nova  Sgr 1987 and  Nova Sco 1989b.  Last row: Nova  Cen 1991,
  Nova Oph 1991b, Nova Sgr 1991 and Nova Sco 1992.}
\label{novae_04}
\end{figure*}

\addtocounter{figure}{-1}

\begin{figure*} 
  \includegraphics[bb= -1.2cm   1.6cm   2cm  8cm, scale=0.65]{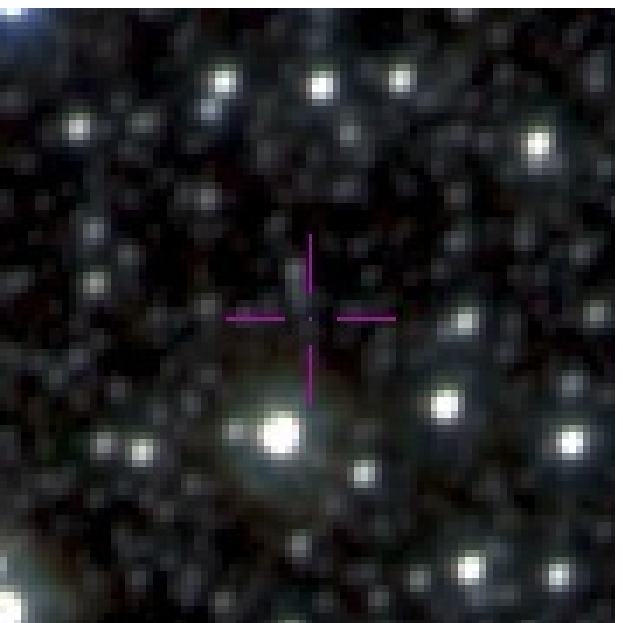} 
  \includegraphics[bb= -4.4cm   1.6cm   0cm  8cm, scale=0.65]{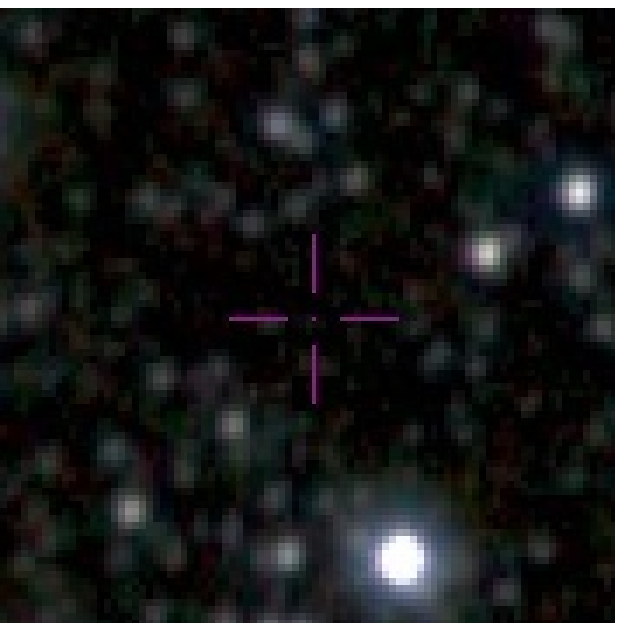}
  \includegraphics[bb= -6.4cm   1.6cm   0cm  8cm, scale=0.65]{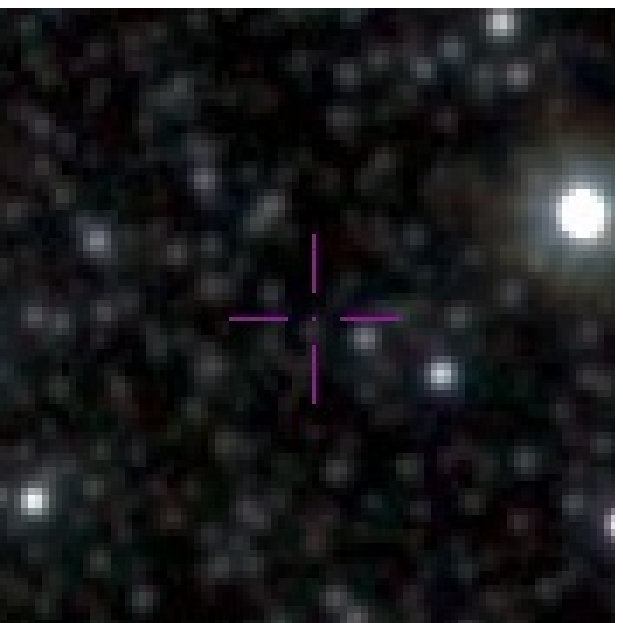} 
  \includegraphics[bb= -6.4cm   1.6cm   0cm  8cm, scale=0.65]{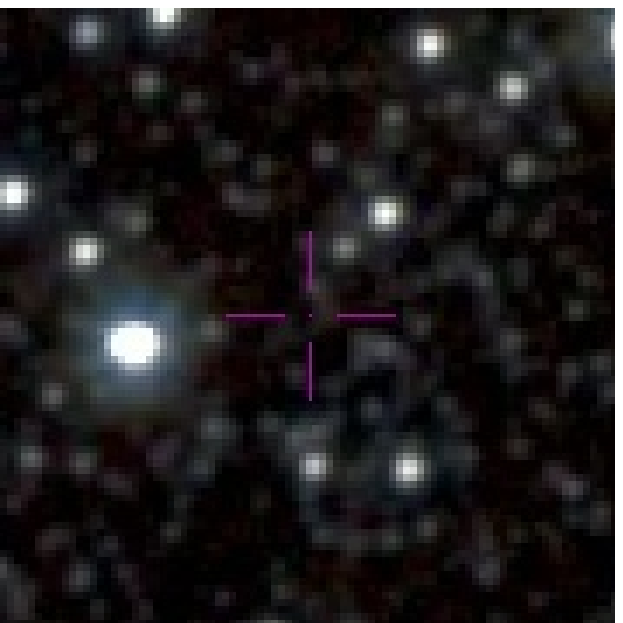}  
                                                            
  \includegraphics[bb= -1.2cm   1.6cm   2cm  8cm, scale=0.65]{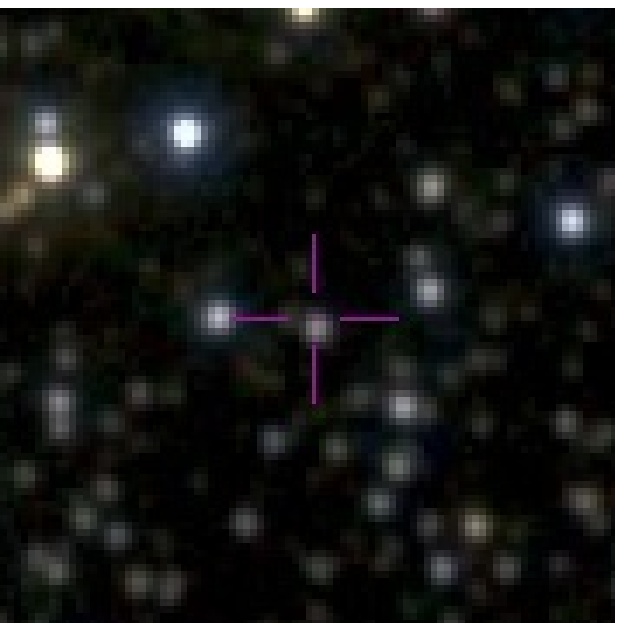}  
  \includegraphics[bb= -4.4cm   1.6cm   0cm  8cm, scale=0.65]{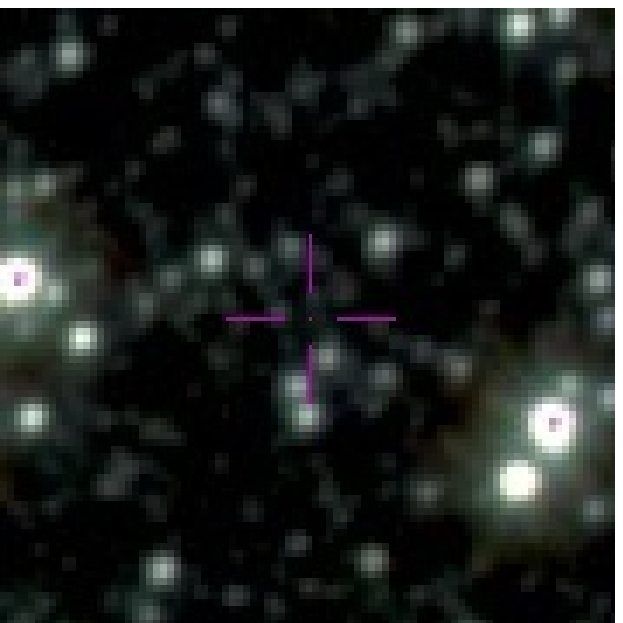}  
  \includegraphics[bb= -6.4cm   1.6cm   0cm  8cm, scale=0.65]{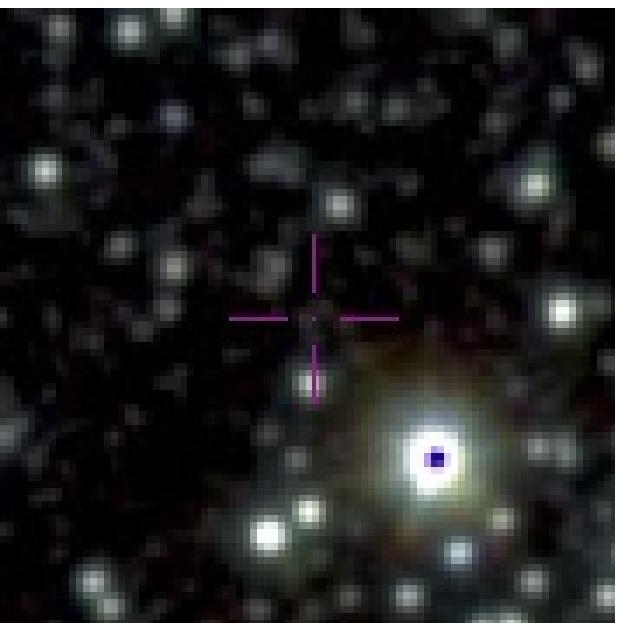}
  \includegraphics[bb= -6.4cm   1.6cm   0cm  8cm, scale=0.65]{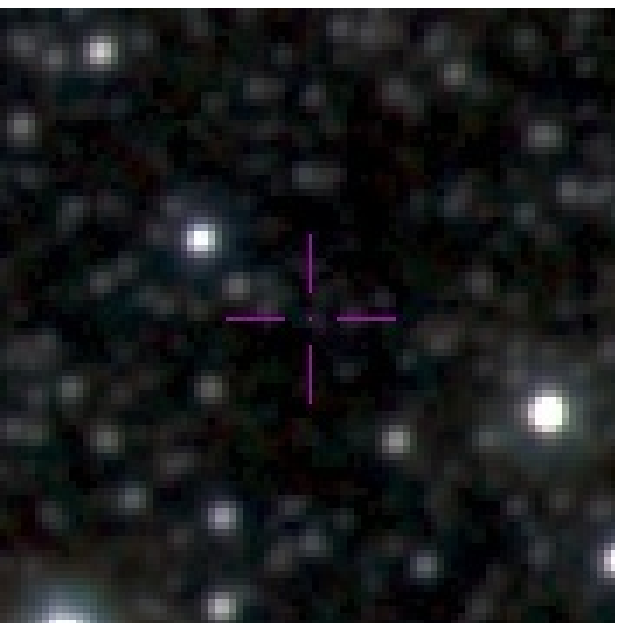}
                                                            
  \includegraphics[bb= -1.2cm   1.6cm   2cm  8cm, scale=0.65]{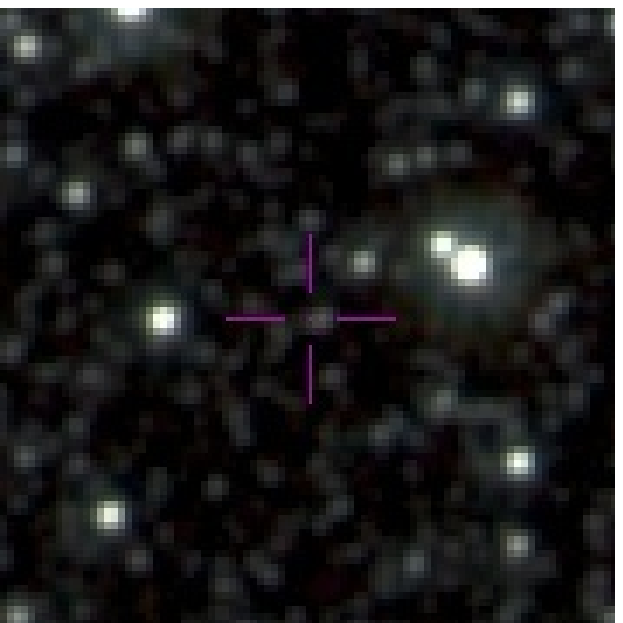}  
  \includegraphics[bb= -4.4cm   1.6cm   0cm  8cm, scale=0.65]{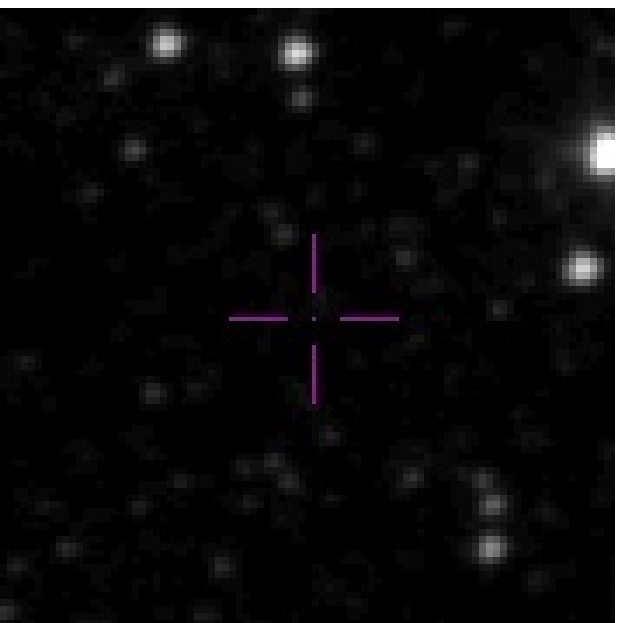}
  \includegraphics[bb= -6.4cm   1.6cm   0cm  8cm, scale=0.65]{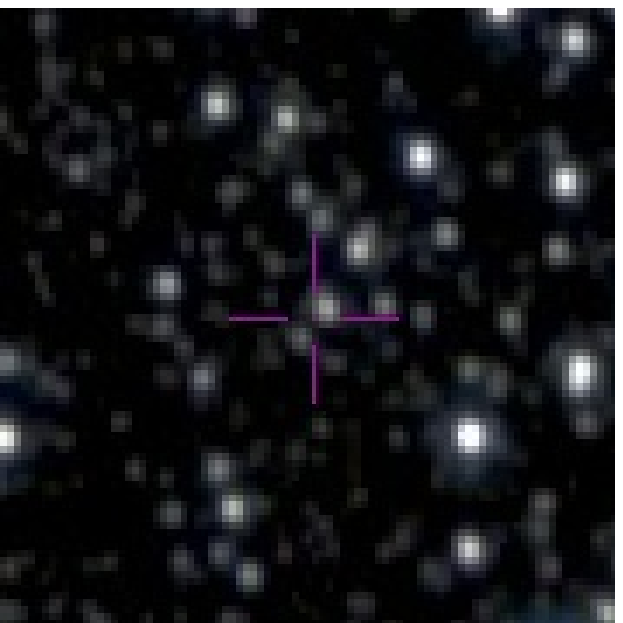}
  \includegraphics[bb= -6.4cm   1.6cm   0cm  8cm, scale=0.65]{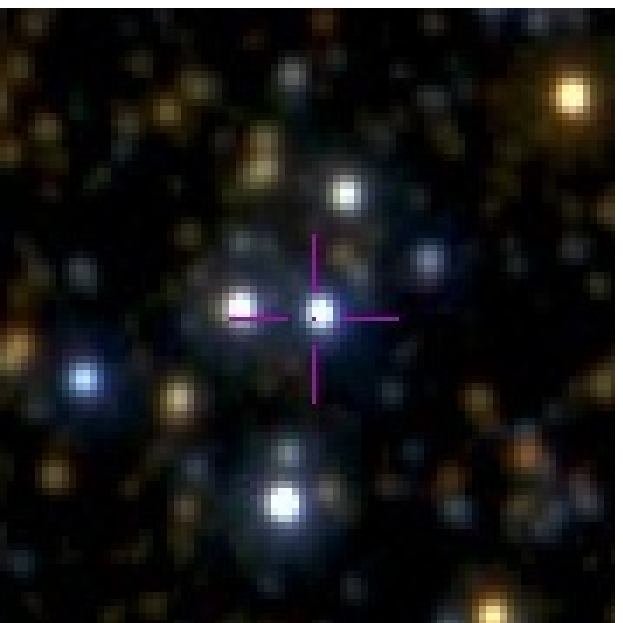}
                                                            
  \includegraphics[bb= -1.2cm   1.6cm   2cm  8cm, scale=0.65]{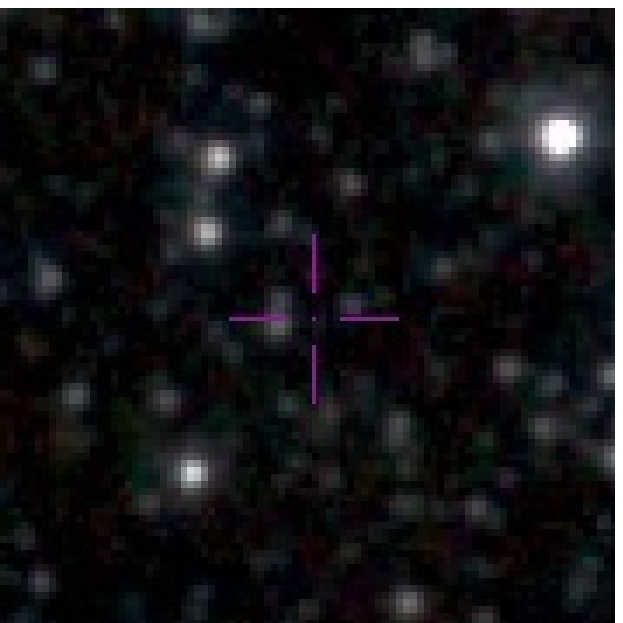}
  \includegraphics[bb= -4.4cm   1.6cm   0cm  8cm, scale=0.65]{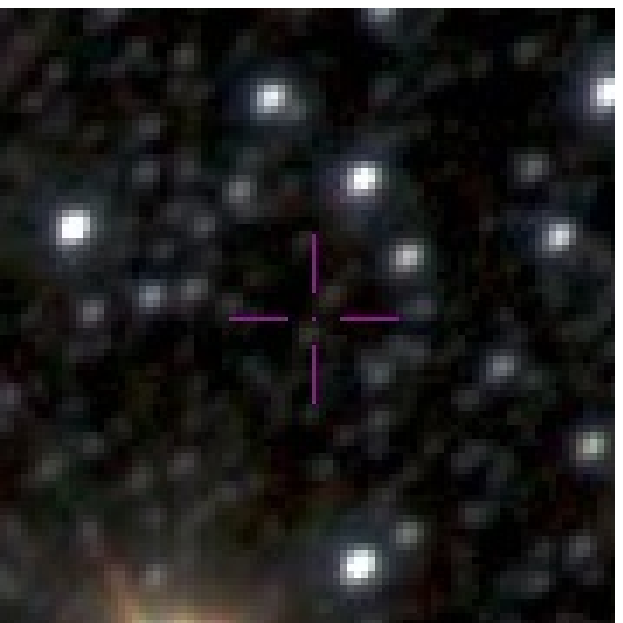} 
  \includegraphics[bb= -6.4cm   1.6cm   0cm  8cm, scale=0.65]{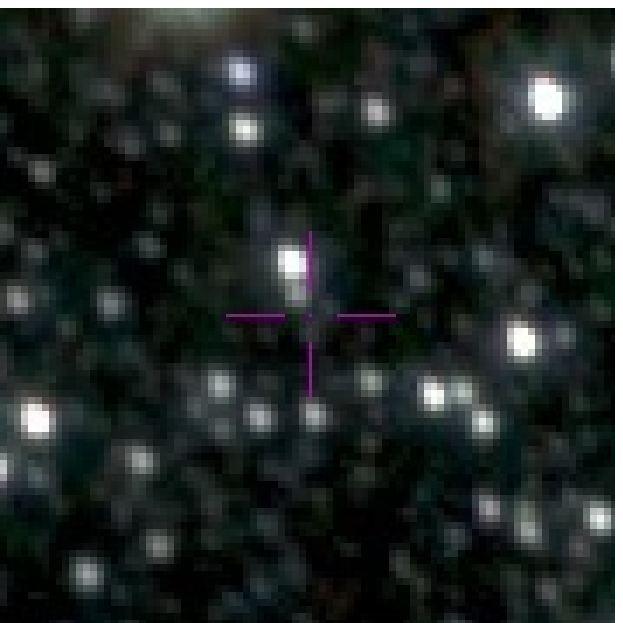}
  \includegraphics[bb= -6.4cm   1.6cm   0cm  8cm, scale=0.65]{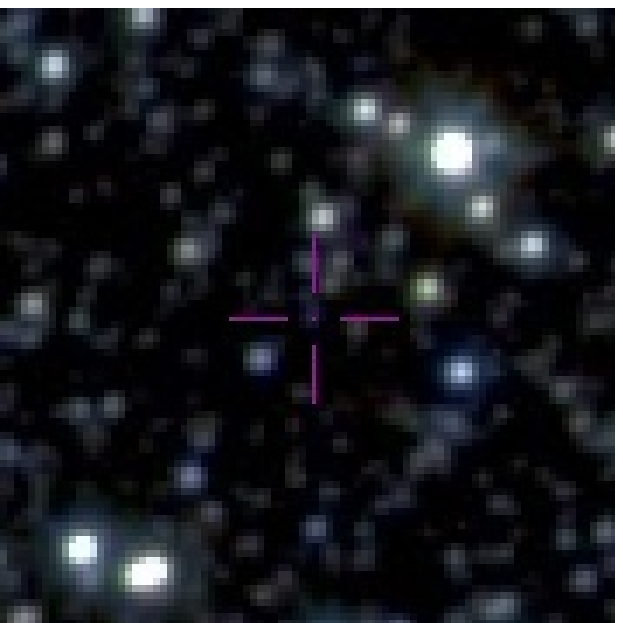} 
                                                            
  \includegraphics[bb= -1.2cm  -0.5cm   2cm  8cm, scale=0.65]{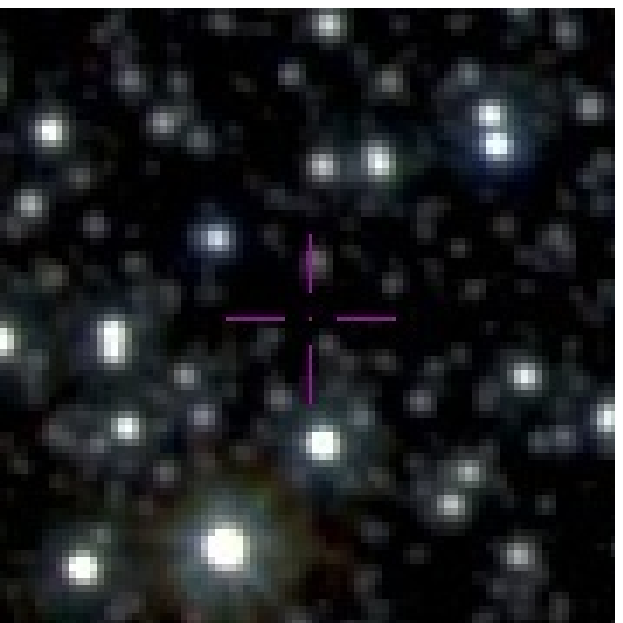} 
  \includegraphics[bb= -4.4cm  -0.5cm   0cm  8cm, scale=0.65]{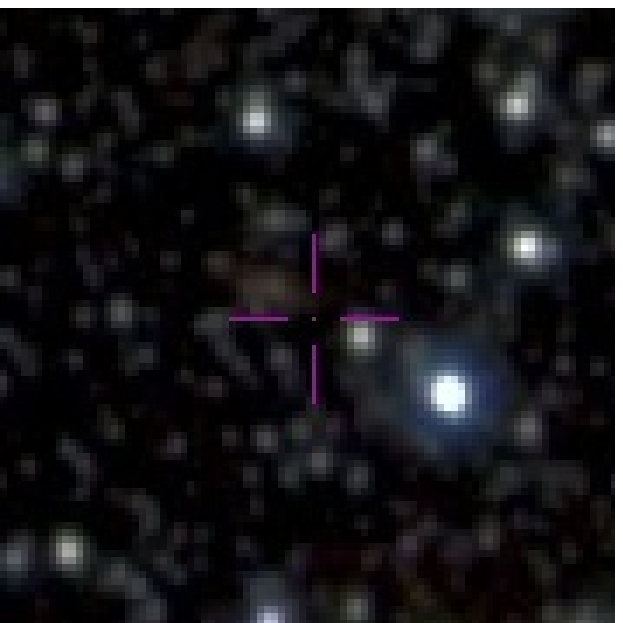} 
  \includegraphics[bb= -6.4cm  -0.5cm   0cm  8cm, scale=0.65]{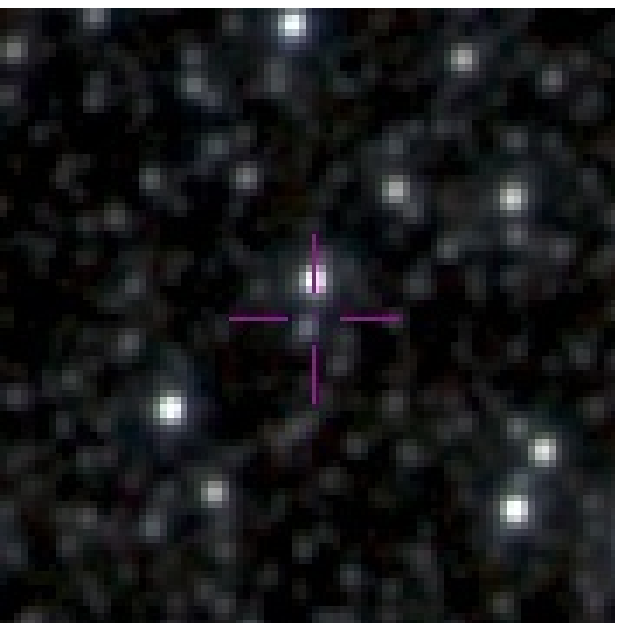}
  \includegraphics[bb= -6.4cm  -0.5cm   0cm  8cm, scale=0.65]{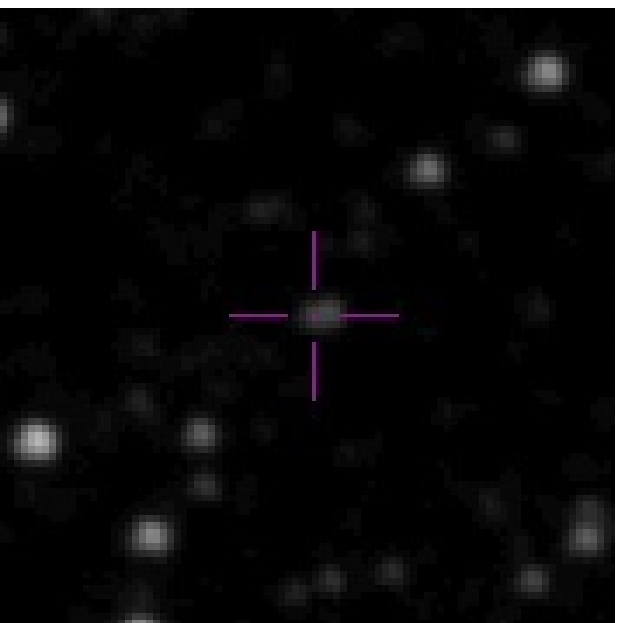}
\caption{{\bf cont.} First  row: Nova Sgr 1992a, Nova  Sgr 1992b, Nova
  Sgr 1992c  and Nova Sgr  1993. Second row:  Nova Cru 1996,  Nova Sco
  1997, Nova  Sco 1998 and  Nova Sgr 1998.  Third row: Nova  Sgr 1999,
  Nova Sgr 2000  ($K_{\rm s}$-band image), Nova Sco  2001 and Nova Sgr
  2001. Fourth row: Nova Sgr 2001b,  Nova Sgr 2001c, Nova Sgr 2002 and
  Nova Sgr 2002b.  Last row: Nova Sgr 2002d, Nova  Sgr 2003b, Nova Oph
  2004 and Nova Sco 2004 ($K_{\rm s}$-band image).}
\label{novae_05}
\end{figure*}

\addtocounter{figure}{-1}

\begin{figure*} 
  \includegraphics[bb= -1.2cm   1.6cm   2cm  8cm, scale=0.65]{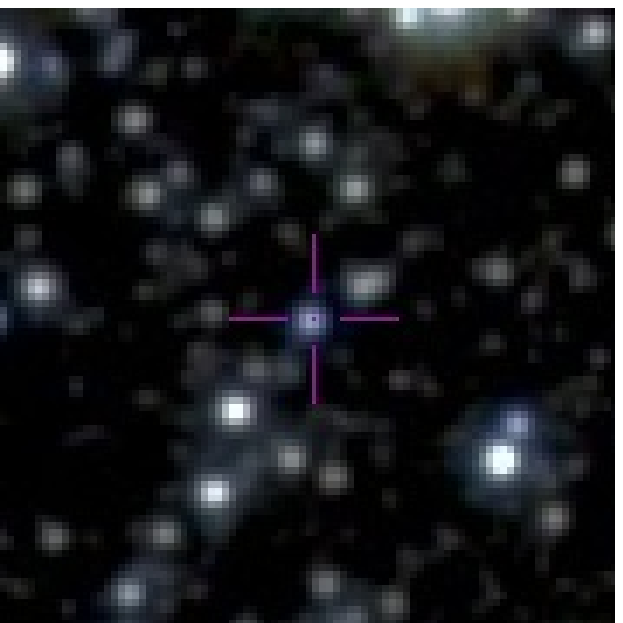} 
  \includegraphics[bb= -4.4cm   1.6cm   0cm  8cm, scale=0.65]{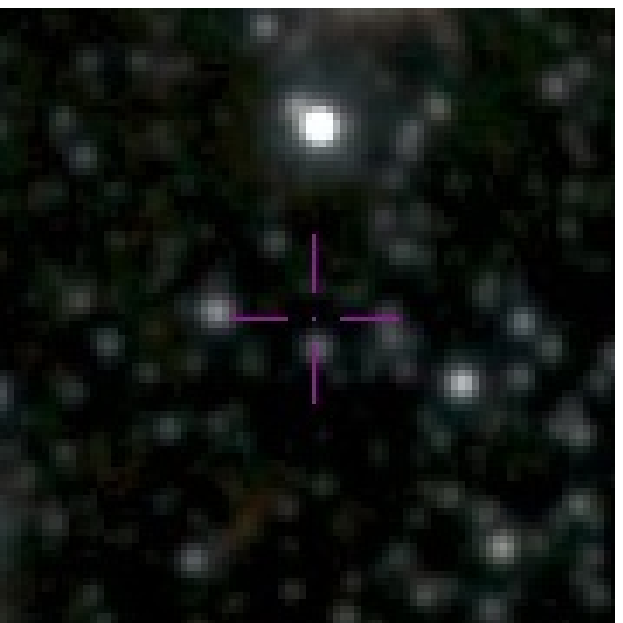}
  \includegraphics[bb= -6.4cm   1.6cm   0cm  8cm, scale=0.65]{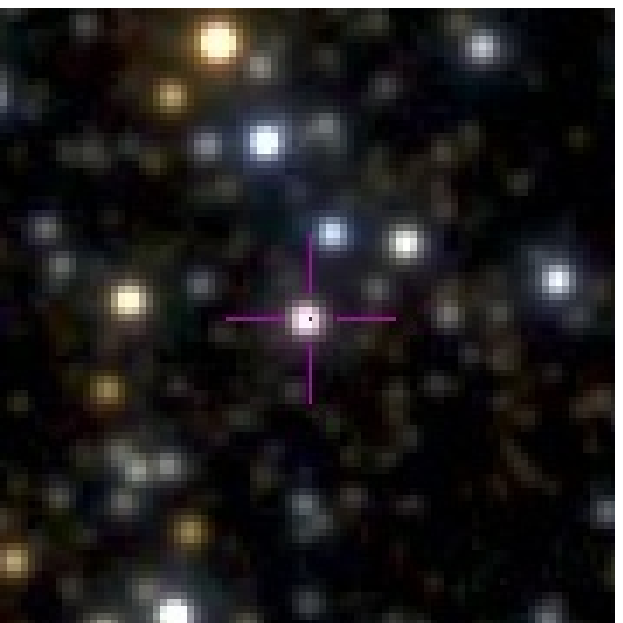}
  \includegraphics[bb= -6.4cm   1.6cm   0cm  8cm, scale=0.65]{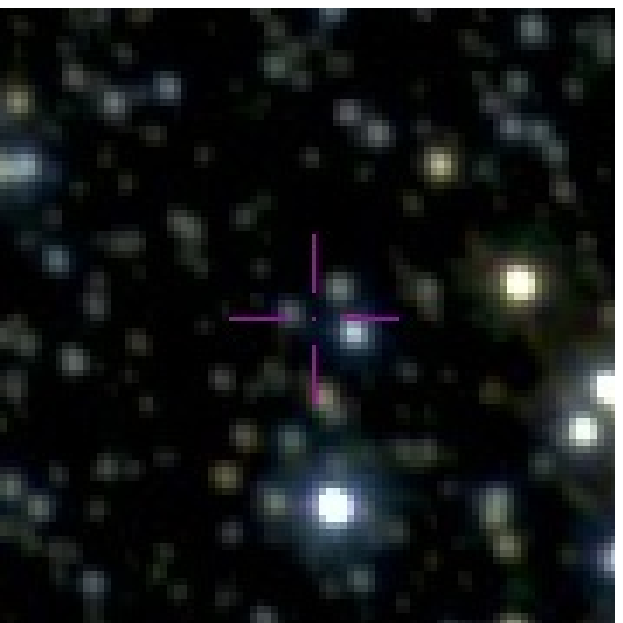} 
                                                            
  \includegraphics[bb= -1.2cm   1.6cm   2cm  8cm, scale=0.65]{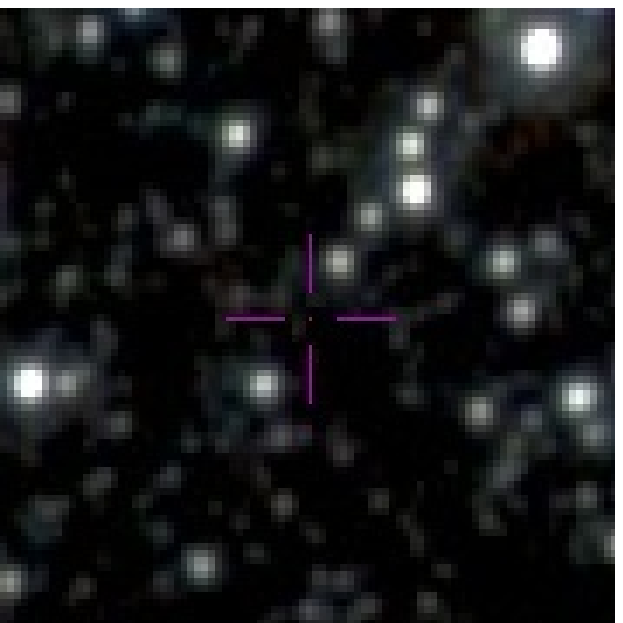} 
  \includegraphics[bb= -4.4cm   1.6cm   0cm  8cm, scale=0.65]{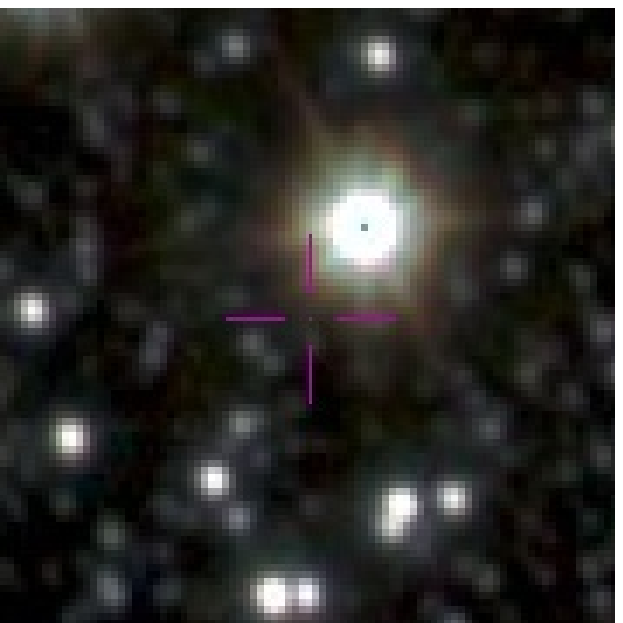}
  \includegraphics[bb= -6.4cm   1.6cm   0cm  8cm, scale=0.65]{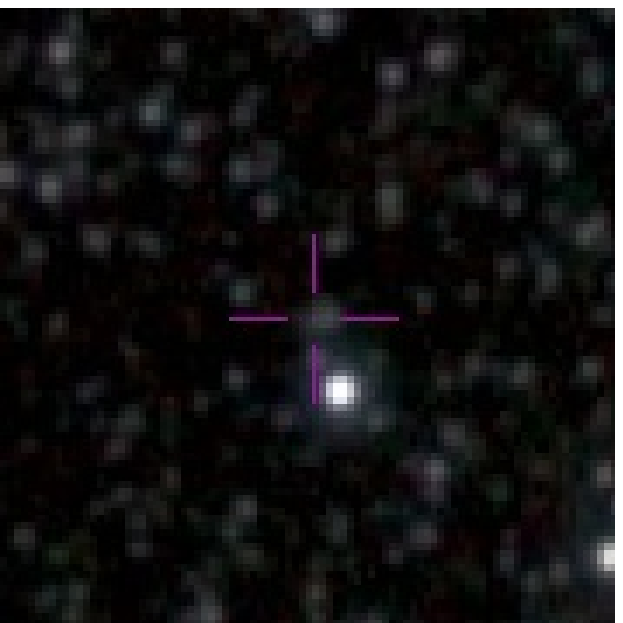}
  \includegraphics[bb= -6.4cm   1.6cm   0cm  8cm, scale=0.65]{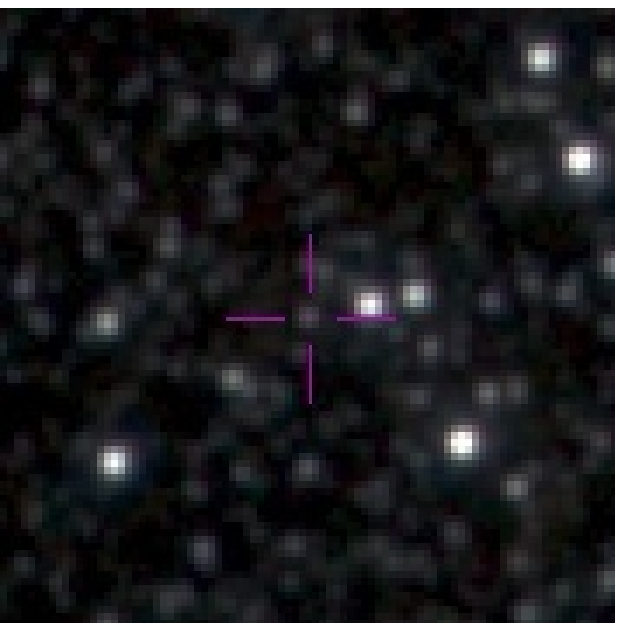} 
                                                            
  \includegraphics[bb= -1.2cm   1.6cm   2cm  8cm, scale=0.65]{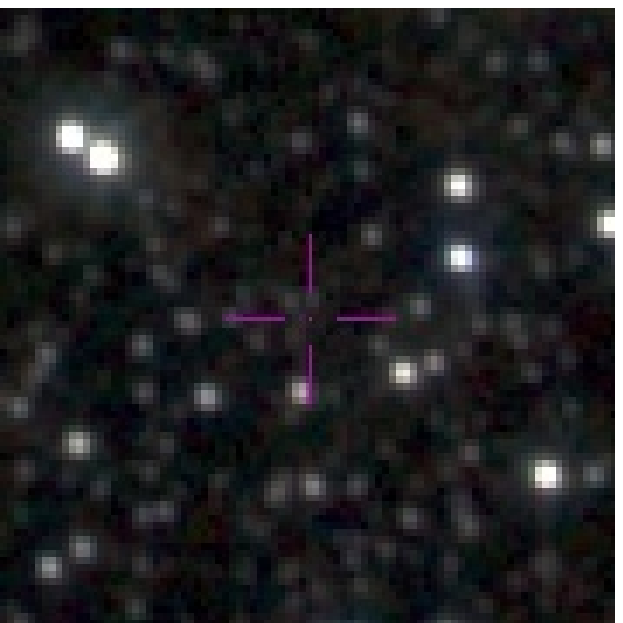}   
  \includegraphics[bb= -4.4cm   1.6cm   0cm  8cm, scale=0.65]{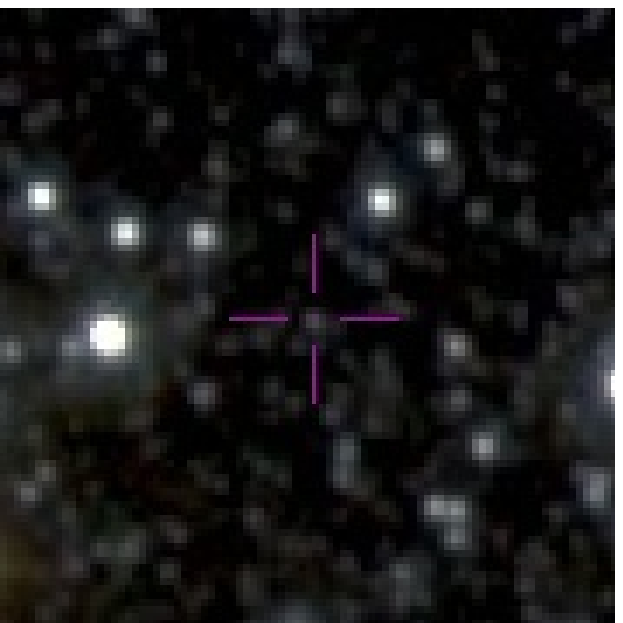}   
  \includegraphics[bb= -6.4cm   1.6cm   0cm  8cm, scale=0.65]{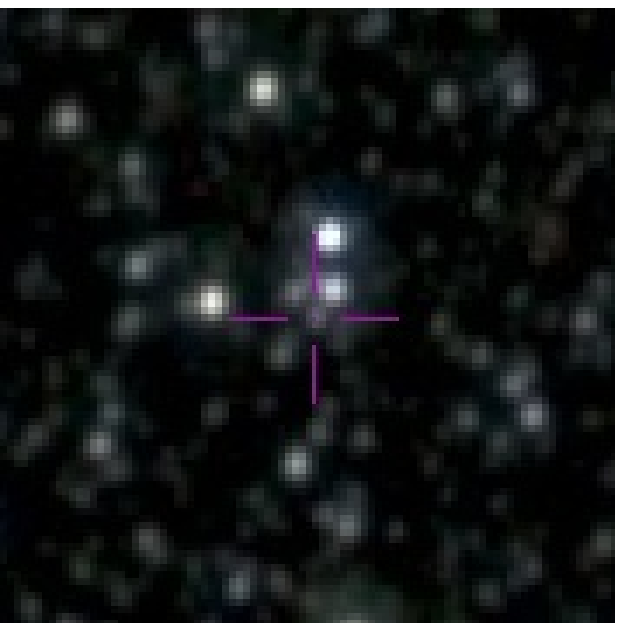}   
  \includegraphics[bb= -6.4cm   1.6cm   0cm  8cm, scale=0.65]{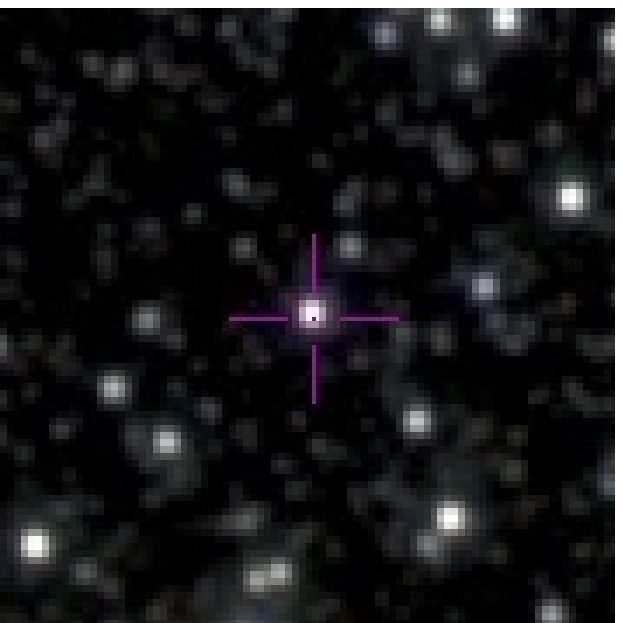}
                                                            
  \includegraphics[bb= -1.2cm   1.6cm   2cm  8cm, scale=0.65]{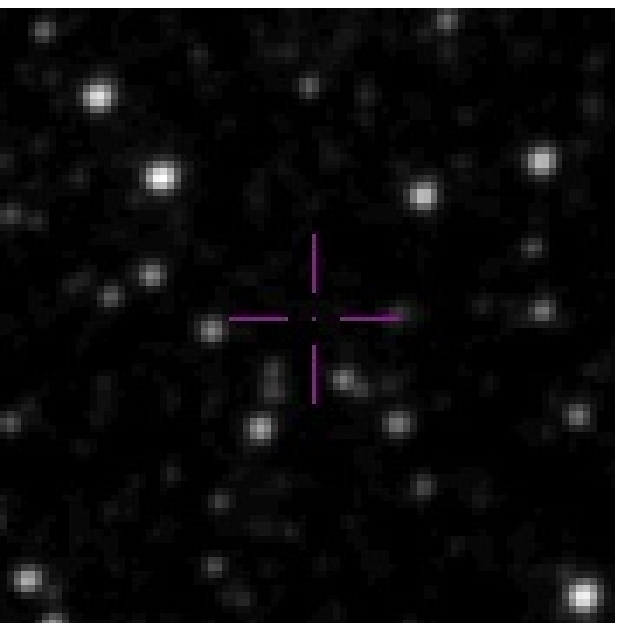}
  \includegraphics[bb= -4.4cm   1.6cm   0cm  8cm, scale=0.65]{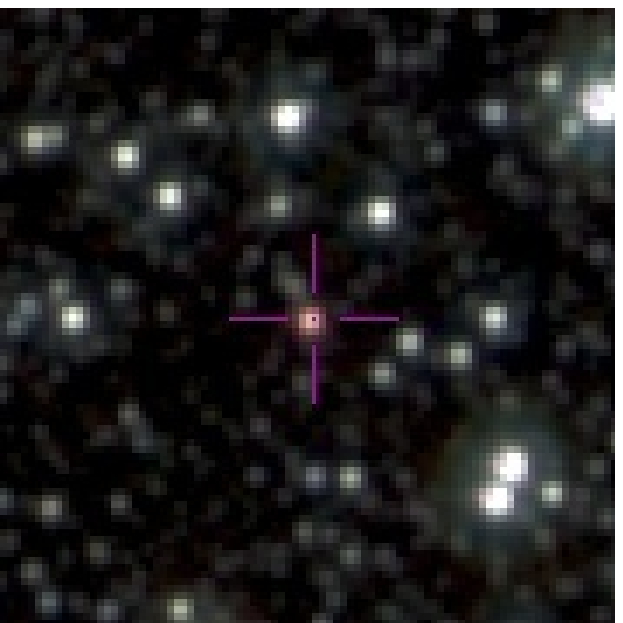}   
  \includegraphics[bb= -6.4cm   1.6cm   0cm  8cm, scale=0.65]{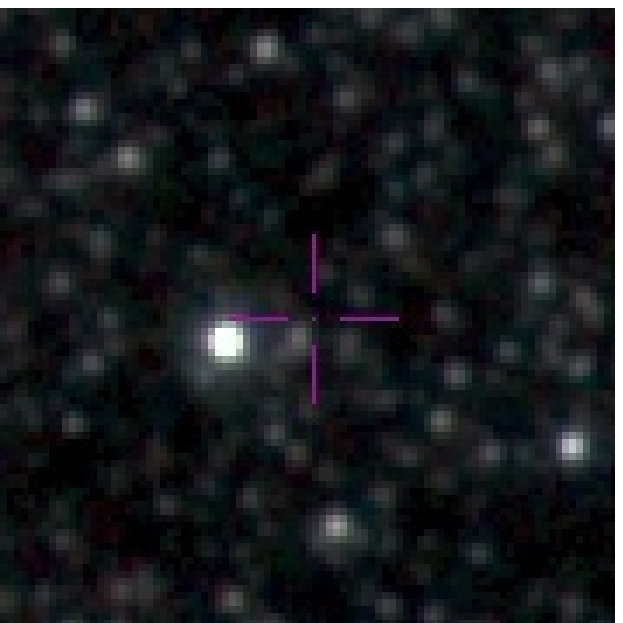}  
  \includegraphics[bb= -6.4cm   1.6cm   0cm  8cm, scale=0.65]{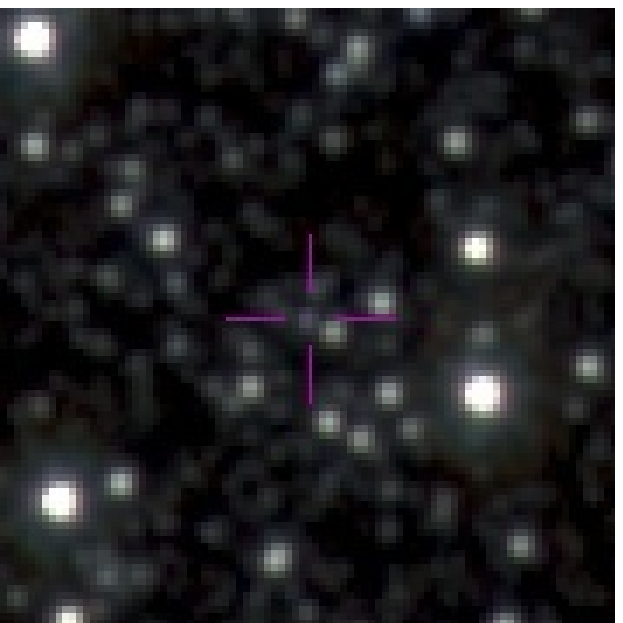} 
                                                            
  \includegraphics[bb= -1.2cm  -0.5cm   2cm  8cm, scale=0.65]{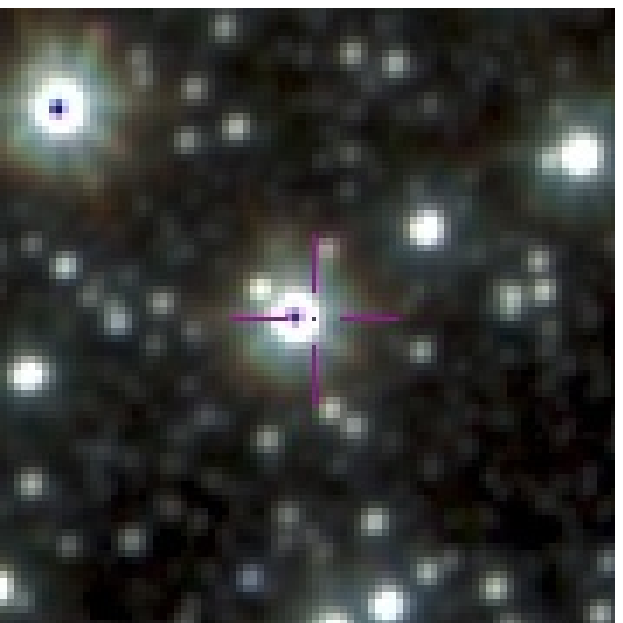}
  \includegraphics[bb= -4.4cm  -0.5cm   0cm  8cm, scale=0.65]{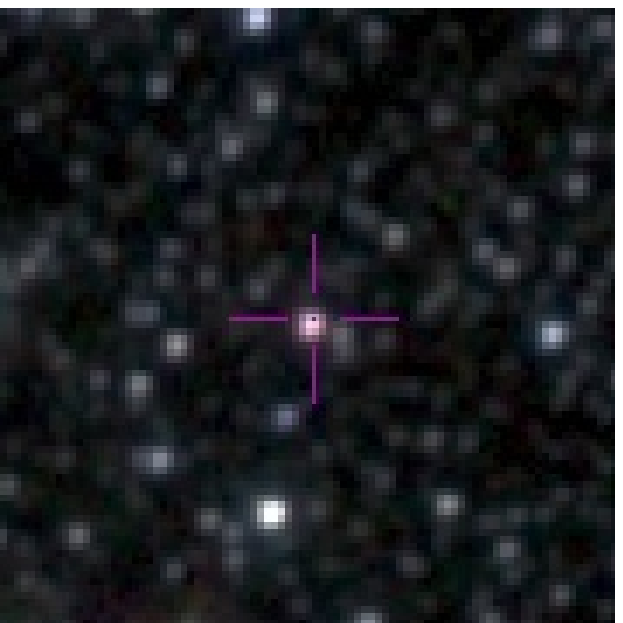} 
  \includegraphics[bb= -6.4cm  -0.5cm   0cm  8cm, scale=0.65]{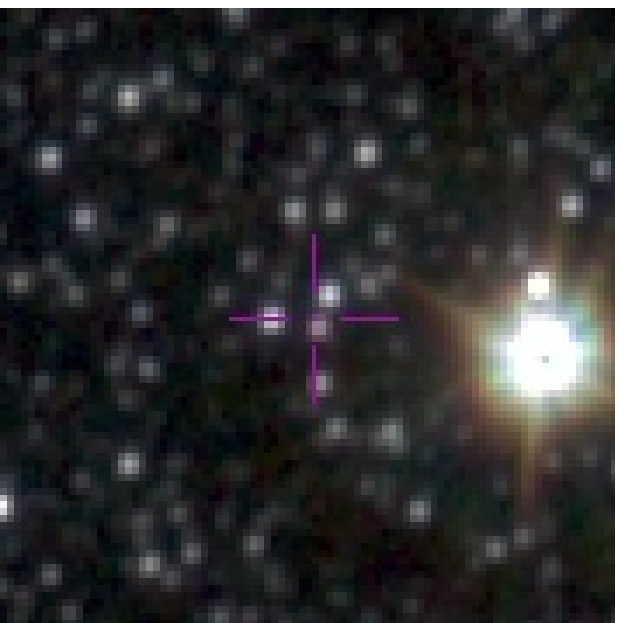} 
  \includegraphics[bb= -6.4cm  -0.5cm   0cm  8cm, scale=0.65]{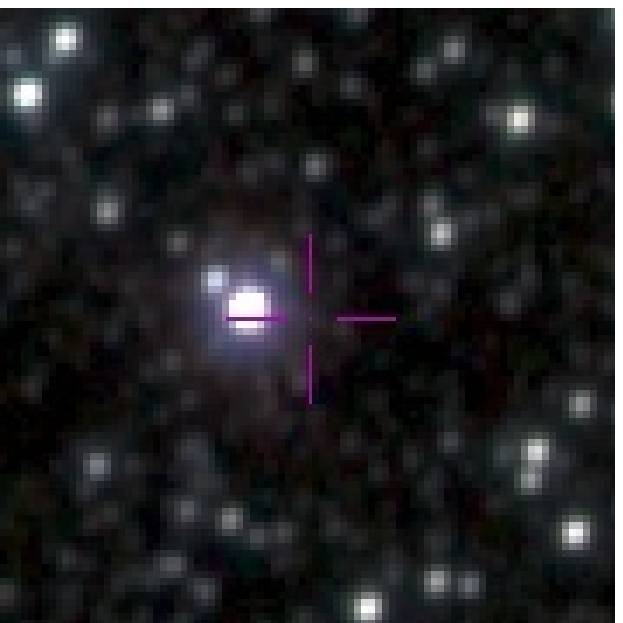} 
\caption{{\bf cont.}  First row: Nova  Sco 2004b, Nova Sgr  2004, Nova
  Cen 2005  and Nova  Nor 2005.  Second row: Nova  Sco 2005,  Nova Sgr
  2005a, Nova Sgr  2005b and Nova Oph 2006. Third  row: Nova Sgr 2006,
  Nova Oph  2007, Nova Nor 2007  and Nova Oph 2008a.  Fourth row: Nova
  Oph 2008b  ($K_{\rm s}$-band image),  Nova Sgr 2008, Nova  Sgr 2008b
  and Nova  Oph 2009. Last row:  Nova Sgr 2009a, Nova  Sgr 2009b, Nova
  Sgr 2009c and Nova Oph 2010.}
\label{novae_06}
\end{figure*}

\addtocounter{figure}{-1}

\begin{figure*}    
  \includegraphics[bb= -1.2cm   1.6cm   2cm  8cm, scale=0.65]{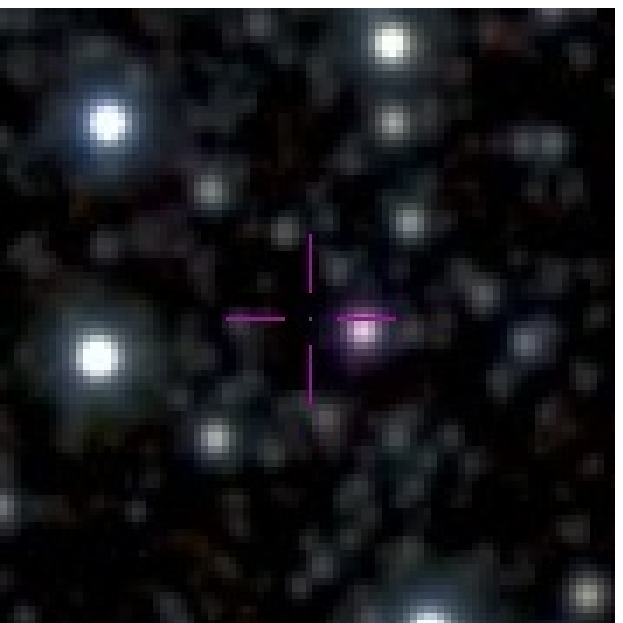} 
  \includegraphics[bb= -4.4cm   1.6cm   0cm  8cm, scale=0.65]{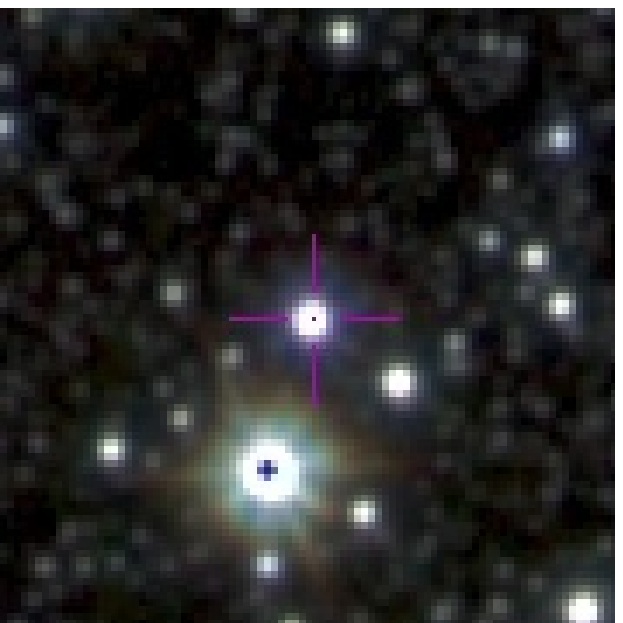} 
  \includegraphics[bb= -6.4cm   1.6cm   0cm  8cm, scale=0.65]{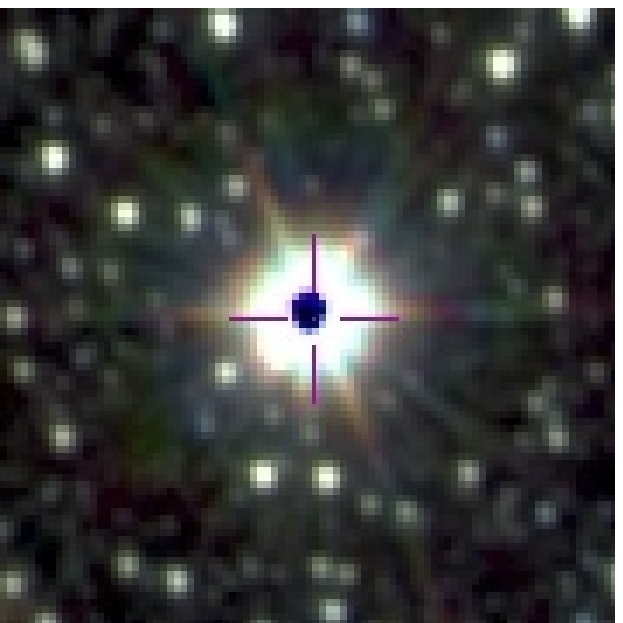} 
  \includegraphics[bb= -6.4cm   1.6cm   0cm  8cm, scale=0.65]{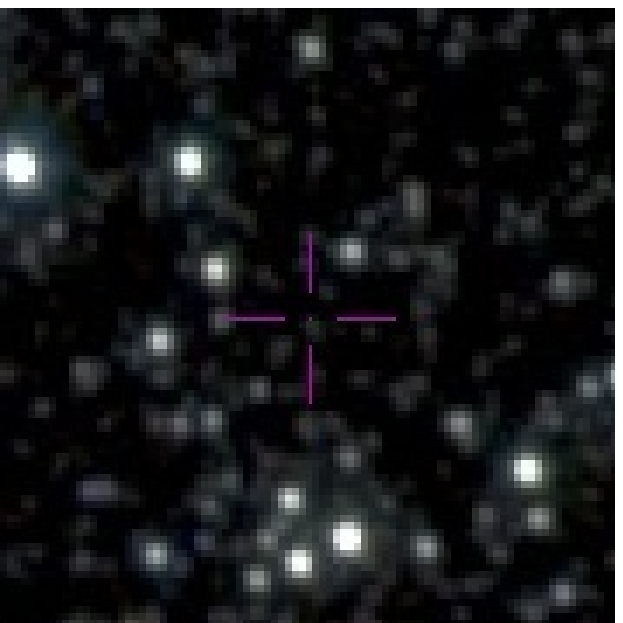}  
                                                             
  \includegraphics[bb= -1.2cm   1.6cm   2cm  8cm, scale=0.65]{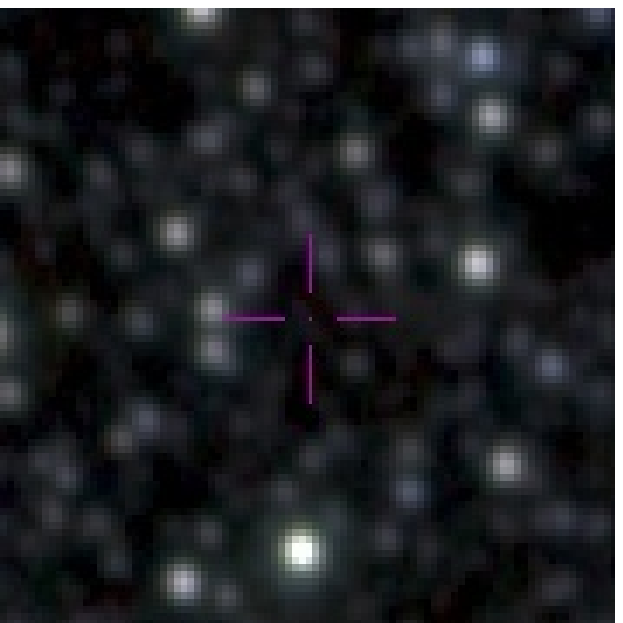} 
  \includegraphics[bb= -4.4cm   1.6cm   0cm  8cm, scale=0.65]{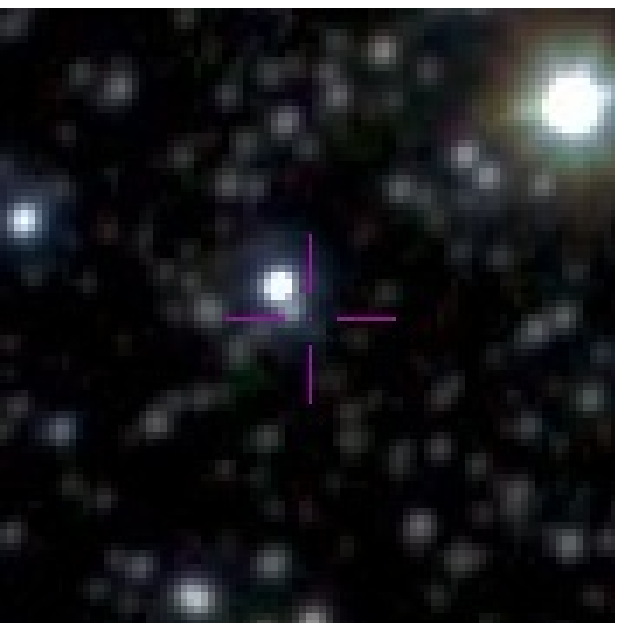} 
  \includegraphics[bb= -6.4cm   1.6cm   0cm  8cm, scale=0.65]{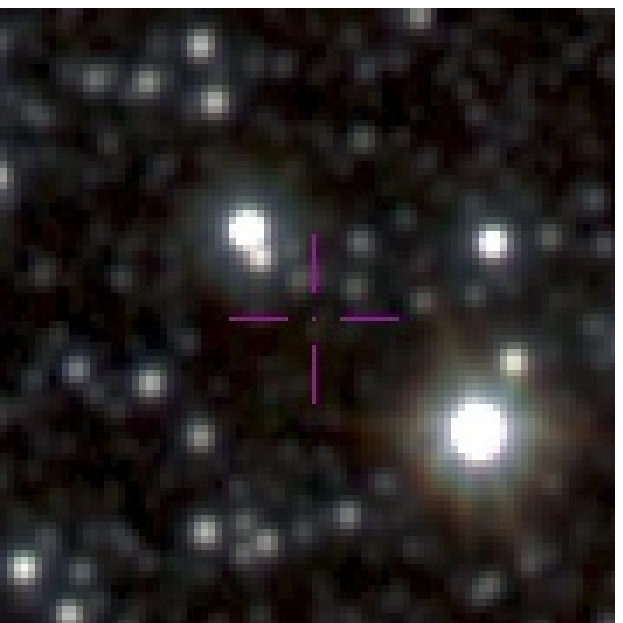} 
  \includegraphics[bb= -6.4cm   1.6cm   0cm  8cm, scale=0.65]{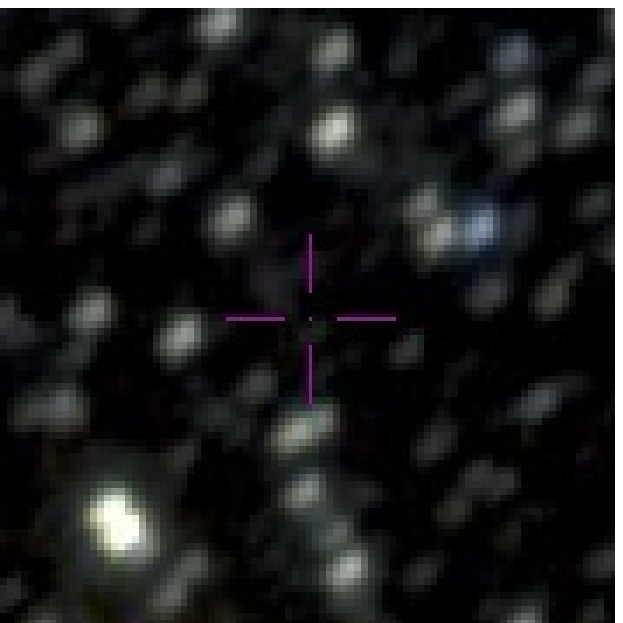}  
                                                             
  \includegraphics[bb= -1.2cm   1.6cm   2cm  8cm, scale=0.65]{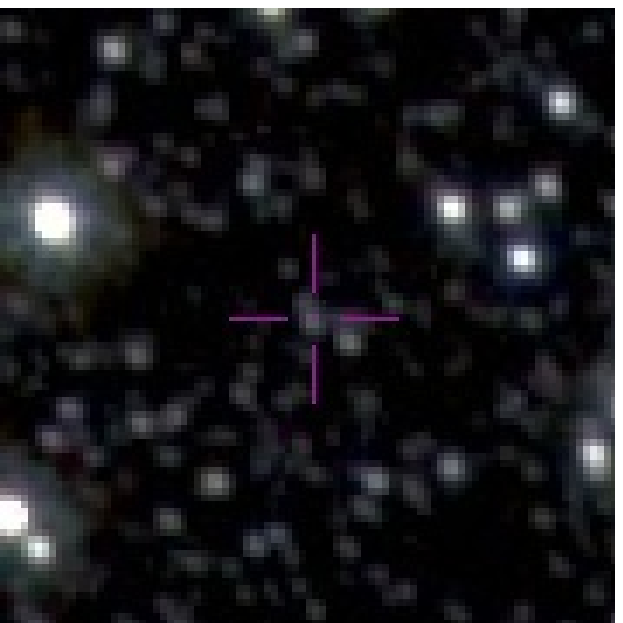}  
  \includegraphics[bb= -4.4cm   1.6cm   0cm  8cm, scale=0.65]{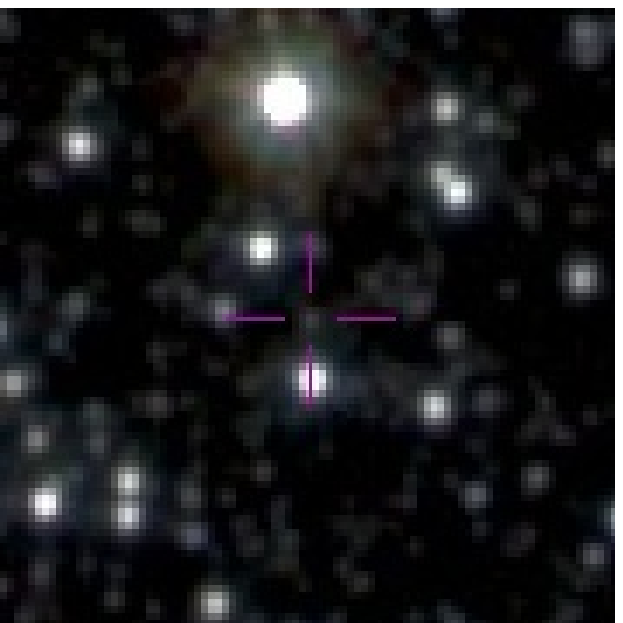}
  \includegraphics[bb= -6.4cm   1.6cm   0cm  8cm, scale=0.65]{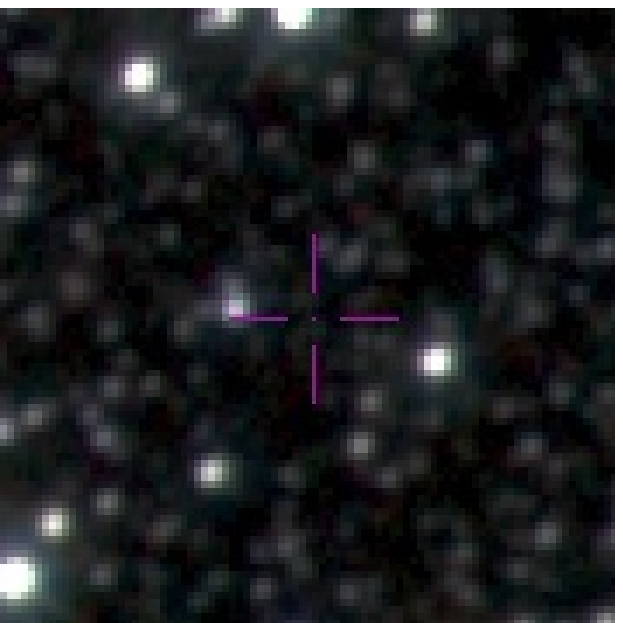} 
  \includegraphics[bb= -6.4cm   1.6cm   0cm  8cm, scale=0.65]{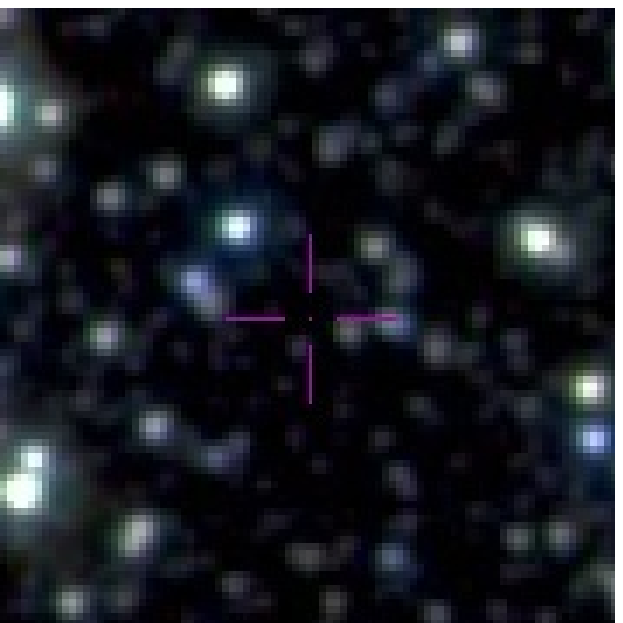}
                                                             
  \includegraphics[bb= -1.2cm   1.6cm   2cm  8cm, scale=0.65]{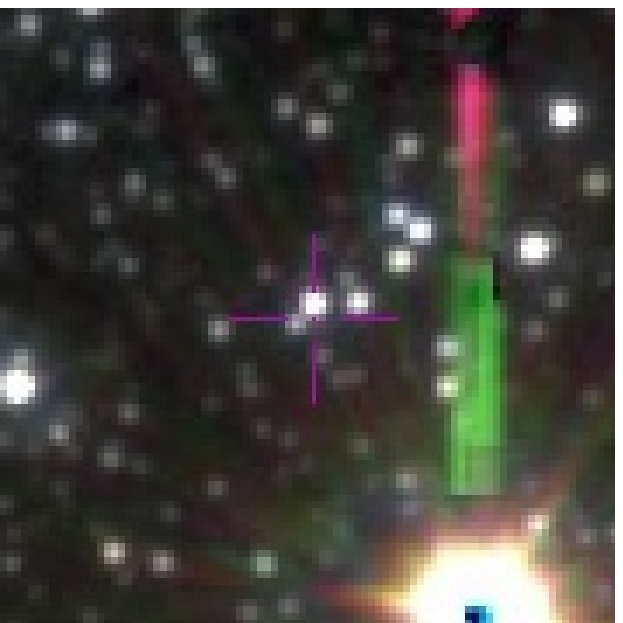}
  \includegraphics[bb= -4.4cm   1.6cm   0cm  8cm, scale=0.65]{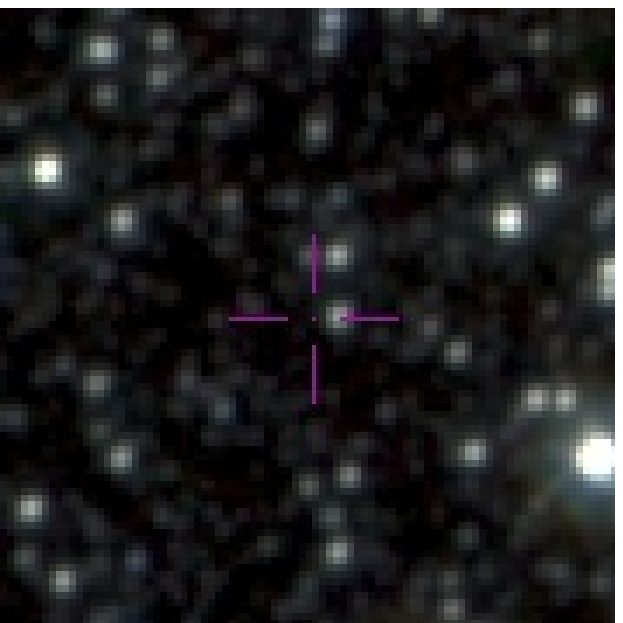}
  \includegraphics[bb= -6.4cm   1.6cm   0cm  8cm, scale=0.65]{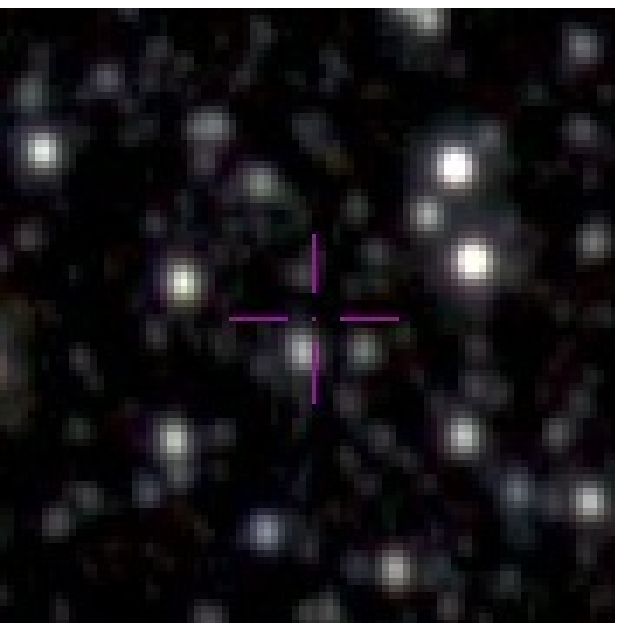} 
  \includegraphics[bb= -6.4cm   1.6cm   0cm  8cm, scale=0.65]{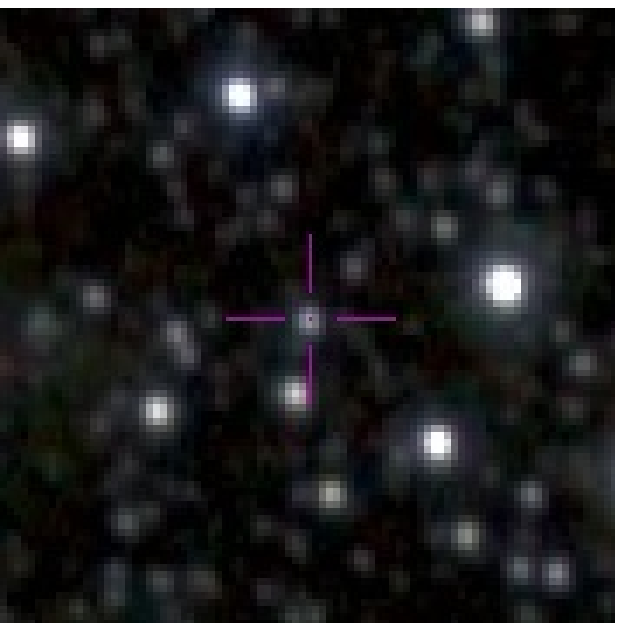}
                                                             
  \includegraphics[bb= -1.2cm  -0.5cm   2cm  8cm, scale=0.65]{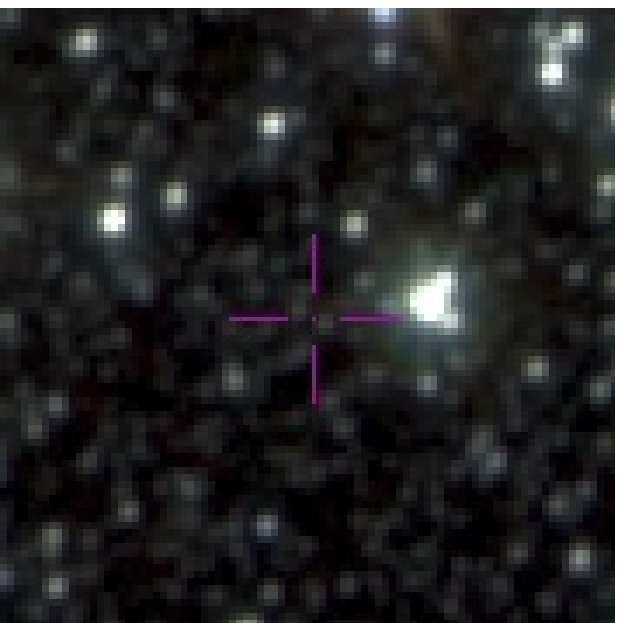}
  \includegraphics[bb= -4.4cm  -0.5cm   0cm  8cm, scale=0.65]{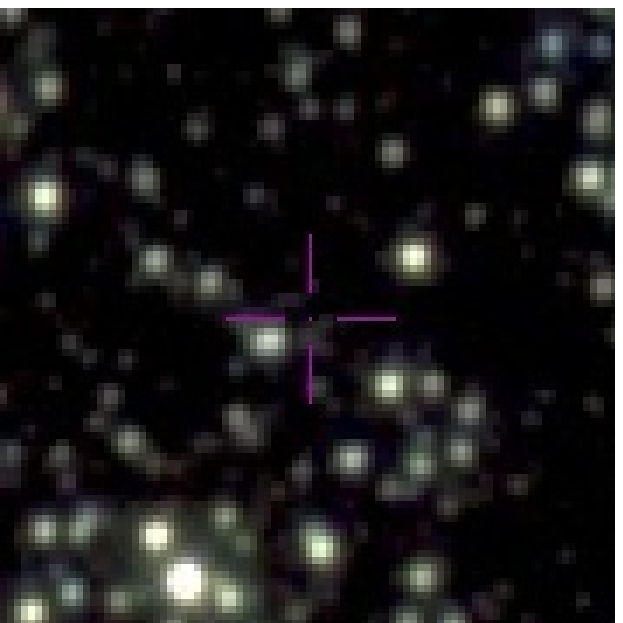}
  \includegraphics[bb= -6.4cm  -0.5cm   0cm  8cm, scale=0.65]{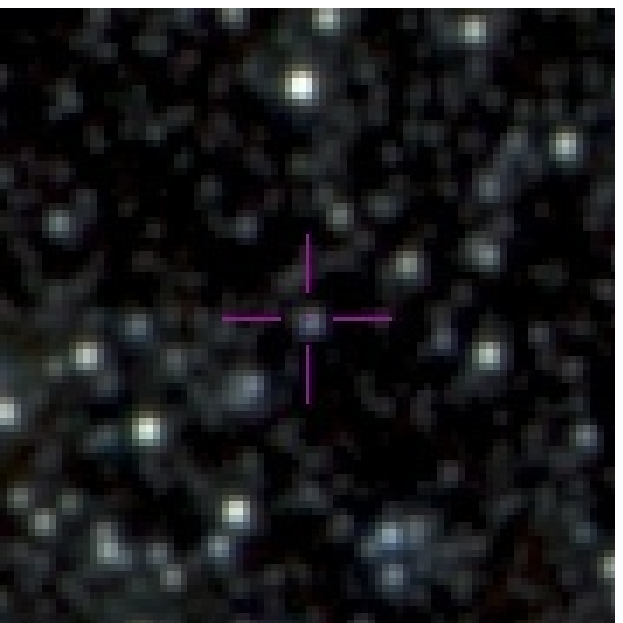}
  \includegraphics[bb= -6.4cm  -0.5cm   0cm  8cm, scale=0.65]{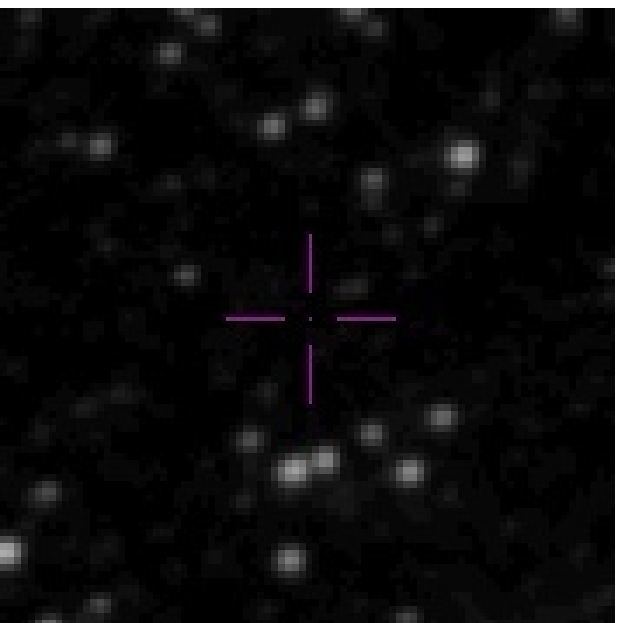}

\caption{{\bf cont.}  First  row: Nova Oph 2010b, Nova  Sgr 2010, Nova
  Sgr 2010b and  Nova Sgr 2011.  Second row: Nova  Sgr 2011b, Nova Cen
  2012, Nova  Oph 2012b and Nova  Sco 2012. Third row:  Nova Sgr 2012,
  Nova Sgt 2012c, Nova Sgt  2012d and OGLE-NOVA-001. Fourth row: V1213
  Cen, PN G002.6,  V0733 Sco and GR Sgr. Last row:  AT Sgr, V2859 Sco,
  KW2003 105 and V0729 Sco.}
\label{novae_07}
\end{figure*}

\end{appendix}

\end{document}